\documentclass[12pt,reprint,aps,groupedaddress,showpacs,nofootinbib,prx,twocolumn,superscriptaddress]{revtex4-2}

\usepackage{graphicx}
\usepackage{dcolumn}
\usepackage{bm}
\usepackage{amssymb}
\usepackage{pifont}
\usepackage{xcolor}
\usepackage{amsmath}
\usepackage{orcidlink}

\begin{document}

\title{Detectability of massive binary black holes with sub-mHz gravitational wave missions}

\author{Renjie Wang\,\orcidlink{0000-0002-9754-3225}}
\email{wangrenjie@ucas.ac.cn}
\affiliation{School of Fundamental Physics and Mathematical Sciences, Hangzhou Institute for Advanced Study, University of Chinese Academy of Sciences, Hangzhou 310024, China}

\author{Yumeng Xu\,\orcidlink{0000-0001-8697-3505}}
\email{yumeng.xu@uib.es}
\affiliation{Departament de F\'isica, Universitat de les Illes Balears, IAC3 -- IEEC, Crta. Valldemossa km 7.5, E-07122 Palma, Spain}

\author{Gang Wang\,\orcidlink{0000-0002-9668-8772}}
\email[Corresponding author: ]{gwanggw@gmail.com}
\affiliation{Institute of Fundamental  Physics and Quantum Technology, Ningbo University, Ningbo, 315211, China}
\affiliation{Shanghai Astronomical Observatory, Chinese Academy of Sciences, Shanghai, 200030, China}

\author{Bin Hu\,\orcidlink{0000-0001-5093-8118}}
\email[Corresponding author: ]{bhu@bnu.edu.cn}
\affiliation{Institute for Frontier in Astronomy and Astrophysics, Beijing Normal University, Beijing, 102206, China}
\affiliation{Department of Astronomy, Beijing Normal University, Beijing 100875, China}

\author{Rong-Gen Cai\,\orcidlink{0000-0002-3539-7103}}
\email[Corresponding author: ]{caironggen@nbu.edu.cn}
\affiliation{Institute of Fundamental  Physics and Quantum Technology, Ningbo University, Ningbo, 315211, China}

\begin{abstract}

Future space-based gravitational wave (GW) observatories beyond LISA aim to explore the sub-millihertz to microhertz frequency band, targeting massive binary black hole (MBBH) coalescences across cosmic distances. In this work, we evaluate the detection and localization capabilities of such missions, taking into account the full galactic foreground. We show that signal-to-noise ratios (SNRs) can reach several thousand over a wide redshift range. Using ASTROD-GW as a representative mission concept, we consider three orbital configurations---non-precessing and precessing with varying inclinations---and assess their performance across different MBBH populations. In the absence of precession, we find that the dominant quadrupole mode alone results in a two-hemisphere degeneracy, which is resolved by including higher-order harmonics. These harmonics equalize localization performance across configurations, reducing the relative advantage of orbital precession. Operating in the 10\,$\mu$Hz--10\,mHz band, sub-mHz detectors partially overlap with LISA’s sensitivity while extending significantly into lower frequencies, enabling more precise localization of high-mass MBBHs. The resulting high SNRs open new opportunities for multi-messenger astronomy, cosmological inference, and tests of gravitational theory. Our results highlight the complementary strengths of sub-mHz missions in advancing the next frontier of GW astrophysics.

\end{abstract}

\maketitle

\section{Introduction}

The direct detection of gravitational waves (GW) by LIGO marked the beginning of GW astronomy \cite{LIGOScientific:2016aoc}. During the O1-O3 observing runs, the advanced LIGO and advanced Virgo observed several dozen compact binary coalescence (CBC) events \cite{LIGOScientific:2018mvr,LIGOScientific:2020ibl,LIGOScientific:2021djp}. Since 2023, KAGRA \cite{KAGRA:2018plz} has joined the global GW detection network, and the LIGO-Virgo-KAGRA (LVK) collaboration has reported more than two hundreds GW events during O4 \cite{gracedb}. Theses ground-based laser interferometers, along with future third-generation detectors such as the Einstein Telescope \cite{Punturo:2010zz} and Cosmic Explorer \cite{Reitze:2019iox}, are designed to observe GWs in the high-frequency band (few Hz to kHz).

Space-borne GW observatories, such as LISA \cite{LISA:2017pwj}, Taiji \cite{Hu:2017mde} and TianQin \cite{Luo:2020bls} will target the millihertz (mHz) band, enabling the detection of sources inaccessible to ground-based instruments. These include massive binary black holes (MBBHs) \cite[and reference therein]{Barausse:2020kjy}, extreme mass-ratio inspirals (EMRIs) \cite[and reference therein]{Babak:2017tow} and stellar mass compact binaries \cite[and reference therein]{Littenberg:2020bxy}, among others. In particular, MBBHs are among the primary target of the space-based GW mission due to their significance in probing galaxy evolution, black hole formation, and cosmology \cite{Schutz:1986gp,Zhu:2021bpp,Yang:2020yoc,Holz:2005df}. At even lower frequencies, in the nanohertz regime, GW signals are expected from the inspiral of supermassive black hole binaries. Detection in this band is pursued by pulsar timing arrays (PTAs), which observe the stochastic GW background by monitoring the correlated timing residuals of millisecond pulsars. Recent results from collaborations, including NANOGrav \cite{NANOGrav:2023gor}, EPTA \cite{EPTA:2023fyk}, PPTA \cite{Reardon:2023gzh}, and CPTA \cite{Xu:2023wog}, have provided compelling evidence for such a background, marking a milestone in low-frequency GW astronomy.

To bridge the gap between the mHz and nanohertz bands, several mission concepts have been proposed to explore the sub-mHz to $\mu$Hz range. These include ASTROD-GW \cite{Ni:2012eh,Ni:2016wcv}, Folkner mission \cite{Baker:2019pnp}, LISAmax \cite{Martens:2023mgm}, and $\mu$Ares \cite{Sesana:2019vho}, as well as Super-ASTROD \cite{Ni:2008bj}. These missions envision triangular interferometers formed by spacecraft (S/C) in heliocentric orbits (e.g., $\sqrt{3}$ AU baselines) or Martian/Jovian orbits to achieve longer interferometric arms and access lower GW frequencies. Such designs are expected to deliver high signal-to-noise ratios (SNRs)—in some cases exceeding $10^4$—for MBBH mergers, enabling high-precision measurements of source parameters and tests of gravitational theory.
However, GW detection in the low-frequency band is challenged by the foreground generated by galactic binaries, particularly double white dwarfs (DWDs) in early inspiral phases emitting below $\sim$10 mHz. LISA is expected to individually resolve thousands to tens of thousands of these systems during its mission lifetime \cite{Vecchio:2004ec,Nissanke:2012eh,Cornish:2017vip,Korol:2017qcx,LISA:2022yao,Korol:2018ulo}. The unresolved portion of these binaries forms a confusion foreground, especially around $\sim$1 mHz, potentially dominating instrumental noise \cite{Nelemans:2001hp}. For sub-mHz missions, the impact of this foreground becomes even more significant, potentially exceeding the instrumental noise by several orders of magnitude \cite{Baker:2019pnp,Wang:2023jct} and substantially affecting GW observability.

In this work, we select ASTROD-GW as a representative sub-mHz mission to investigate the detectability and localization of MBBH coalescences, accounting for the effects of the galactic foreground. We first evaluate the SNRs of equal-mass MBBHs across different distances. Then, we assess the sky localization capability using three orbital configurations: non-precessing, and precessing orbits inclined by $1^\circ$ and $3^\circ$ relative to the ecliptic plane. We consider three source scenarios with varying total masses and explore the effects of waveform content and mission configuration on parameter estimation. Specifically, we analyze (i) detection with only the dominant (2,2) mode, (ii) detection including higher harmonic modes [(2,1), (3,3), (3,2), (4,4)], and (iii) joint observation with two identical detectors. Our results show that when only the (2,2) mode is used, the non-precessing orbit leads to degeneracies in sky location, inclination, and polarization angle. These degeneracies can be broken either by introducing orbital precession or by including higher-order modes. When higher-order modes are present, the localization performance is significantly improved, especially for sources near the ecliptic plane, and the differences between orbital configurations become negligible. Furthermore, joint observations with two detectors can further enhance detectability, depending on the mass of the binary. Compared to LISA, which has limited localization capability in the mHz band, sub-mHz missions with longer baselines offer superior directional sensitivity to high-mass MBBHs.

This paper is organized as follows. In Sec. \ref{sec:mission}, we describe the ASTROD-GW orbital configurations considered, including both non-precessing and precessing cases. We discuss the detector response using time-delay interferometry (TDI) and evaluate the averaged sensitivity curves including galactic foreground noise. In Sec. \ref{sec:detectability}, we assess SNR distributions for MBBH systems and demonstrate how higher-order GW modes and precessing orbits help break parameter degeneracies. Sec. \ref{sec:localization} presents our sky localization analysis for three representative MBBH mergers and discusses the roles of orbital design and harmonic content. We conclude and summarize our findings in Sec. \ref{sec:conclusion}.

\section{SUB-MILLIHZ MISSIONS} \label{sec:mission}

\subsection{Orbit Configurations} \label{subsec:orbit}

\begin{figure}[htbp]
    \centering
    \includegraphics[width=0.7\linewidth]{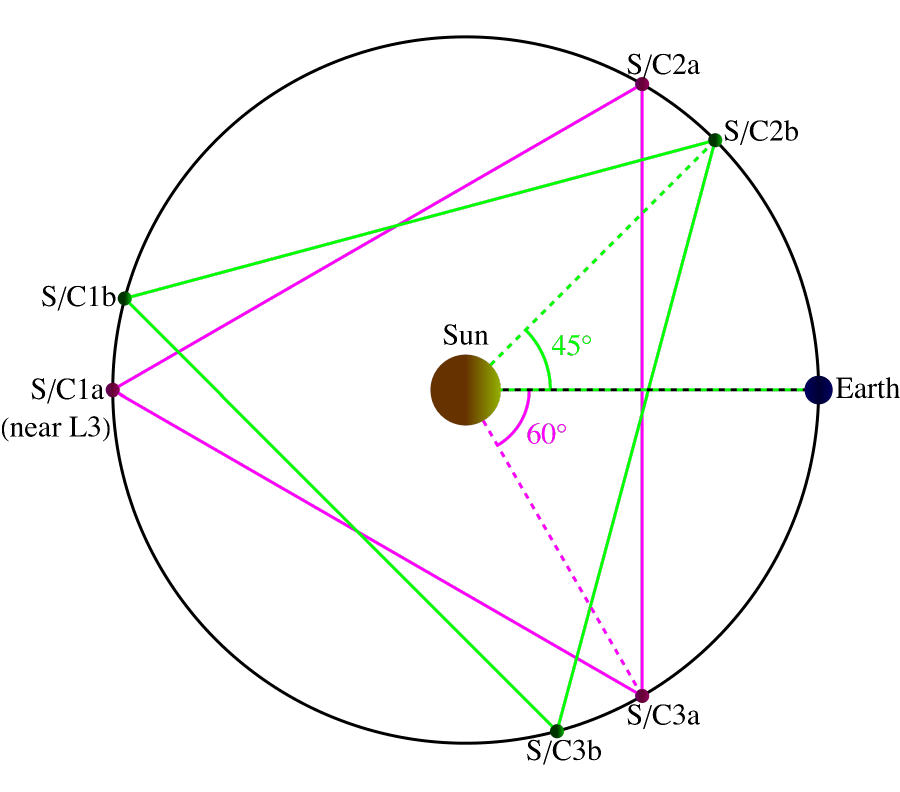}
    \includegraphics[width=0.84\linewidth]{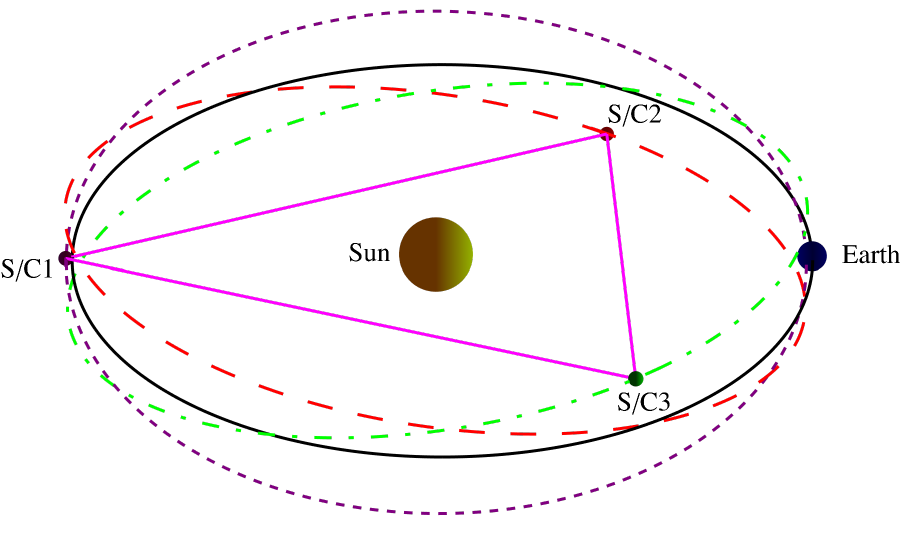}
    \caption{The orbital configurations of sub-mHz missions. The upper diagram shows ASTROD-GW mission (magenta) and Folkner mission (green) without inclination, the lower plot shows the orbit of ASTROD-GW with inclination and precession with respect to ecliptic plane.}
    \label{fig:orbits}
\end{figure}

The baseline configuration of ASTROD-GW consists of three spacecraft positioned near the Sun-Earth Lagrange points $L_3$, $L_4$ and $L_5$ \cite{MEN:2010434,WANG:2012211,Ni:2012eh,Ni:2016wcv}.
Three spacecraft follow heliocentric orbits, forming a nearly equilateral triangular interferometer with arm lengths of approximately $2.6\times 10^8$ km. The orbits lie nearly in the ecliptic plane and are expected to be dynamically stable, with arm-length variations of less than $\sim 0.04\%$ \cite{MEN:2010434,Wang:2011tlj}. 
Owing to the gravitational stability of the Sun-Earth Lagrange points, the constellation can maintain its configuration for over 10 years \cite{Wang:2012fqs,Wang:2014uaw}. Such long-term stability and near-equilateral arm geometry are favorable for implementing time-delay interferometry (TDI), which is essential for suppressing laser frequency noise in space-based GW detection.
In this study, we consider four mission orbit configurations:
\begin{itemize}
    \item \textbf{Orbit 0a}: A non-inclined configuration where the three spacecraft are located near $L_3, L_4$, and $L_5$, forming a triangle in the ecliptic plane (illustrated as the magenta triangle in the upper panel of Fig. \ref{fig:orbits}).
    \item \textbf{Orbit 0b}: Similar to Orbit 0a, but with the constellation rotated by $15^\circ$ such that the angle between the nearest spacecraft, the Sun, and the Earth is $45^\circ$ \cite{Baker:2019pnp} (shown as the green triangle in the upper panel of Fig. \ref{fig:orbits}).
    \item \textbf{Orbit 1}: A mildly inclined configuration, with a $1^\circ$ inclination of the constellation plane relative to the ecliptic, achieved by tilting the orbits of the three spacecraft around $L_3, L_4$, and $L_5$ toward the ecliptic pole \cite{Wang:2014uaw} (depicted in the lower panel of Fig. \ref{fig:orbits}).
    \item \textbf{Orbit 3}: A configuration similar to Orbit 1, but with a higher inclination of $3^\circ$ relative to the ecliptic plane \cite{Wang:2014uaw}.
\end{itemize}

The non-inclined ASTROD-GW configuration follows an orbit that lies in the ecliptic plane. In the spherical coordinates $(r,\theta,\varphi)$, the circular orbit of each spacecraft can be described as:
\begin{equation}
    r=a,\ \theta=90^{\circ},\ \varphi = \omega t+\varphi_0,
\end{equation}
where $a=1\ \rm{AU}$ is orbital radius, $\omega=2\pi/T$ is angular velocity, $T$ is one sidereal year, and $\varphi_0= [0, 2\pi/3, -2\pi/3]$ represents the initial phases offsets of the three S/C, respectively. In Cartesian coordinates, the spacecraft position are given by:
\begin{equation}
\begin{bmatrix}
x_{\mathrm{S/C}k} \\
y_{\mathrm{S/C}k} \\
z_{\mathrm{S/C}k}
\end{bmatrix} 
= a
\begin{bmatrix}
\cos \varphi_k \\
\sin \varphi_k \\
0
\end{bmatrix}.
\end{equation}
In practice, however, the orbits will deviate from perfect circles. More realistic configurations must be obtained through numerical ephemeris-based optimization, as specified in \cite{MEN:2010434,WANG:2012211,Wang:2012te}.

For the orbit with a small inclination angle $\theta_\mathrm{incl}$, a useful expansion parameter is defined as 
\cite{Wang:2014uaw}
\begin{equation}
    \xi = 1 - \cos \theta_\mathrm{incl} \simeq \frac{\theta_\mathrm{incl}^2}{2} + \mathcal{O} (\theta_\mathrm{incl}^4).
\end{equation}
Using this approximation, the initial positions of the spacecraft in a heliocentric ecliptic coordinate system can be expressed as follows. 
For S/C1: 
\begin{equation*}
\begin{bmatrix}
x_\mathrm{S/C1} \\
y_\mathrm{S/C1} \\
z_\mathrm{S/C1}
\end{bmatrix} 
= a
\begin{bmatrix}
\cos \omega t - \xi \cos \omega t \\
\sin \omega t \\
\sin \lambda \cos \omega t
\end{bmatrix}.
\end{equation*}
For S/C2:
\begin{equation*}
\begin{bmatrix}
x_\mathrm{S/C2} \\
y_\mathrm{S/C2} \\
z_\mathrm{S/C2}
\end{bmatrix} 
= a
\begin{bmatrix}
\cos (\omega t + \frac{2}{3} \pi) - \frac{1}{2} \xi \cos (\omega t - \frac{2}{3} \pi ) \\
\sin ( \omega t + \frac{2}{3} \pi) + \frac{\sqrt{3}}{2} \xi \sin ( \omega t - \frac{1}{6} \pi) \\
\sin \lambda \cos (\omega t - \frac{2}{3} \pi)
\end{bmatrix}.
\end{equation*}
For S/C3:
\begin{equation*}
\begin{bmatrix}
x_\mathrm{S/C3} \\
y_\mathrm{S/C3} \\
z_\mathrm{S/C3}
\end{bmatrix} 
= a
\begin{bmatrix}
\cos (\omega t - \frac{2}{3} \pi) - \frac{1}{2} \xi \cos (\omega t - \frac{2}{3} \pi ) \\
\sin ( \omega t - \frac{2}{3} \pi) + \frac{\sqrt{3}}{2} \xi \sin ( \omega t - \frac{1}{3} \pi) \\
\sin \lambda \cos (\omega t + \frac{2}{3} \pi)
\end{bmatrix}.
\end{equation*}
The spacecraft velocities can be readily obtained by taking time derivatives of the respective position vectors. A detailed discussion of the orbital design and optimization can be found in \cite{Wang:2014uaw}.

\subsection{GW Response of TDI}

For a GW source located at direction $(\lambda,\beta)$, the GW propagation vector $\hat{k}$ is defined as 
\begin{equation}
  \hat{k} = -\left( \cos \lambda \cos \beta, \ \sin \lambda \cos \beta, \ \sin \beta \right),
\end{equation}
where $\lambda$ and $\beta$ are the ecliptic latitude and longitude, respectively, in the solar system barycentric (SSB) coordinate system. In the transverse-traceless gauge, the GW strain can be decomposed into two polarizations $h_+$ and $h_{\times}$ as,
\begin{equation}
  H = h_+ P_+ + h_{\times} P_{\times},
\end{equation}
where $P_{+}, \ P_{\times}$ are polarization tensors defined as \cite{Vallisneri:2012np}
\begin{eqnarray}
  P_+ = O_1 \cdot  \begin{pmatrix} 1&0&0 \\  0 & -1 & 0 \\ 0 & 0 & 0 \end{pmatrix} \cdot O_1^T , \\
  P_{\times} = O_1 \cdot \begin{pmatrix}  0 & 1 & 0 \\ 1 & 0 & 0 \\ 0 & 0 & 0 \end{pmatrix} \cdot O_1^T\;,
\end{eqnarray}
with
\begin{widetext}
\begin{equation}
O_1 = 
  \begin{pmatrix}
  \sin \lambda \cos\psi-\cos \lambda \sin \beta \sin \psi & - \sin \lambda \sin \psi - \cos \lambda \sin \beta \cos \psi & - \cos \lambda \cos \beta \\
  -\cos \lambda \cos \psi - \sin \lambda \sin \beta \sin \psi  & \cos \lambda \sin \psi - \sin \lambda \sin \beta \cos \psi & -\sin \lambda \cos \beta \\
  \cos \beta \sin \psi & \cos \beta \cos \psi & -\sin \beta
  \end{pmatrix},
\end{equation}
\end{widetext}
where $\psi$ is polarization angle. 
The response of a single-link measurement from transmitting S/C$i$ to the receiving S/C$j$ is given by \cite{Vallisneri:2004bn}
\begin{equation}
  y_{ij}\left(t\right) = \frac{1}{2}\frac{\hat{n}_{ij} \cdot [H(p_i,t-L_{ij})-H(p_j,t)]\cdot \hat{n}_{ij}}{1-\hat{n}_{ij} \cdot \hat{k}}
\end{equation}
where $\hat{n}_{ij}$ is the unit vector from S/C$i$ to S/C$j$, and $p_i, \ p_j$ are the respective positions of the S/C.

For a GW mode labeled by harmonic indices $(\ell,m)$, the source-frame waveform in frequency domain is given by 
\begin{equation}
\tilde{h}_{\ell m}(f) = A_{\ell m}(f) e^{-\mathrm{i} \Psi_{\ell m}(f)}\;.
\end{equation}
where $A_{\ell m}(f)$ and $\Psi_{\ell m}(f)$ denote the amplitude and phase, respectively. Under the (quasi-)monochromatic approximation, the frequency-domain response of a single link can be formulated as \cite{Vallisneri:2007xa}
\begin{equation}
    \tilde{y}_{ij}(f, t) = \mathcal{T}_{ij} (f, t) \ast \tilde{h}_{\ell m}(f),
\end{equation}
where the transfer function $\mathcal{T}_{ij} (f, t)$ is given by
\begin{equation}
\begin{split}
\mathcal{T}_{ij} (f, t) =&  \frac{ (1 + \cos^2 \iota ) \hat{n}_{ij} \cdot {\rm e}_+ \cdot \hat{n}_{ij} + i (- 2 \cos \iota ) \hat{n}_{ij} \cdot {\rm e}_\times \cdot \hat{n}_{ij} }{4 (1 - \hat{n}_{ij} \cdot \hat{k} ) } \\
  &  \times \left[  \exp( 2 \pi i f (L_{ij} + \hat{k} \cdot p_i ) ) -  \exp( 2 \pi i f  \hat{k} \cdot p_j )  \right] ,
\end{split}
\end{equation}
where $\iota$ is the inclination angle of the binary, and $L_{ij}$ is the arm length. The time-frequency correspondence for each harmonic mode is defined via the stationary phase approximation 
\begin{equation}
t_{\ell m}(f) = - \frac{1}{2\pi}\frac{\mathrm{d} \Psi_{\ell m}(f)}{\mathrm{d} f}.
\end{equation}

TDI will be employed by space-based GW interferometers to suppress overwhelming laser frequency noise \cite{Ni:1996ns,1999ApJ...527..814A}. In recent works, we proposed two alternative second-generation TDI configurations to enhance the data analysis, demonstrating superior performance compared to the fiducial Michelson TDI scheme \cite{Wang:2024alm,Wang:2024hgv,Wang:2025mee}.
In this study, we utilize the PD4L configuration including three ordinary channels (PD4L1, PD4L2, PD4L3) for our evaluations \cite{Wang:2025mee}. 
Their (quasi-)orthogonal observables A, E and T can be generated from the linear combination,
\begin{equation} \label{eq:abc2AET}
\begin{bmatrix}
\mathrm{A}  \\ \mathrm{E}  \\ \mathrm{T}
\end{bmatrix}
 =
\begin{bmatrix}
-\frac{1}{\sqrt{2}} & 0 & \frac{1}{\sqrt{2}} \\
\frac{1}{\sqrt{6}} & -\frac{2}{\sqrt{6}} & \frac{1}{\sqrt{6}} \\
\frac{1}{\sqrt{3}} & \frac{1}{\sqrt{3}} & \frac{1}{\sqrt{3}}
\end{bmatrix}
\begin{bmatrix}
\mathrm{PD4L1} \\ \mathrm{PD4L2}  \\ \mathrm{PD4L3}
\end{bmatrix}.
\end{equation}
The A and E serve as science channels and can effectively response to GW signals, while the T channel functions as a null stream dominated by instrumental noise at lower frequencies. 

\subsection{Average Sensitivities}

Assuming that laser frequency noise is effectively suppressed by TDI, the dominant noise sources become the acceleration noise and optical metrology system (OMS) noise. As discussed in \cite{Wang:2023jct}, we consider two noise configurations. The first corresponds to an elementary ASTROD-GW (eASTROD-GW) setup, with its noise budget based on the requirements for the LISA mission \cite{LISA:2017pwj}.
For eASTROD-GW, the amplitude of the acceleration noise is assumed to match that of LISA. However, since the ASTROD-GW arm length is 104 times longer than LISA’s, the OMS noise is scaled up by the same factor.
Thus, the upper limits for the acceleration noise $S_\mathrm{acc}$ and OMS noise $S_\mathrm{oms}$ are expected to be \cite{,Ni:2016wcv,Ni:2012eh}:
\begin{eqnarray}
  S_{\rm acc,\ eASTROD-GW}^{1/2} &= 3 \frac{\rm fm/s^2}{\sqrt{\rm Hz}}\sqrt{1+\left(\frac{0.1 \ \rm mHz}{f}\right)^2}, \\
S_{{\rm oms,\ eASTROD-GW}}^{1/2} & = 1040 \frac{\rm pm}{\sqrt{\rm Hz}}\sqrt{1+\left(\frac{0.2 \ \rm mHz}{f}\right)^4}.
\end{eqnarray}
The second configuration represents an optimistic scenario for an advanced ASTROD-GW (aASTROD-GW), where both instrumental noises are reduced by one order of magnitude compared to the eASTROD-GW setup:
\begin{eqnarray}
  S_{\rm acc,\ aASTROD-GW}^{1/2} & = 0.3 \frac{\rm fm/s^2}{\sqrt{\rm Hz}}\sqrt{1+\left(\frac{0.1 \ \rm mHz}{f} \right)^2}, \\
  S_{{\rm oms,\ aASTROD-GW}}^{1/2} & = 104 \frac{\rm pm}{\sqrt{\rm Hz}}\sqrt{1+\left(\frac{0.2 \ \rm mHz}{f} \right)^4} .
\end{eqnarray}
Given these noise budgets, the power spectral density (PSD) of optimal TDI channels will be
\begin{align}
\centering
S_\mathrm{A}  = S_\mathrm{E} = & 64 \sin ^4 \frac{x}{2}  (2 \cos x+\cos 2 x+3) S_\mathrm{acc} \notag \\ 
&+32 \sin ^4 \frac{x}{2} (\cos x +2)  S_\mathrm{oms}, \\
S_\mathrm{T}  = & 128 \sin ^4 \frac{x}{2}  (\cos x+2)^2 S_\mathrm{acc} \notag \\
                           &  +8 \left(3 \sin \frac{x}{2}+\sin \frac{3x}{2} \right)^2 S_\mathrm{oms}.
\end{align}
where $x = 2\pi f L/c$.

\begin{figure*}[htbp]
    \centering
    \includegraphics[width=0.48\linewidth]{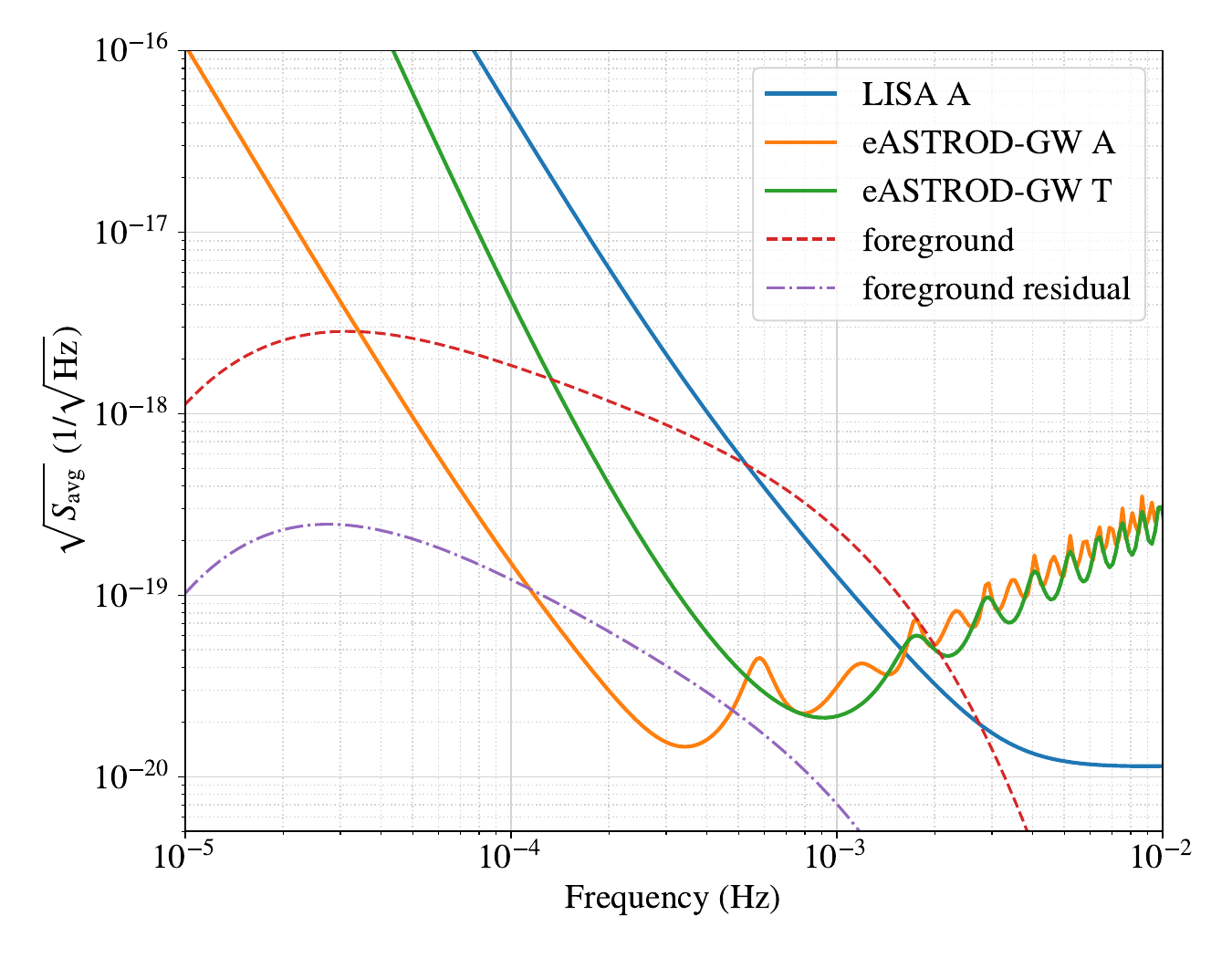}
    \includegraphics[width=0.48\linewidth]{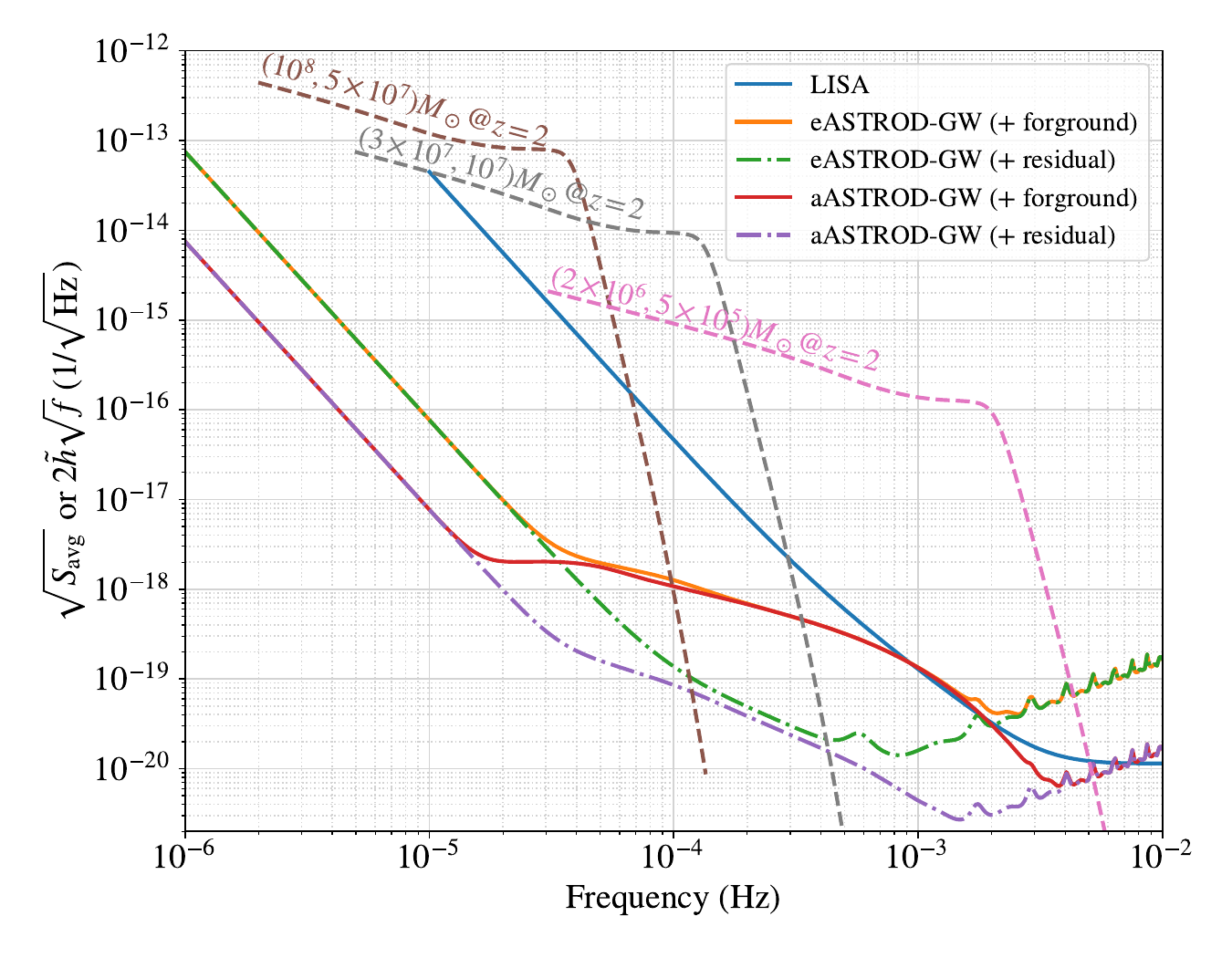}
    \caption{The averaged sensitivity of ASTROD-GW without (left panel) and with (right panel) the galactic foreground. In the left plot, the sensitivity of LISA is shown by solid blue for comparison, the sensitivity of A (or E) and T for eASTROD-GW setup are depicted by solid orange and green curves, respectively. The dashed red and dashdotted curves represent the foreground and its residual utilized in this work. In the right plot, both eASTROD-GW and aASTROD-GW sensitivities are present, including the effects of foreground (solid) and residual foreground (dashdotted). The dashed lines represent frequency-domain waveforms from MBBHs with three different total masses at redshift $z=2$, and the components masses are labeled along each line.}
    \label{fig:sensitivity}
\end{figure*}

To evaluate the detectability of GW signal of TDI channels, we average the GW response over sky position and polarization. The average response of a TDI channel is calculated by implementing
\begin{equation}
  \mathcal{R}_{\rm TDI}(f) = \frac{1}{4\pi^2} \int_{0}^{2\pi} \int_{-\frac{\pi}{2}}^{\frac{\pi}{2}} \int_{0}^{\pi} |\mathcal{T}(f, \lambda, \beta, \psi)|^2 \cos \beta \mathrm{d} \psi \mathrm{d} \beta \mathrm{d} \lambda.
\end{equation}
The corresponding averaged sensitivity in a TDI channel is obtained by dividing the noise PSD by the averaged response:
\begin{equation}
    S_{\rm avg, TDI} = S_{\rm n,TDI}/\mathcal{R}.
\end{equation}
The left panel of Fig.~\ref{fig:sensitivity} shows the averaged sensitivity of eASTROD-GW considering only instrumental noise. For comparison, the LISA sensitivity curve is shown in blue. The sensitivity of the A channel for eASTROD-GW is shown by the solid orange curve. Since the E channel has an identical response, it is not plotted. The T channel (a null stream) is shown in green; its sensitivity is significantly lower than that of A/E channels at frequencies below 0.3 mHz and becomes comparable at higher frequencies.
The aASTROD-GW configuration offers two orders of magnitude lower noise PSDs, resulting in approximately one order of magnitude improvement in sensitivity compared to eASTROD-GW.
However, space-based detectors targeting the $\mu$Hz to sub-mHz band will be affected by the galactic foreground. The impact of foreground on the sensitivity is assessed in the following subsection.

\subsection{Galactic Foreground} \label{subsec:foreground}

A large population of galactic binaries, particularly double white dwarfs (DWDs), are expected to be in the early inspiral stage, continuously emitting GWs at frequencies below $\sim$10 mHz. The LISA could resolve tens of thousands binaries in its mission duration, which is a small fraction of the galactic binary population \cite{Nelemans:2001hp,Nissanke:2012eh,Lamberts:2019nyk,Li:2020voo,Zhang:2021htc}. The unresolved majority will form a foreground or confuse noise which can degrade the detector's sensitivity. \citet{Thiele:2021yyb} modeled this foreground for LISA mission using a polynomial fit in log-log space: 
\begin{equation}
    {\rm log}_{10} S_{\rm conf} = \sum^{4}_{k=0}a_k ({\rm log}_{10}f)^k .
\end{equation}
This functional form was adopted in \cite{Wang:2023jct} to characterize the unresolved galactic foreground in the sub-mHz frequency range. For the fiducial population model, the fitted coefficients are: $ a_0=-180.460, a_1=-145.710, a_2=-56.2753, a_3=-9.80524, a_4=-0.64482$. The amplitude spectral density (ASD) of the resulting foreground is shown as the dashed red curve in the left panel of Fig.~\ref{fig:sensitivity}. In the most sensitive band of the ASTROD-GW mission, this foreground exceeds the instrumental noise by more than an order of magnitude.

Under an optimistic scenario, if the foreground can be accurately modeled, it may be subtracted from the data, thereby enhancing the sensitivity to astrophysical signals. As analyzed in \cite{Wang:2023jct}, the foreground residual is estimated as the difference between the median and the $1\sigma$ uncertainty bounds, assuming the foreground behaves as Gaussian noise. The residual spectrum is depicted by the dashdotted purple curve in the left panel of Fig. \ref{fig:sensitivity}. With both the full foreground and its residual included, the sensitivities of eASTROD-GW and aASTROD-GW are presented in the right panel of Fig. \ref{fig:sensitivity}. As shown, after foreground subtraction, the residual can be reduced by approximately an order of magnitude in the mission’s most sensitive frequency band.

\section{Detectability and parameter estimation to massive binary black holes} \label{sec:detectability}

\subsection{SNRs for MBBHs}

\begin{figure}[htbp]
    \centering
    \includegraphics[width=0.48\textwidth]{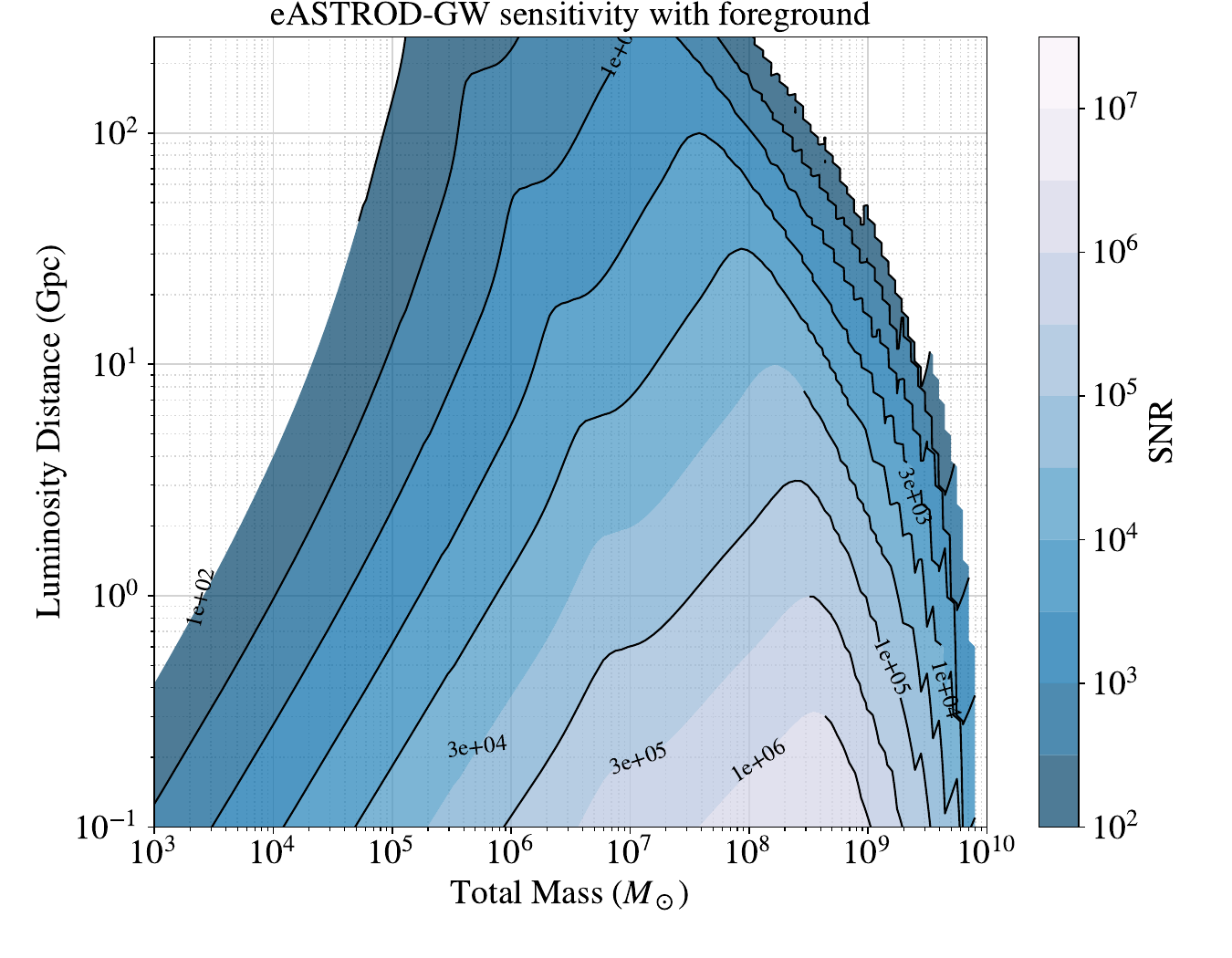}
    \includegraphics[width=0.48\textwidth]{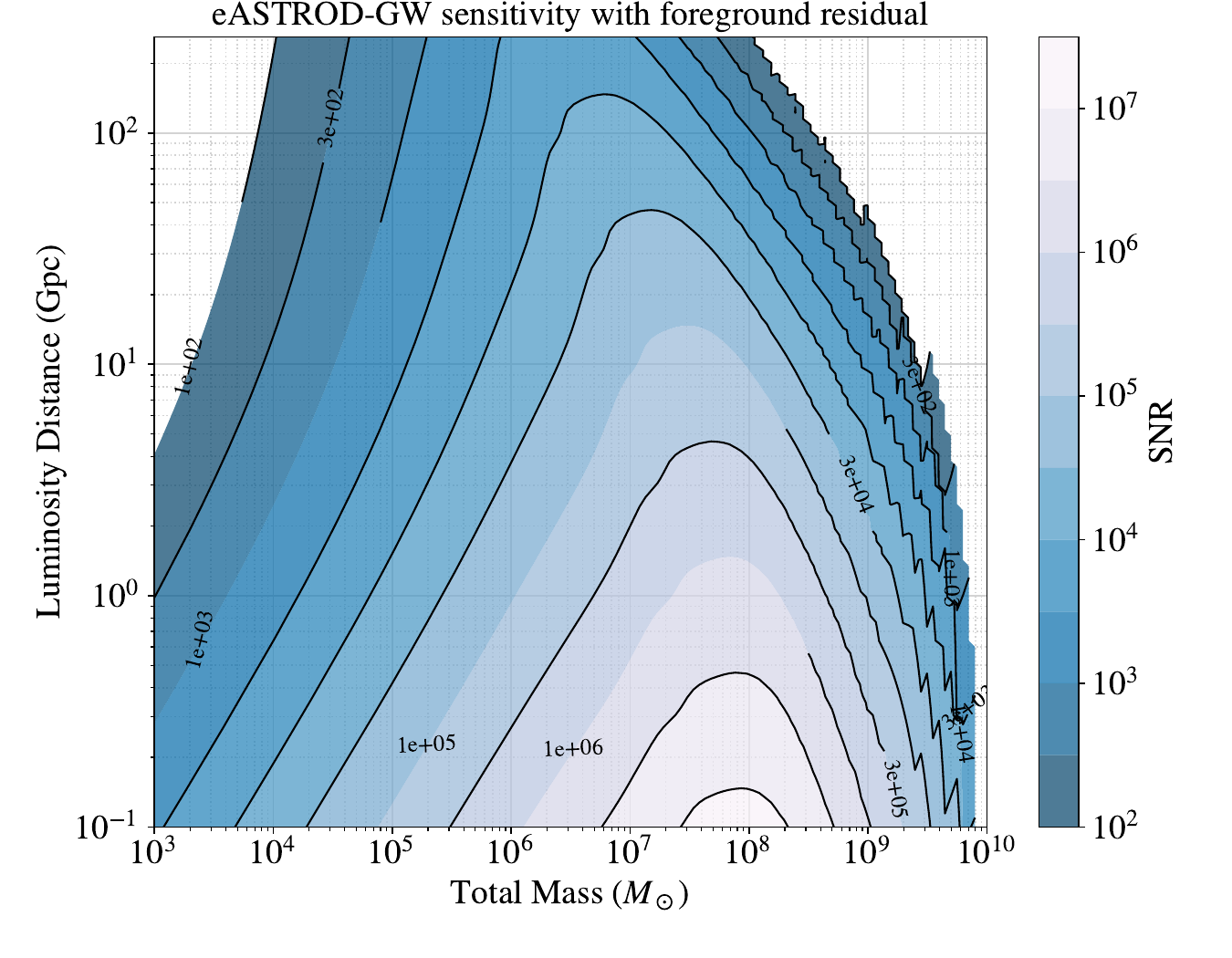}
    \includegraphics[width=0.48\textwidth]{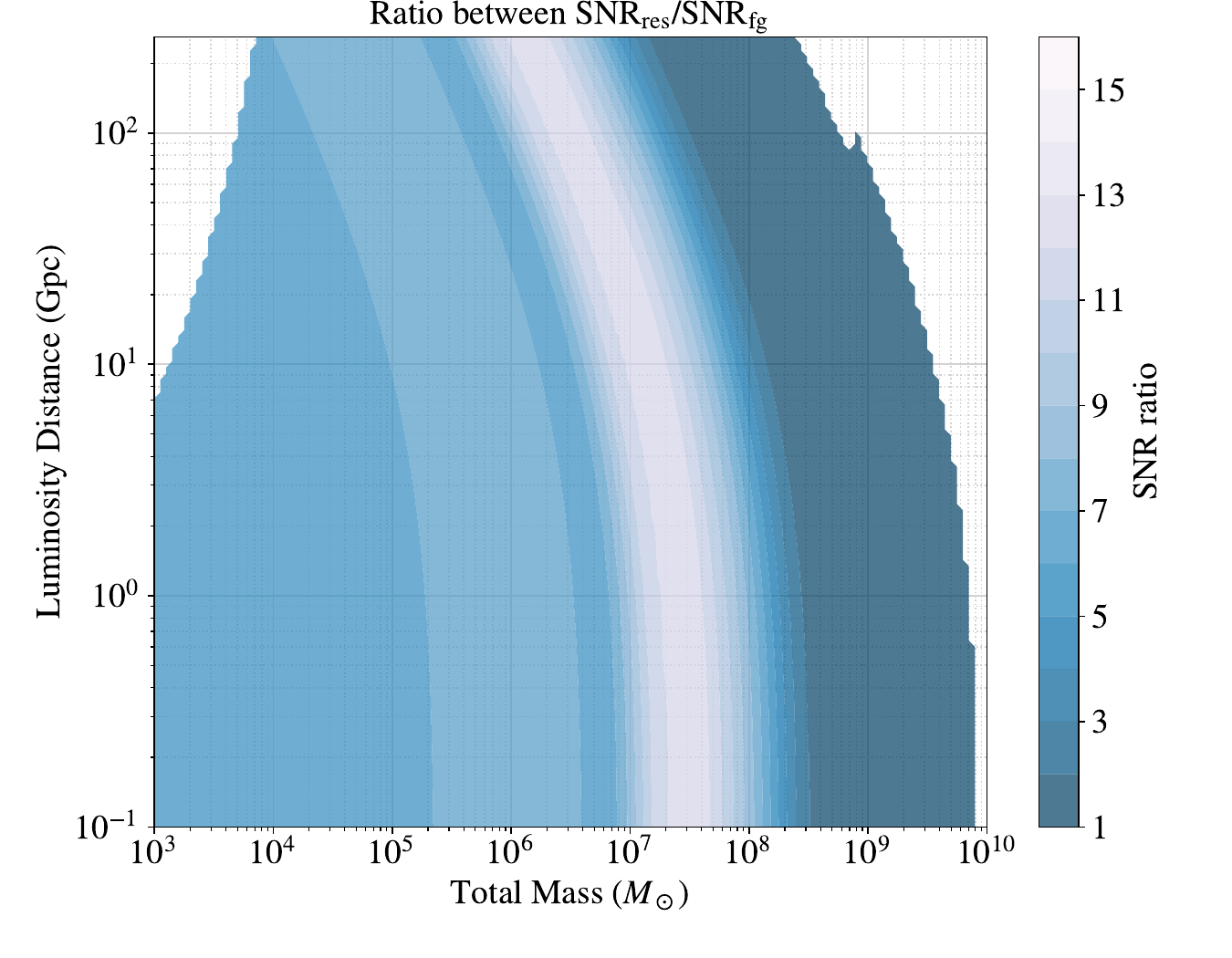}
    \caption{SNRs of equal-mass MBBH mergers as observed by eASTROD-GW, plotted as a function of total source-frame mass (horizontal axis) and luminosity distance (vertical axis). Upper: SNRs with galactic foreground included. Middle: SNRs with only the foreground residual. Lower: Ratio between SNRs with residual and those with foreground.}  \label{fig:SNR}
\end{figure}

MBBHs are among the most promising sources GW detection in the $\mu$Hz to sub-mHz band. Compared to LISA, which is sensitive primarily to the mHz band, ASTROD-GW's sensitivity peak at frequencies roughly two orders of magnitude lower, making it especially well-suited for observing active galactic nuclei (AGNs) \cite{Wu:2022knm,LISA:2024hlh,2023OJAp....6E..49F}. 
The optimal SNR for an MBBH detected by ASTROD-GW is computed using matched filtering across three (quasi-)orthogonal TDI channels (A, E and T), 
\begin{equation} \label{eq:SNR}
    \rho^2_{\rm opt} = \sum_\mathrm{A, E, T}  \left( h_\mathrm{TDI}|h_\mathrm{TDI} \right),
\end{equation}
where the inner product is defined as
\begin{equation}
\left( g | h\right)_{\rm TDI} = 4 \mathrm{Re} \int^{\infty}_0 \frac{g^{\ast} (f) h(f)}{S_{\rm n, TDI} (f) + S_{\rm fg, TDI} } \mathrm{d} f,
\end{equation}
with $S_\mathrm{n, TDI}(f)$ and $S_{\rm fg,TDI}(f)$ denoting the PSDs of instrumental noise and foreground noise, respectively.
Figure~\ref{fig:SNR} shows the SNRs computed for an equal-mass MBBH population at various luminosity distances under the eASTROD-GW configuration. The analysis assumes a 10-year observation ending at merger. The lower frequency cutoff is set to 1 $\mu$Hz or the binary’s starting frequency for a 10-year inspiral, whichever is higher. The top-left panel includes the galactic foreground in the sensitivity curve, while the middle panel shows SNRs with the foreground residual included. As illustrated, even without foreground subtraction, eASTROD-GW can achieve strong SNRs for MBBHs at high redshifts.

To illustrate the improvement achieved by foreground substraction, the bottom panel of Fig. \ref{fig:SNR} presents the ratios of SNRs with residual foreground to those with full foreground. The most significant enhancement occurs for MBBHs with total masses in the range of [$10^7, 10^8$] M$_\odot$, which also represents the most promising source population for sub-mHz missions. For binaries more massive than $\sim 5 \times 10^8 M_\odot$, its merger frequencies drops below $\sim 30 \mu$Hz, where the galactic foreground becomes negligible.

\subsection{Parameter Inference with (2,2) mode}

The optimal SNR calculation employs the average sensitivity and not be affected by different orbital configurations. However, orbit configuration can significantly influence a mission's parameter inference capability and source detectability. The primary motivation for evolving orbit design from non-inclined to inclined is to introduce time-varying antenna patterns. This modulation helps break parameter degeneracies and enhances accuracy of GW measurements. To illustrate the impact of different orbital configurations on parameter estimation, we simulate the detection of a GW signal from a non-spinning MBBH using SATDI \cite{Wang:2024ssp}. The binary has component masses of $m_1=2 \times 10^6 M_\odot, m_2 = 5 \times 10^5 M_\odot$, located at a redshift $z=2$. The source is positioned at an ecliptic latitude $\beta=\pi/10$, cliptic longitude of source is $\lambda=4.603$, with an inclination angle $\iota=\pi/3$ between the line-of-sight and binary's orbital angular momentum. and a polarization angle is set to be $\psi=0.55$. The signal spans 180 days of uninterrupted observation. The time-domain waveform is generated by using the SEOBNRv5HM model \cite{Pompili:2023tna}.  

The data observed in a TDI channel containing a GW signal $h_\mathrm{TDI}(t)$ can be described as: 
\begin{equation}
 d_\mathrm{TDI} (t) = h_\mathrm{TDI}(t) + n_\mathrm{TDI} (t),
\end{equation}
where $n_\mathrm{TDI}(t)$ denotes noise(s). Bayesian inference yields the posterior distributions of the source parameters:
\begin{equation}
p(\vec{\theta}|d) = \frac{\mathcal{L}( d | \vec{\theta} ) p( \vec{\theta} )}{p( d ) },
\end{equation}
where $\vec{\theta}$ represents the waveform parameters, $p(\vec{\theta})$ is the prior, $ \mathcal{L}( d | \vec{\theta} )$ is the likelihood function, and $p(d|\Lambda)$ is the evidence.
The likelihood function used in parameter estimation is:
\begin{equation} \label{eq:likelihood}
\begin{aligned}
    \ln \mathcal L(d|\vec{\theta}) = \sum_{f_i} \bigg[ & -\frac{1}{2} (\mathbf{\tilde{d}} - \mathbf{\tilde{h}} )^T \mathbf{C}^{-1} (\mathbf{\tilde{d}} - \mathbf{\tilde{h}} )^\ast
    - \frac{1}{2} {\rm ln} (\det 2\pi \mathbf{C} ) \bigg]
\end{aligned}
\end{equation}
where $\mathbf{\tilde{d}}$ is the Fourier-transformed data vector, $\mathbf{\tilde{h}}$ is the frequency-domain TDI waveforms, and $\mathbf{C}$ is the noise covariance matrix among the three TDI channels:
\begin{equation}
\mathbf{C}= \frac{T_\mathrm{obs}}{4}
\begin{pmatrix}
\begin{smallmatrix}
S_{\mathrm{n}, \mathrm{A}}+ S_\mathrm{fg, A} & 0 & 0 \\
0 & S_{\mathrm{n}, \mathrm{E}}+ S_\mathrm{fg, E} & 0 \\
0 & 0 & S_{\mathrm{n}, \mathrm{T}}+ S_\mathrm{fg, T}
\end{smallmatrix}
\end{pmatrix}
\end{equation}
where $T_\mathrm{obs}$ is the duration of data, and $S_\mathrm{fg, TDI}$ representing the galactic foreground noise in the corresponding TDI channel. The waveform in the frequency-domain is generated by using the reduced-order model of SEOBNRv5HM \cite{Pompili:2023tna}. Parameter inference is performed using the \textsf{MultiNest} \cite{Feroz:2008xx,Buchner:2014nha}. To reduce degeneracies due to the polarization symmetry and accelerate convergence, the prior range for $\psi$ is limited to $[0, \pi/2]$. The source direction $(\lambda, \beta)$ is sampled uniformly over the celestial sphere, and the inclination $\iota$ follows a sinusoidal prior. 

\begin{figure*}[htbp]
    \centering
    \includegraphics[width=0.96\linewidth]{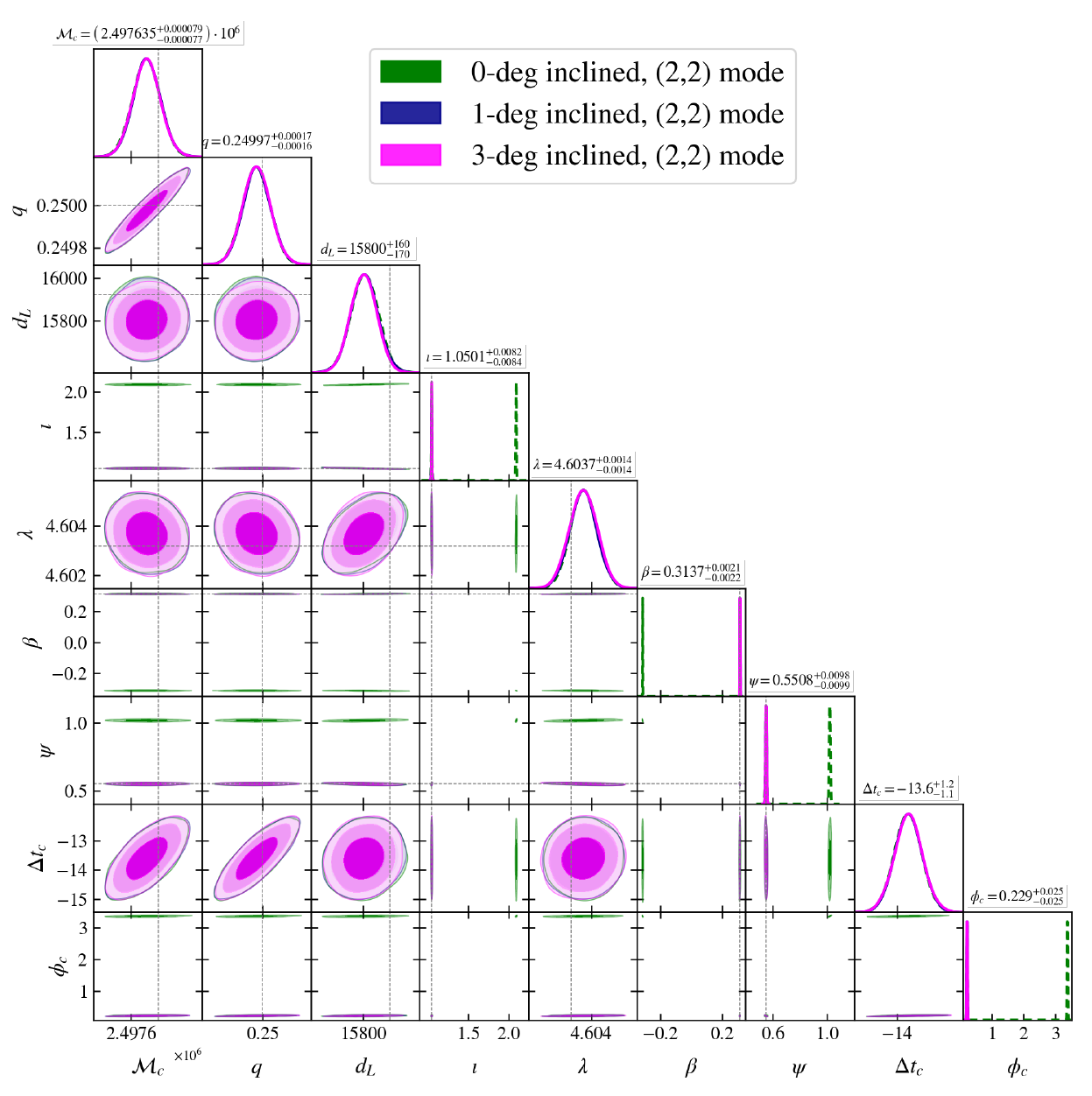}
    \caption{Posterior distributions for $m_1=2 \times 10^6 M_\odot, m_2 = 5 \times 10^5 M_\odot$ at redshift $z=2$, using only the dominant GW (2,2) mode. The green, blue, and magenta colors represent the orbital inclination of $0^\circ$, $1^\circ$, and $3^\circ$, respectively. The $0^\circ$ configuration could not resolve the degeneracies in inclination $\iota$, and the ecliptic latitude $\beta$, which are better constrained in the inclined orbit configurations. The polarization angle $\psi$ is also be better resolved with inclined orbits. The parameter distributions for $\iota$, $\beta$, $\psi$, and $\phi_c$ in the $0^\circ$ case are bimodal and partially overlap with those from $1^\circ$, and $3^\circ$. The inferred values with $3\sigma$ uncertainties from the $1^\circ$ inclined orbit are shown at the top of each column. }
    \label{fig:corner_fg_22_mode}
\end{figure*}

Figure~\ref{fig:corner_fg_22_mode} displays the posterior distributions for the (2,2) GW mode under three orbital configurations, incorporating the galactic foreground in the analysis. The green curves correspond to the non-inclined ($0^\circ$) orbit, while blue and magenta represent configurations with $1^\circ$ and $3^\circ$ inclinations, respectively. The non-inclined orbit yields poor resolution of the inclination $\iota$ and ecliptic latitude $\beta$, due to parameter degeneracies. Introducing a $1^\circ$ orbital inclination significantly breaks these degeneracies and improves inference. Increasing the inclination further to $3^\circ$ does not yield noticeable improvements, as the uncertainties in key parameters remain nearly the same.

\subsection{Parameter Inference with higher-order modes} \label{subsec:pe_hm}

GW signals are generally consist of a superposition of multiple harmonic modes \cite{Thorne:1980ru}, with the the dominant contribution arising from the $(\ell,|m|)=(2,2)$ mode. Higher-order modes are generally than the dominant (2,2) mode; however, incorporating them in waveform models is important, as they help break the degeneracies particularly between the luminosity distance and the inclination angle \cite{Ohme:2013nsa, Usman:2018imj}. In this subsection, we examine how including higher-order modes affects parameter determination under different orbital configurations.

The GW polarizations $h_+$ and $h_{\times}$ can be decomposed into the spherical harmonic modes $h_{\ell m}$ using spin-weighted spherical harmonics $\ _{-2}Y_{\ell m}(\iota,\phi_c)$ which depend on the inclination angle $\iota$ and coalescence phase $\phi_c$ \cite{Mills:2020thr}.
The full waveform can be expressed as 
\begin{equation}
    h_+ -i h_{\times} = \sum_{\ell\geq2}\sum_{m=-\ell}^{\ell}\ _{-2}Y_{\ell m}(\iota,\phi_c)h_{\ell m}\;.
\label{decompose}
\end{equation}
All modes beyond the dominant harmonic $h_{22}$ are labeled as higher-order modes (HM), despite their mode number $\ell$ and/or $m$ could be less than 2.
The response in individual observables $y_{ij}$ can be modeled with a transfer function applied to each harmonic mode:
\begin{equation}
    \tilde{y}_{ij}(f)= \sum_{\ell,m} \mathcal{T}_{ij}^{\ell m}(f)\tilde{h}_{\ell m}(f)\;.
\end{equation}
Due to the motion of the detector constellation, the transfer function is both time- and frequency-dependent and is defined as:
\begin{equation}
  \mathcal{T}_{ij}^{\ell m}\left(f\right) = G^{\ell m}_{ij}\left(f,t^{\ell m}_{f}\right) ,
\end{equation}
The signal of each harmonic in each TDI channel is given by
\begin{equation}
\tilde{h}_{A,E,T}^{\ell m}(f)= \mathcal{T}_{A,E,T}^{\ell m}(f,t_{\ell m}(f))\tilde{h}_{\ell m}(f)
\end{equation}
By incorporating the full multi-mode waveform into the likelihood function in Eq.~\eqref{eq:likelihood}, we can perform parameter inference that includes contributions from higher-order harmonics.

\begin{figure*}[htbp]
    \centering
    \includegraphics[width=0.96\linewidth]{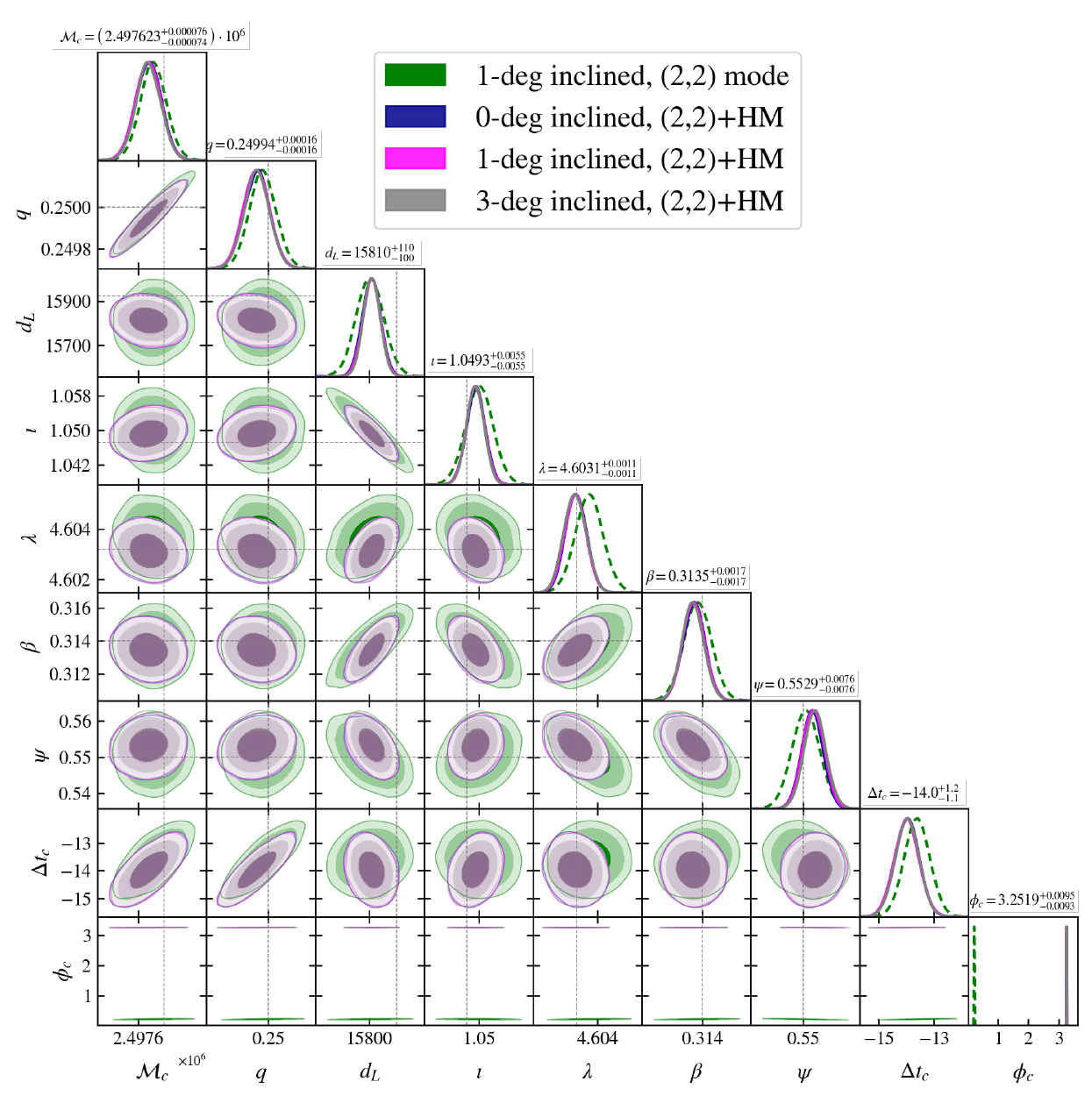}
    \caption{Posterior distributions for source with $m_1=2 \times 10^6 M_\odot, m_2 = 5 \times 10^5 M_\odot$ at redshift $z=2$, including (2,2), (3,3) and (4,4) modes. Blue, magenta and grey correspond to orbital inclinations of $0^\circ$, $1^\circ$, and $3^\circ$, respectively. The result from 1$^\circ$ inclined orbit using only the (2,2) mode is shown by the green colors for comparison. The inferred values with $3\sigma$ uncertainties from the $1^\circ$ inclined configuration including higher modes are indicated at the top of each column. }
    \label{fig:corner_higher_mode}
\end{figure*}

The posterior distributions for the same signal in the previous subsection, now including [(2,2), (3,3), (4,4)] modes, are shown in Fig. \ref{fig:corner_higher_mode}. Remarkably, all three orbit configurations yield the consistent results when higher-order modes are included, and their posterior distributions nearly identical and fully overlap. This contrasts with the (2,2)-only case, where degeneracies arise in the non-inclined orbit configuration. The inclusion of higher-order modes helps resolve degeneracies in the inclination angle $\iota$, ecliptic latitude $\beta$, and polarization angle $\psi$.
For comparison, the results from the 1$^\circ$-inclined orbit using only the (2,2) mode is shown in green. While the inclusion of higher-order modes does not significantly improve the estimation of intrinsic parameters such as the component masses, it substantially enhances the resolution of extrinsic parameters. Notably, the coalescence phase $\phi_c$ experiences a near phase reversal (approximately $\pi$ difference) between analyses with and without higher-order modes.

At first glance, the inclusion of higher-order modes may appear to diminish the benefit of inclined or precessing orbital designs. However, when the binary system is nearly face-on or face-off (i.e., $\iota=0$ or $\pi$), higher-order modes are significantly suppressed, and their utility in parameter inference may be reduced. In such cases, the time-varying antenna pattern from an inclined orbit remains essential for breaking degeneracies. Therefore, to ensure robust and consistent detectability across a wide range of orientations, inclined (precessing) orbital configurations remain advantageous.

\section{Sky localization to MBBHs} \label{sec:localization}

\subsection{Formalism of Fisher information matrix}

Sky localization is demanded for GW detectors to enable multi-messenger astronomy. To systematically evaluate the angular resolution for binary sources, the Fisher information matrix (FIM) is employed to efficiently estimate the uncertainties of the parameters \cite[and references therein]{1994PhRvD..49.2658C,Cutler:1997ta,Vallisneri:2007ev,Kuns:2019upi}.
For a single mission with all six links operational, the FIM is constructed using three optimal channels (A, E, and T):
\begin{equation}
\Gamma_{ij}  = \sum_{\rm A,E,T}  \left( \frac{\partial \tilde{h}_\mathrm{TDI} }{ \partial {\theta_i} } \bigg\rvert \frac{\partial \tilde{h}_\mathrm{TDI} }{ \partial {\theta_j} }  \right),
\end{equation}
where $\theta_i$ denotes the $i$-th parameter. The formalism for a joint observation is obtained by summing the FIMs from the two detectors.

Since higher harmonics play a crucial role in the analysis, we employ the \texttt{IMRPhenomXHM} waveform model \cite{Garcia-Quiros:2020qpx} from the LALSuite software package \cite{lalsuite} to include various modes in the inspiral-merger-ringdown phases. The calculation is performed in the frequency domain for spinless black hole binaries. Beyond the dominant (2,2) mode, higher-order harmonics, $(\ell, |m|) = (2,1), (3,3), (3,2), (4,4)$, are incorporated. 
The FIM formalism for multiple harmonic modes is expressed as
\begin{equation}
\Gamma_{ij}  = \sum_{\rm A,E,T}   \left( \sum_{\ell, m}  \frac{\partial \tilde{h}^{lm}_\mathrm{TDI} }{ \partial {\theta_i} } \bigg\rvert \sum_{\ell, m} \frac{ \partial \tilde{h}^{lm}_\mathrm{TDI} }{ \partial {\theta_j} }  \right).
\end{equation}
The standard deviations $\sigma_i$ and covariances $\sigma^2{ij}$ of the parameters, for $\rho \gg 1$, can be approximated as
\begin{eqnarray}
 \sigma_i &= \sqrt{(\Gamma^{-1})_{ii} } \\
 \sigma^2_{ij} &= (\Gamma^{-1})_{ij}
\end{eqnarray}
The uncertainty in sky localization for a single source is evaluated by
\begin{equation} \label{eq:delta_Omega}
 \Delta \Omega \simeq 2 \pi \cos \beta \sqrt{ \sigma_{\lambda } \sigma_{\beta } -  \sigma^2_{\lambda \beta}   }.
\end{equation}

In this investigation, three sources with different masses but located at the same redshift ($z=2$) are used to evaluate the precision of sky localization: 1)	a light source with component masses $(10^6, 5 \times 10^5)\ M_\odot$ in the source frame, 2)	a medium source with component masses $(3 \times 10^7, 10^7)\ M_\odot$ in the source frame, and 3)	a heavy source with component masses $(10^8, 5 \times 10^7)\ M_\odot$ in the source frame.
The typical strains of the three sources over 180 days are shown in the right panel of Fig. \ref{fig:sensitivity}.
Their frequency ranges in the solar-system barycentric frame are approximately [30 $\mu$Hz, 50 mHz], [5, 400] $\mu$Hz and [2, 100] $\mu$Hz, respectively.
For each source, 5000 binaries are simulated for each orbital configuration, with randomized sky location, polarization, and inclination.
We emphasize that the current results, based on the FIM, do not account for the ambiguity in sky localization between the two hemispheres in the case of non-inclined orbits. The true localization uncertainty should therefore be approximately twice the value reported here.

\subsection{Angular resolutions with (2,2) mode}

\begin{figure*}[htbp]
     \centering
     \includegraphics[width=0.42\textwidth]{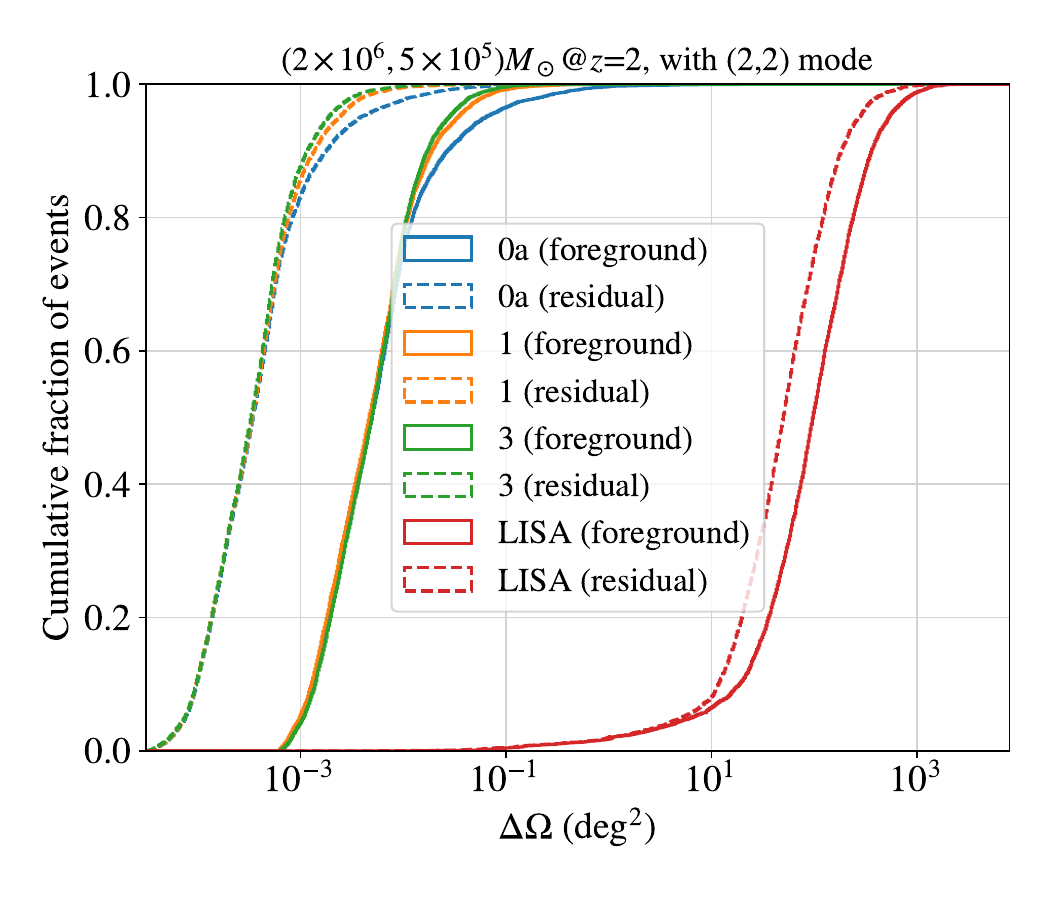}
     \includegraphics[width=0.54\textwidth]{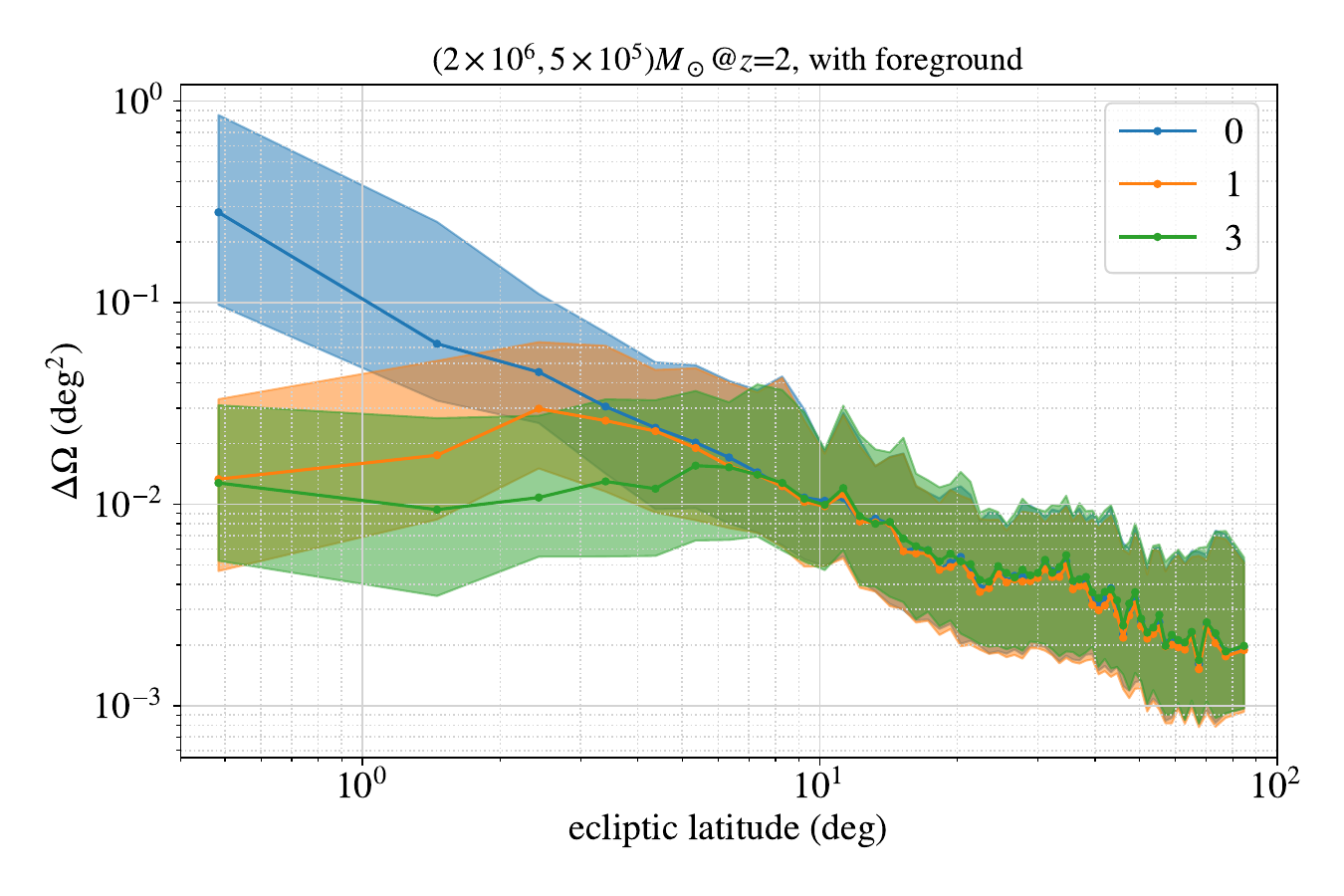}
     \includegraphics[width=0.42\textwidth]{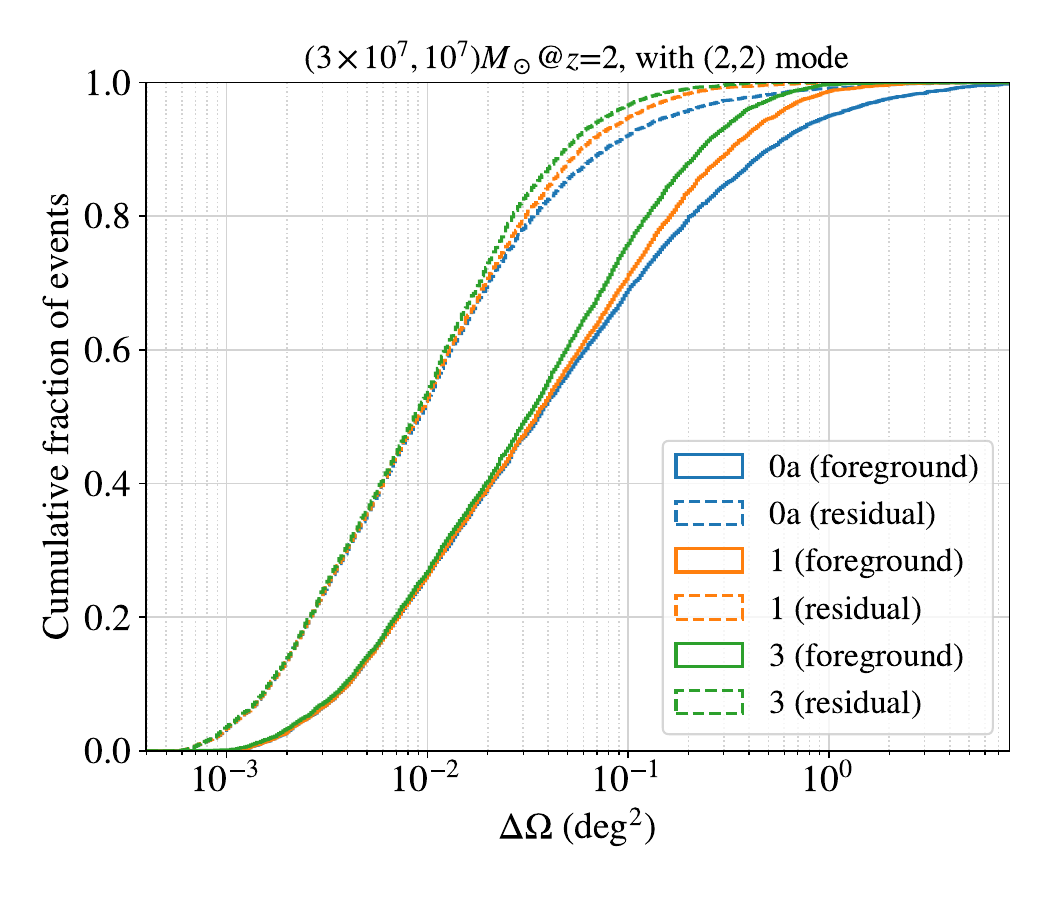}
     \includegraphics[width=0.54\textwidth]{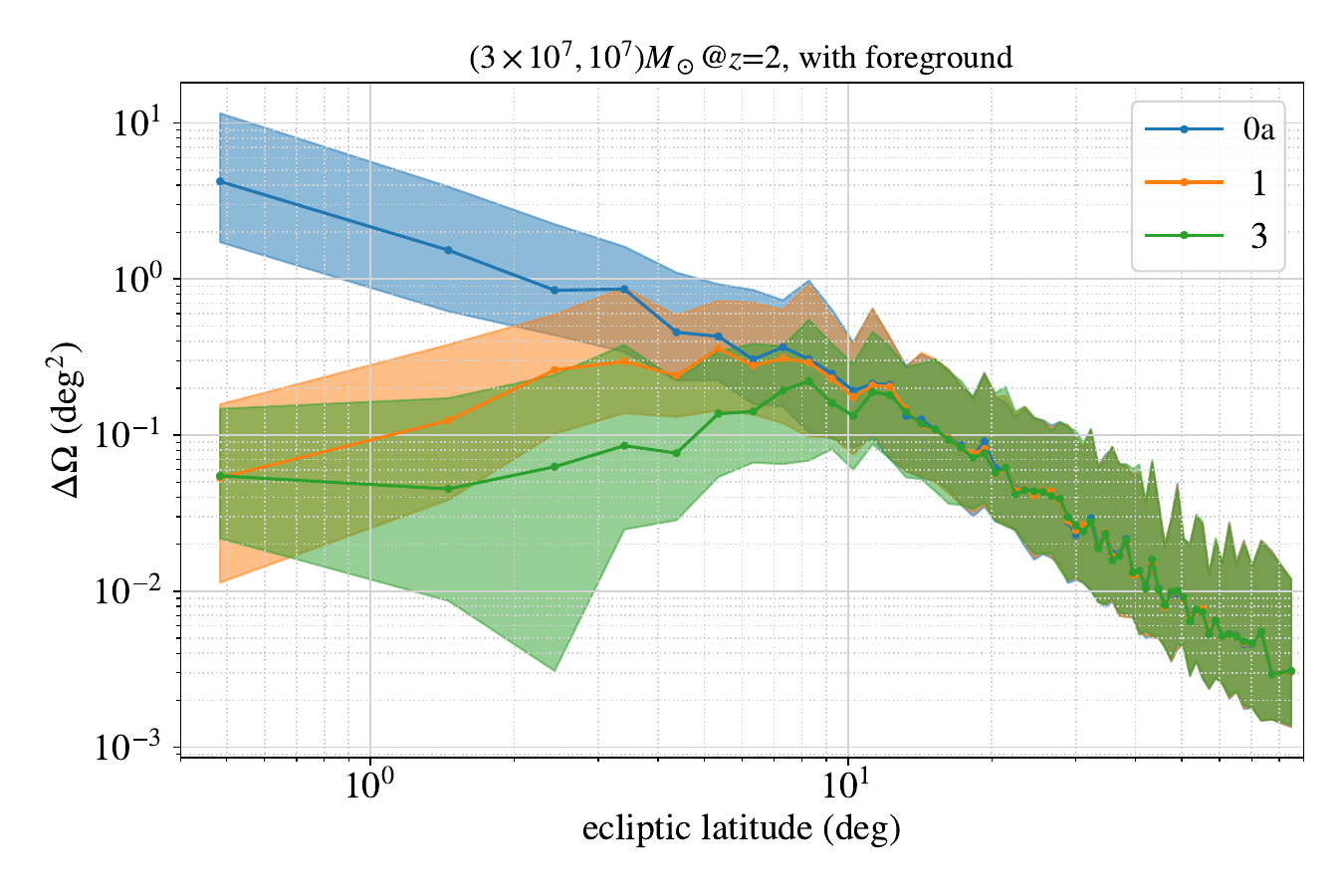}
     \includegraphics[width=0.42\textwidth]{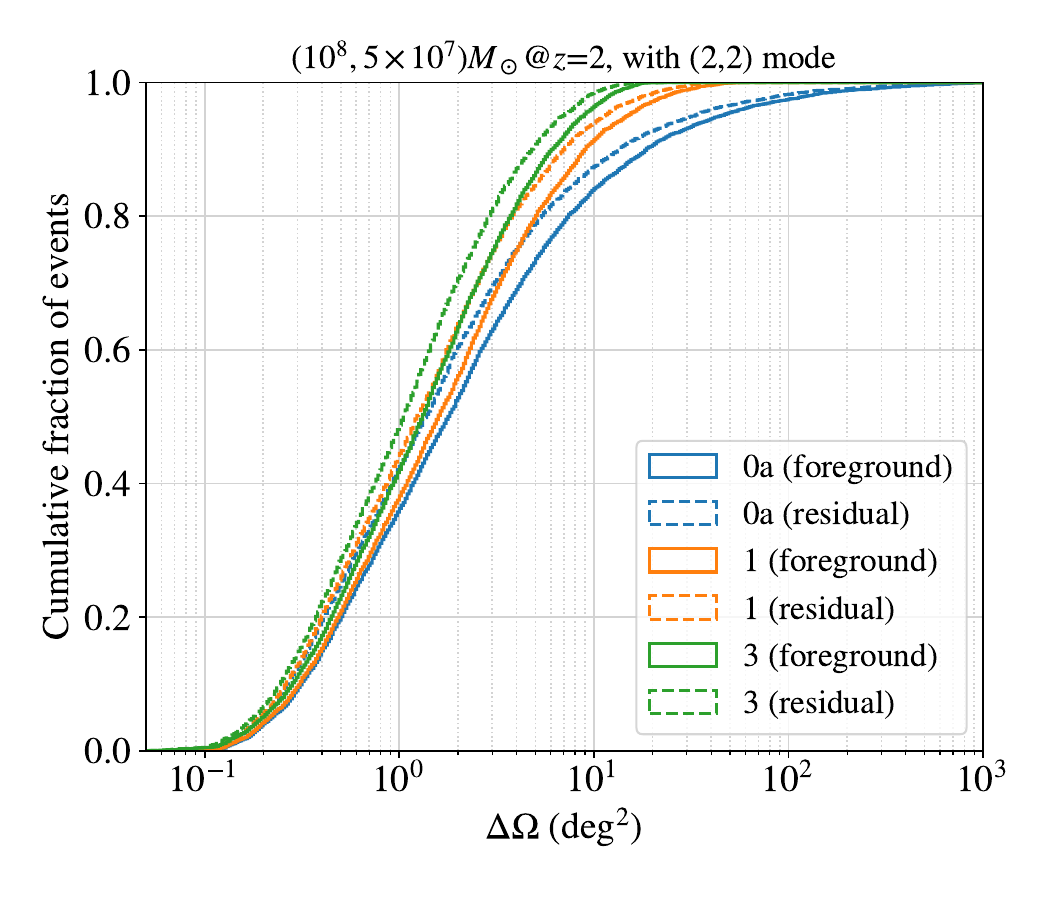}
     \includegraphics[width=0.54\textwidth]{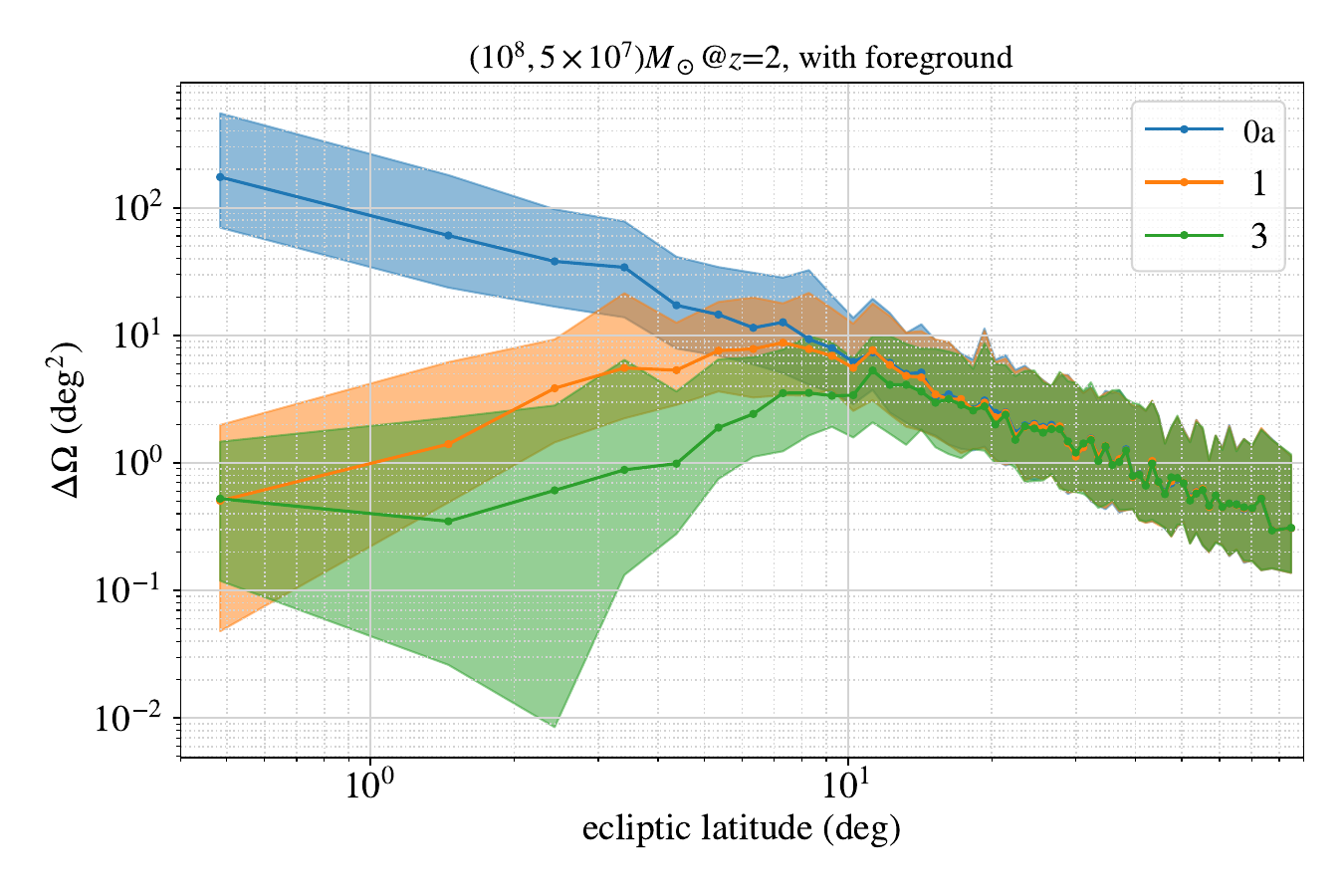}
     \caption{The angular resolutions for the GW (2,2) mode are shown for three sources under different orbital configurations. The three rows correspond to the light, medium, and heavy sources, respectively. The left column presents cumulative histograms of angular resolution under two noise scenarios: one including the original foreground (solid curves) and the other using the foreground residual after subtraction (dashed curves). The legend labeled ``0a'' represents the mission orbit without inclination, while ``1'' and ``3'' correspond to orbits with 1$^\circ$ and 3$^\circ$ inclinations, respectively. The right column depicts how angular resolution varies with ecliptic latitude under full foreground noise. For comparison, angular resolutions from the LISA detector are shown by the red curves in the upper-left panel of the left column. In the right column, solid curves represent the median angular resolution within each latitude bin, and the shaded regions of the same color indicate the corresponding $1\sigma$ ranges. }
     \label{fig:histogram_22_single}
 \end{figure*}

Fig. \ref{fig:histogram_22_single} illustrates the angular resolutions of GW signals from the (2,2) mode for three distinct orbital configurations.
The left column presents cumulative histograms of angular resolution under two noise scenarios: the original foreground noise (solid curves) and the foreground residuals after subtraction (dashed curves).
The legends labeled 0a, 1, and 3 correspond to the orbital designs described in Section \ref{subsec:orbit}.
The right column of Fig. \ref{fig:histogram_22_single} displays how angular resolution varies with ecliptic latitude for the $1^\circ$ inclination orbit.
The three rows, from top to bottom, represent results for the light, medium, and heavy source populations, respectively.

The angular resolutions for the light source are illustrated in the first row of Fig.  \ref{fig:histogram_22_single}.
When the full foreground noise is considered (solid curves), the directions of MBBHs can mostly be determined within 0.1 deg$^2$. If the foreground is subtracted and only the residual is considered (dashed curves), the angular resolutions improve by approximately one order of magnitude. This improvement arises because the light source is significantly affected by foreground noise, as shown in Fig. \ref{fig:sensitivity}.
Mitigating the foreground can therefore enhance the observational precision for this source. Differences in capability among the three orbital configurations become most evident in the worst-performing 20\% of cases.
The influence of the mission with orbital precession is illustrated in the right panel by the angular resolution as a function of ecliptic latitude. The solid curves represent the median angular resolution within latitude bins, while the shaded regions of the same color indicate the corresponding $1\sigma$ ranges. For the mission orbit without inclination, the resolution worsens with increasing latitude, from the ecliptic plane to the poles. When the orbit is inclined, the angular resolution near the ecliptic plane improves significantly, with the degree of improvement increasing for larger orbital inclinations.

\begin{figure}[htbp]
     \centering
     \includegraphics[width=0.48\textwidth]{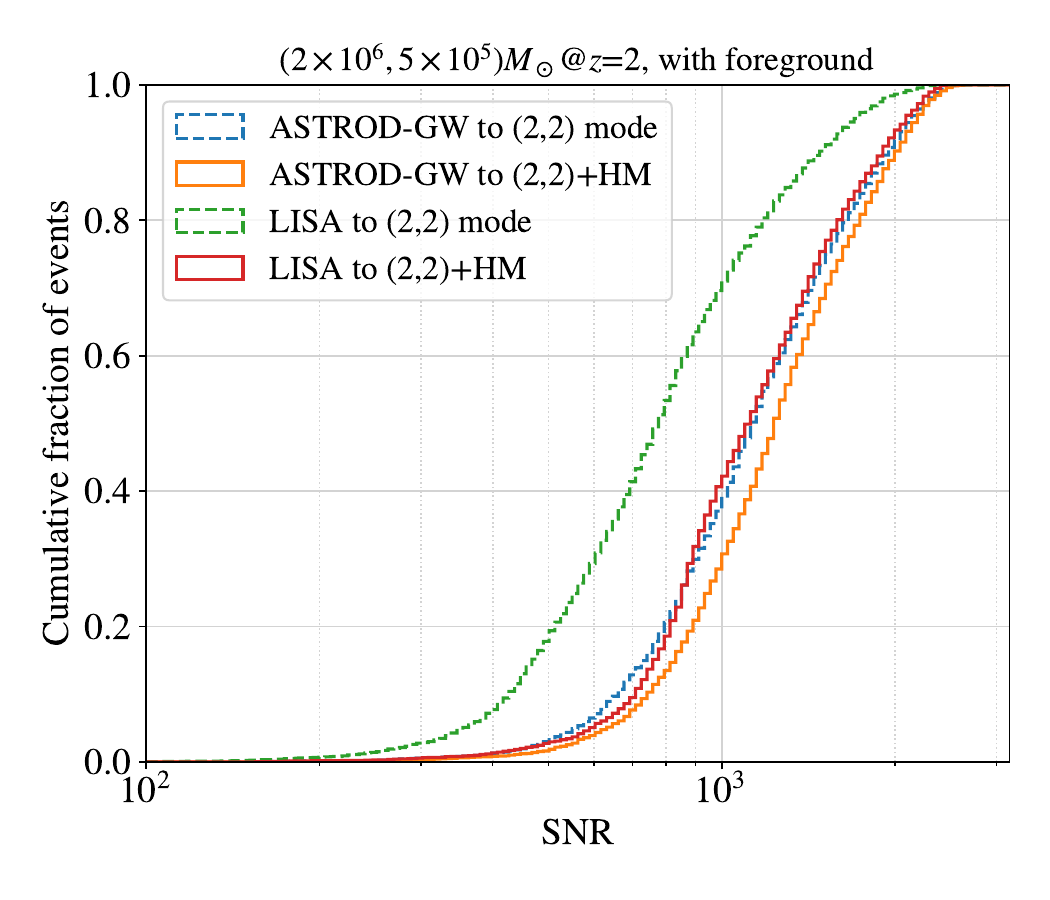}
     \caption{Histograms of the SNR for the light-mass MBBH are shown for both ASTROD-GW and LISA, using the full foreground noise. The dashed curves represent the SNR calculated using only the GW (2,2) mode, while the solid curves show the results obtained when including both the (2,2) and higher-order modes.}
    \label{fig:hist_snr_ASTRODGW_LISA}
 \end{figure}

For comparison, the angular resolutions obtained with the LISA detector are shown by the red curves in the upper-left panel.
The reduction of the foreground has a moderate impact on its performance.
Compared to the ASTROD-GW detector, LISA’s angular resolutions (red curves) are worse by approximately four orders of magnitude, despite LISA’s SNR being only about half that of ASTROD-GW, as shown in Fig. \ref{fig:hist_snr_ASTRODGW_LISA}.
We infer that this disparity arises because the angular resolution of a detector significantly degrades in the long-wavelength regime, and the optimal antenna arm length should be comparable to the targeted wavelengths.

The angular resolutions for the three mission orbits, evaluated for MBBHs, are shown in the middle row of Fig. \ref{fig:histogram_22_single}. The angular localization for this population is worse than that of the light source over the same observation period. Furthermore, the improvement in directional constraints due to foreground subtraction is less significant than for the light source, with gains of less than one order of magnitude. When comparing how angular resolution varies with ecliptic latitude, the inclined orbits show greater capability in localizing sources near low latitudes. 
The bottom row of Fig. \ref{fig:histogram_22_single} presents the angular resolutions for the heavy source population. As shown in the left panel, the localization precision is worse than for the lighter sources when using a single detector. This degradation is attributed to two primary factors: 1) the reduced response of a shorter antenna to long-wavelength signals, and 2) the fewer GW cycles emitted by heavy MBBHs over a fixed observation period.
When comparing results with and without foreground subtraction, little improvement is observed in localization accuracy. 
This is because the foreground is negligible in this more sensitive frequency band, as shown in Fig. \ref{fig:sensitivity}. Inclination of the mission orbit also improves angular resolution near the ecliptic plane, as shown in the lower-right panel. This improvement is primarily confined to the latitude range of $[-10^\circ, 10^\circ]$, and the performance of all three orbit configurations converges for absolute latitudes greater than $10^\circ$.

\subsection{Angular resolution with (2,2) and higher-order modes}

\begin{figure*}[htbp]
     \centering
     \includegraphics[width=0.42\textwidth]{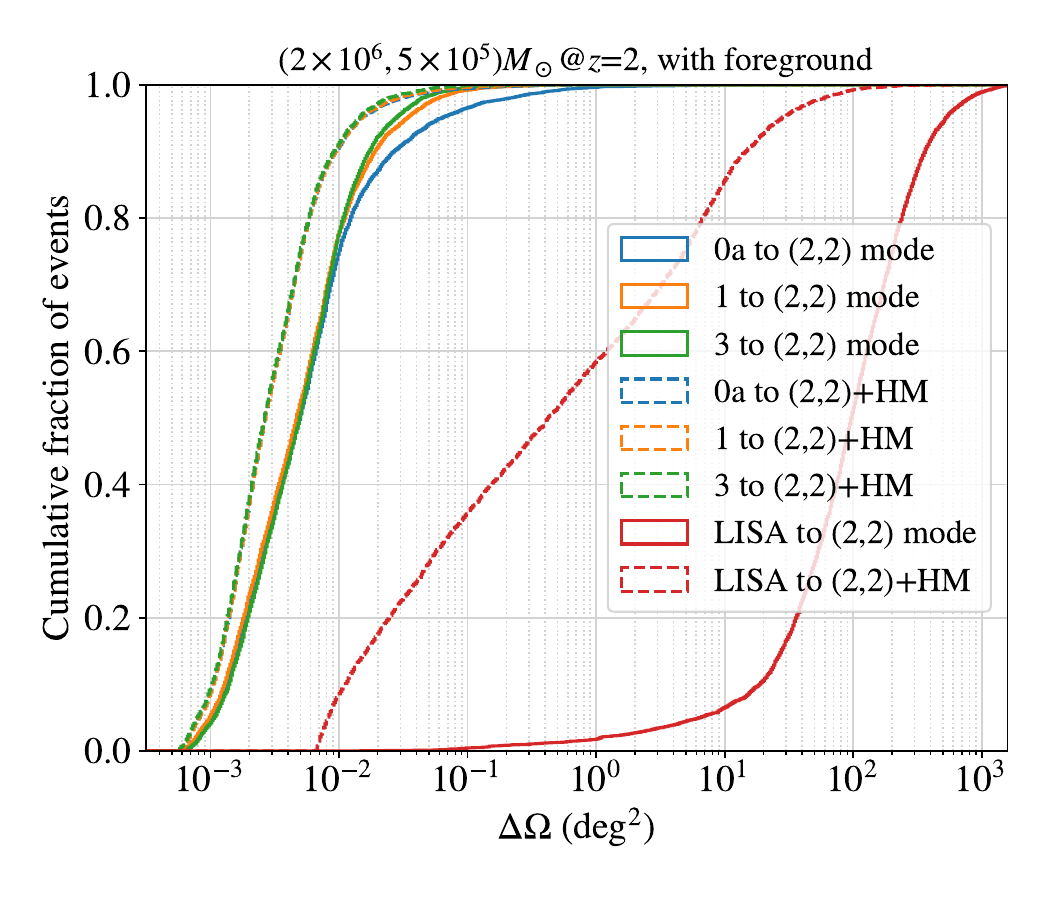}
     \includegraphics[width=0.54\textwidth]{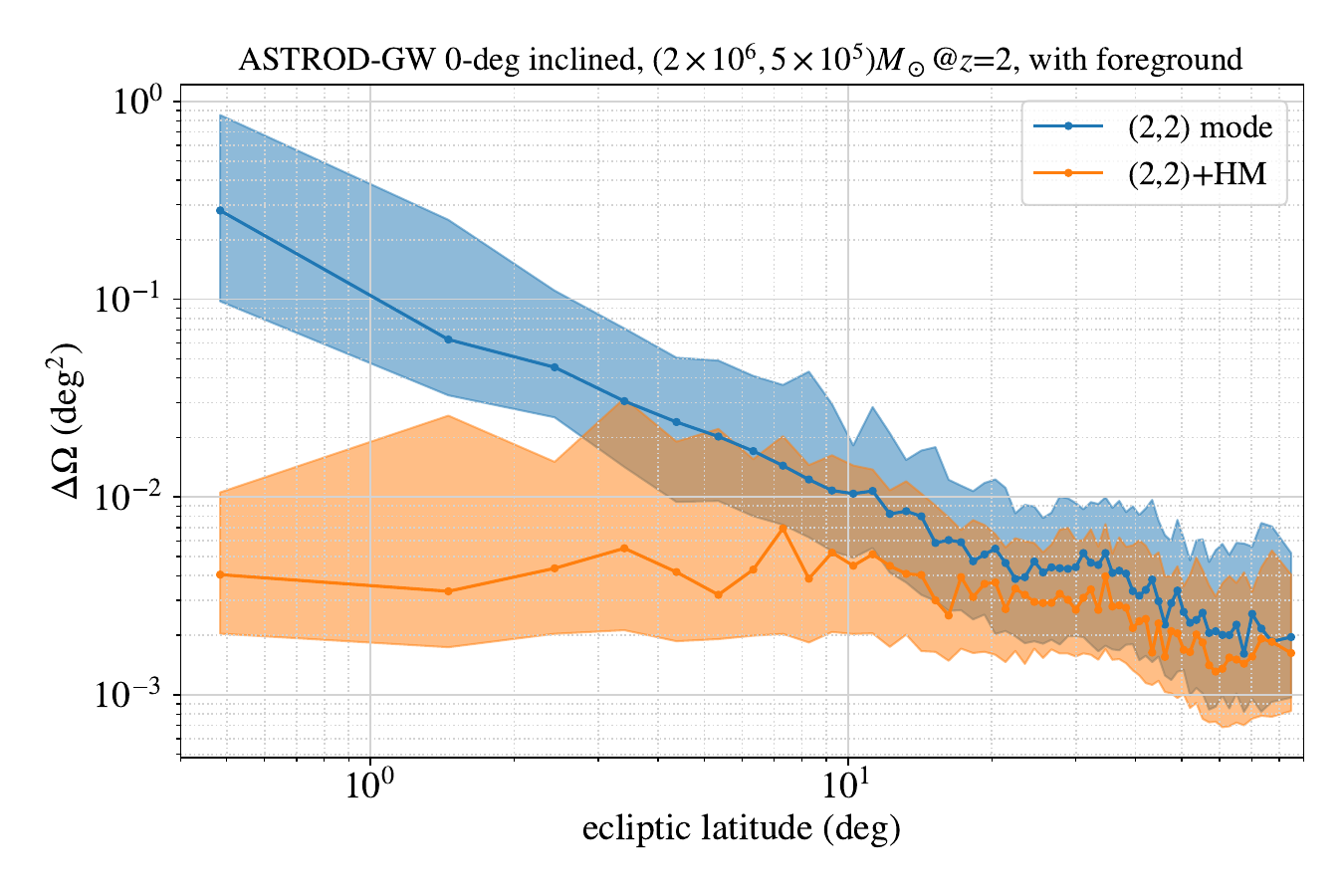}
     \includegraphics[width=0.42\textwidth]{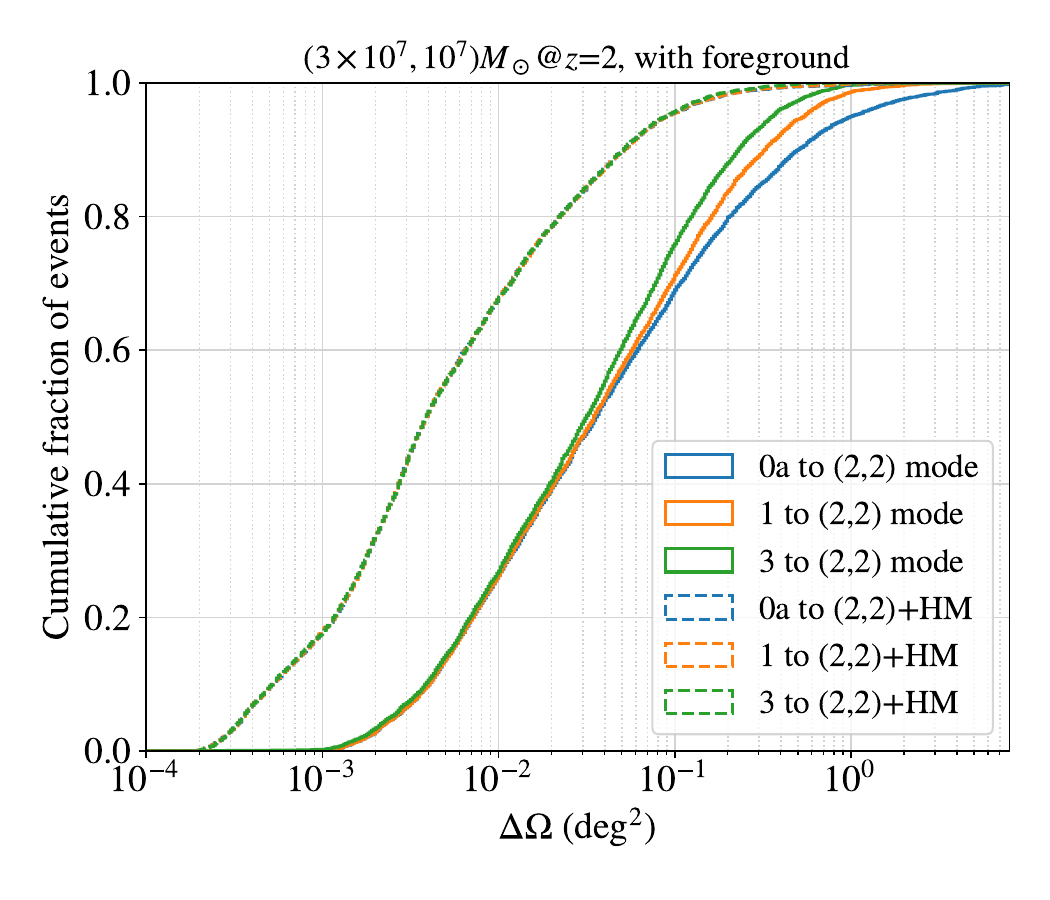}
     \includegraphics[width=0.54\textwidth]{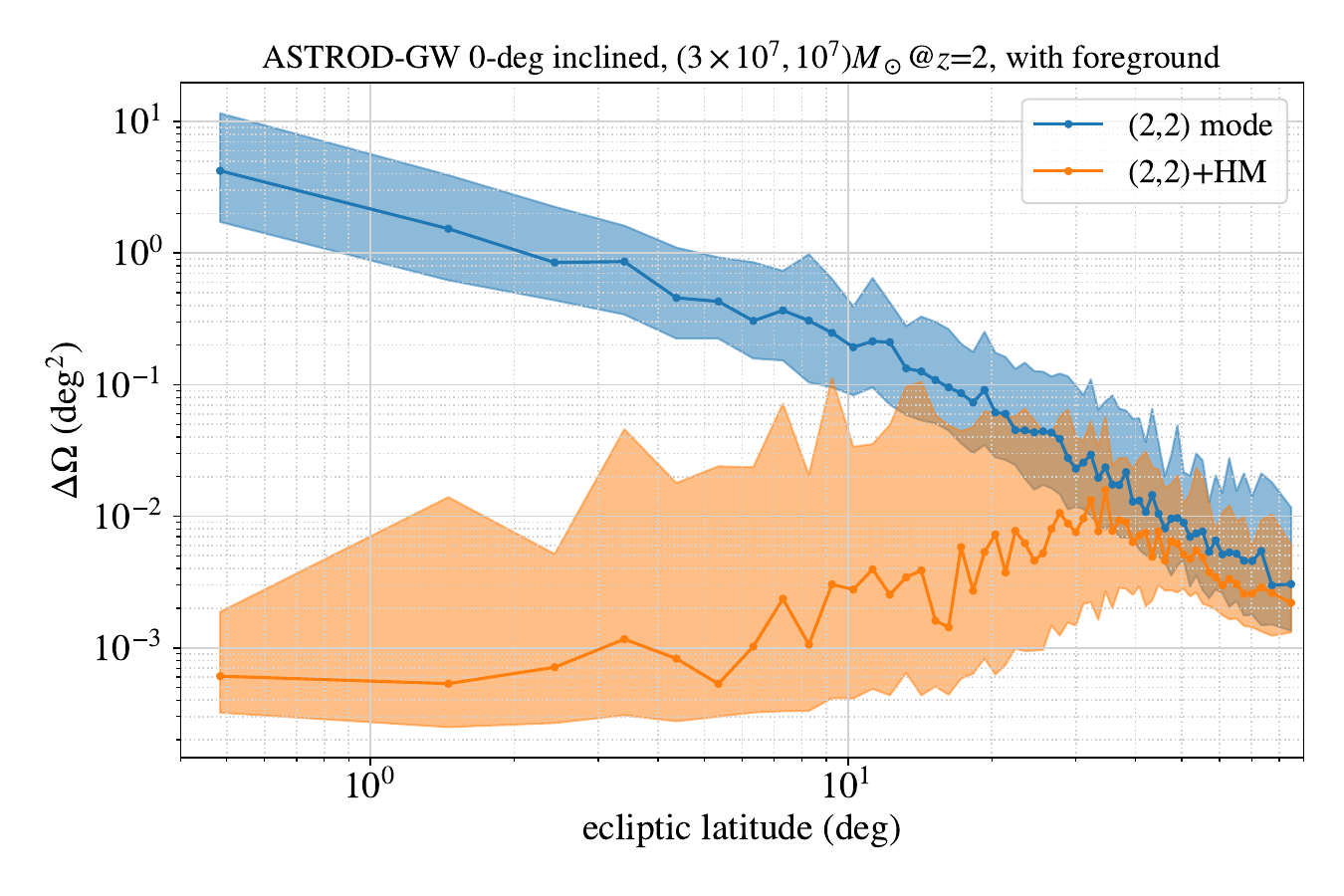}
     \includegraphics[width=0.42\textwidth]{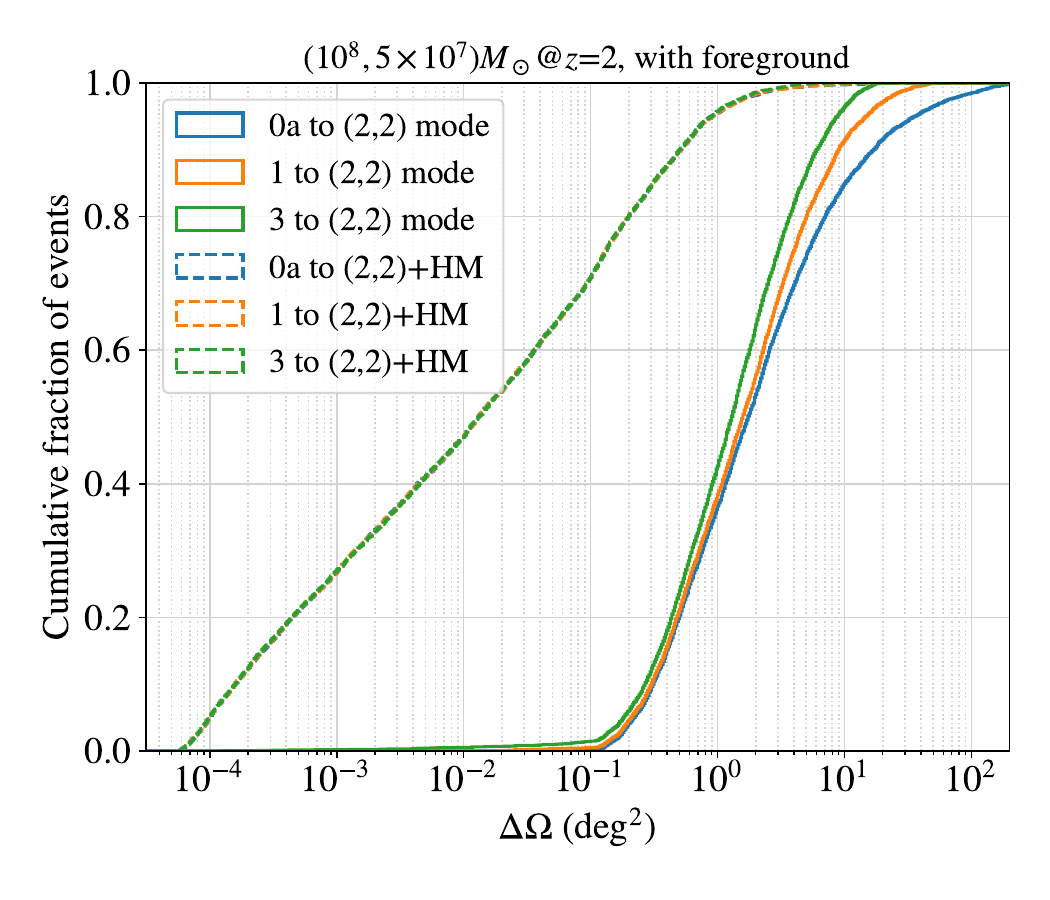}
     \includegraphics[width=0.54\textwidth]{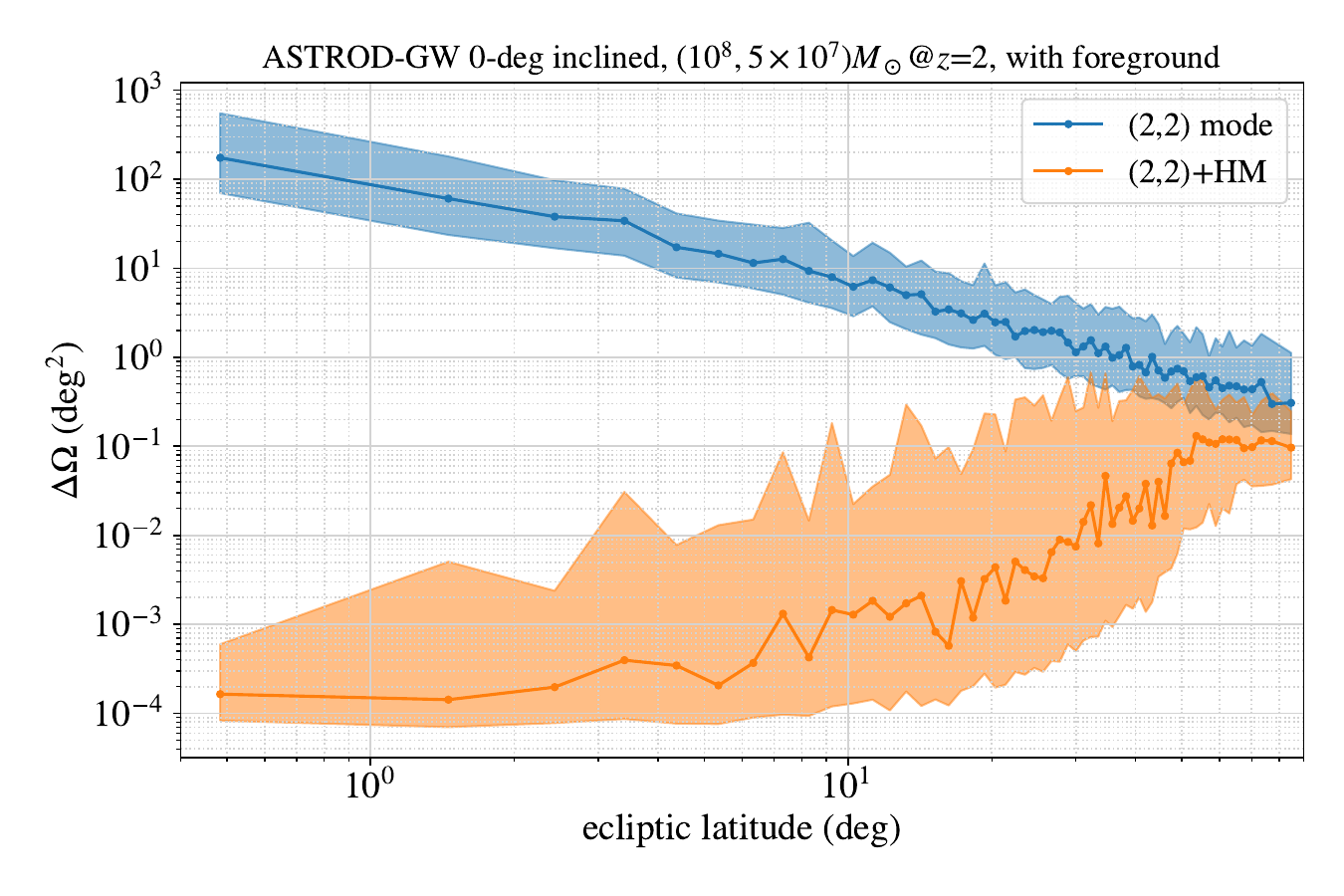}
     \caption{Angular resolutions of ASTROD-GW for three source populations using GW waveforms that include the dominant (2,2) mode and additional higher-order modes (2,1), (3,3), (3,2), (4,4), obtained under the full foreground noise scenario. The left column shows cumulative histograms of angular resolution for different orbital configurations, while the right column presents angular resolution as a function of ecliptic latitude using the non-inclined orbital configuration. For comparison, results from the LISA detector are shown by the red curves in the upper-left panel. }
    \label{fig:histogram_22_vs_hm_single}
 \end{figure*}

In this subsection, we further analyze the improvement in angular resolution resulting from the inclusion of GW higher harmonic modes, $(\ell,|m|) = (2,1), (3,3), (3,2), (4,4)$, in addition to the dominant $(2,2)$ mode. Fig. \ref{fig:histogram_22_vs_hm_single} illustrates the angular resolutions of the three ASTROD-GW orbital configurations for three source populations, with and without the inclusion of higher-order modes.
The left column presents cumulative histograms of angular resolution, where solid curves correspond to results using only the (2,2) mode, and dashed curves include the additional higher harmonics. The layout of Fig. \ref{fig:histogram_22_vs_hm_single} follows the same structure as Fig. \ref{fig:histogram_22_single}. The right column compares angular resolution as a function of ecliptic latitude, with and without higher-order modes, using the $1^\circ$ inclination orbital configuration.

The angular resolutions for the light source are shown in the first row of Fig. \ref{fig:histogram_22_vs_hm_single}. For ASTROD-GW, the localization precision improves by a factor of approximately 2 when higher-order modes are included in addition to the dominant (2,2) mode. The performance of the three orbital configurations is largely comparable, as indicated by the overlapping dashed curves, which is consistent with the conclusion in Section \ref{subsec:pe_hm}. Notably, in the left panel, the inclusion of higher-order modes leads to an improvement in angular resolution for the LISA mission by one or more orders of magnitude. This enhancement occurs because the higher-order modes fall within LISA’s more responsive and sensitive frequency band, contributing substantially to the SNR, as shown in Fig. \ref{fig:hist_snr_ASTRODGW_LISA}. For ASTROD-GW, the improvement due to higher-order modes is most significant for sources located near the ecliptic plane, as shown in the upper-right panel.

The angular resolutions for the medium-mass source are shown in the middle row of Fig. \ref{fig:histogram_22_vs_hm_single}. Compared to the light source, the inclusion of higher-order modes more significantly reduces localization uncertainty for heavier MBBHs, as illustrated in the middle and bottom plots of the left column. One contributing factor is that the characteristic frequency of ASTROD-GW is $f_c = 1/L \simeq \frac{1}{864 \ \mathrm{s}} \simeq 1.16$ mHz. For heavier sources, the (2,2) mode lies at frequencies lower than this characteristic frequency, while their higher-order modes are more optimally matched to the detector’s response and therefore contribute more effectively to constraining extrinsic parameters.
For the medium-mass source shown in the middle-left panel of Fig. \ref{fig:histogram_22_vs_hm_single}, the inclusion of higher-order modes improves angular resolution by approximately an order of magnitude. For the heavy source (bottom-left panel), the improvement is even more pronounced, with gains ranging from a factor of $\sim$10 to $\sim$1000. 
An important conclusion from the left column of Fig. \ref{fig:histogram_22_vs_hm_single} is that the inclusion of higher-order modes reduces the dependence of angular resolution on the source’s orbital inclination, suggesting that these modes yield consistent localization performance regardless of orbital geometry. As a result, the impact of mission orbital inclination on the angular resolution of MBBHs is substantially diminished.

To further investigate the impact of higher-order modes on angular resolution, the right column of Fig. \ref{fig:histogram_22_vs_hm_single} presents how localization precision varies with ecliptic latitude, using the non-inclined orbital configuration. The comparison is made between angular resolutions obtained using only the (2,2) mode and those including additional higher-order modes.
Notably, the improvements are especially pronounced for sources located near the ecliptic plane.
For light sources, the angular resolution improves by approximately two orders of magnitude for binaries situated close to the ecliptic. For medium-mass binaries, localization precision improves by multiple orders of magnitude within the latitude range of [-20$^\circ$, 20$^\circ$]. The angular resolution for heavy sources exhibits the most substantial enhancement, improving by nearly six orders of magnitude at low latitudes.
Another noteworthy observation is that, when using only the (2,2) mode, the most precisely localized sources tend to lie near the polar regions. However, with the inclusion of higher-order modes, binaries near the ecliptic plane become more precisely localized particularly for heavier sources.

\begin{figure}[htbp]
     \centering
     \includegraphics[width=0.48\textwidth]{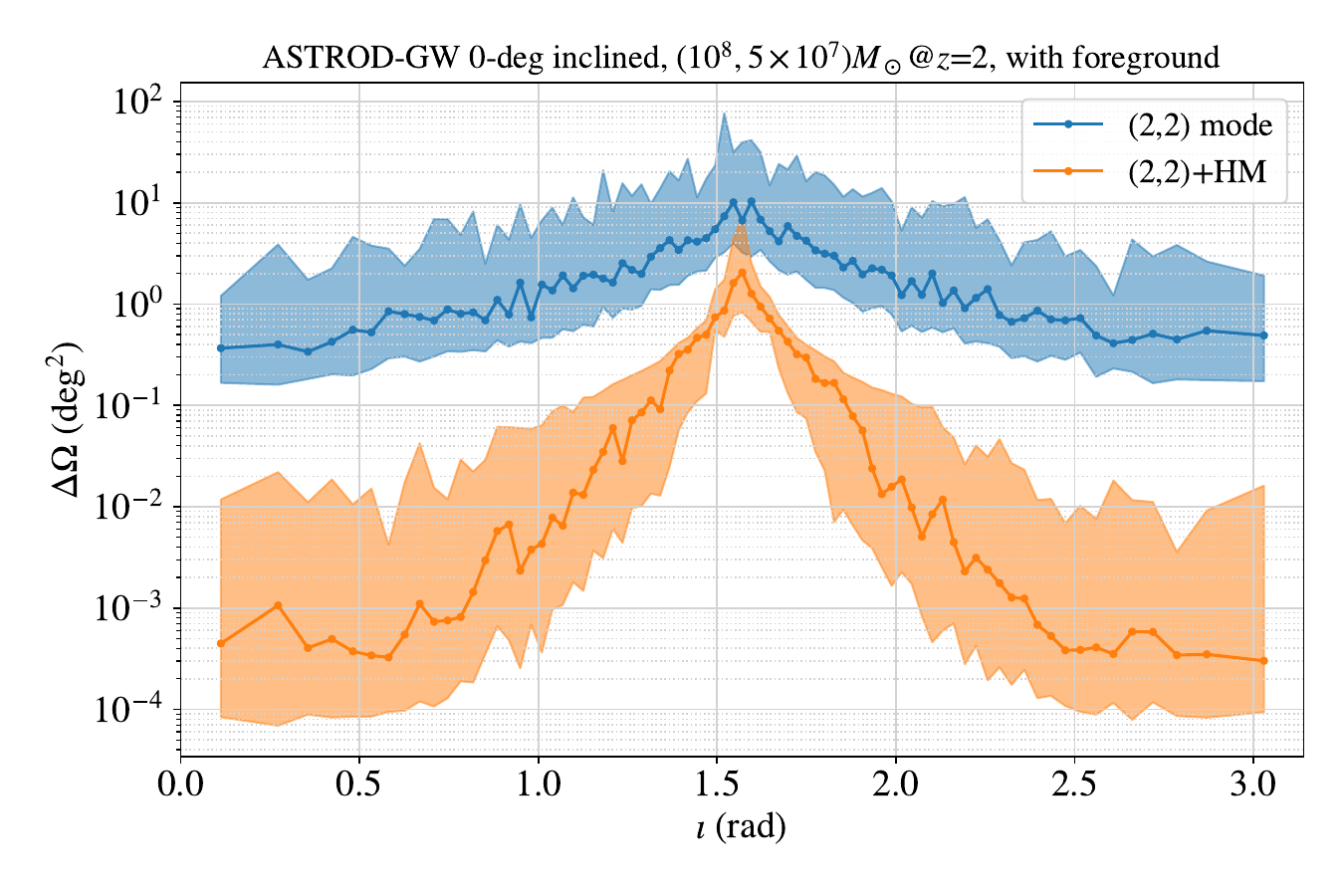}
     \caption{Angular resolution as a function of inclination angle ($\iota$) for three sources, obtained using the full foreground noise model and the non-inclined orbital configuration of ASTROD-GW. Results using only the (2,2) mode are shown in blue, while those including both the (2,2) and higher-order modes are shown in orange.}
    \label{fig:iota_vs_resolution_ASTRODGW0}
 \end{figure}

The amplitudes of GW higher-order modes depend on the inclination angle ($\iota$) between the line of sight and the orbital angular momentum of the binary. These higher-order modes we considered are significantly suppressed when the binary is nearly face-on ($\iota=0$) or face-off ($\iota=\pi$). Fig. \ref{fig:iota_vs_resolution_ASTRODGW0} shows how angular resolution varies with the inclination angle for a heavy source, using the non-inclined orbital configuration of ASTROD-GW. The results obtained using only the GW (2,2) mode are shown in blue, while those including additional higher-order modes are shown in orange. In the (2,2)-only case, the angular resolution is better for binaries that are face-on or face-off, since the $+$ and $\times$ GW polarizations have stronger amplitudes in these orientations, yielding a higher signal-to-noise ratio (SNR). Near the edge-on orientation ($\iota=\pi/2$), the amplitudes of two polarizations are reduced, resulting in poorer parameter estimation. When higher-order modes are included, the most substantial improvements also occur near face-on and face-off orientations. However, because higher-order modes are themselves suppressed at these inclinations, the angular resolution exhibits greater variance around these orientations as shown by the orange region in Fig. \ref{fig:iota_vs_resolution_ASTRODGW0}.

\begin{figure}[htbp]
     \centering
     \includegraphics[width=0.46\textwidth]{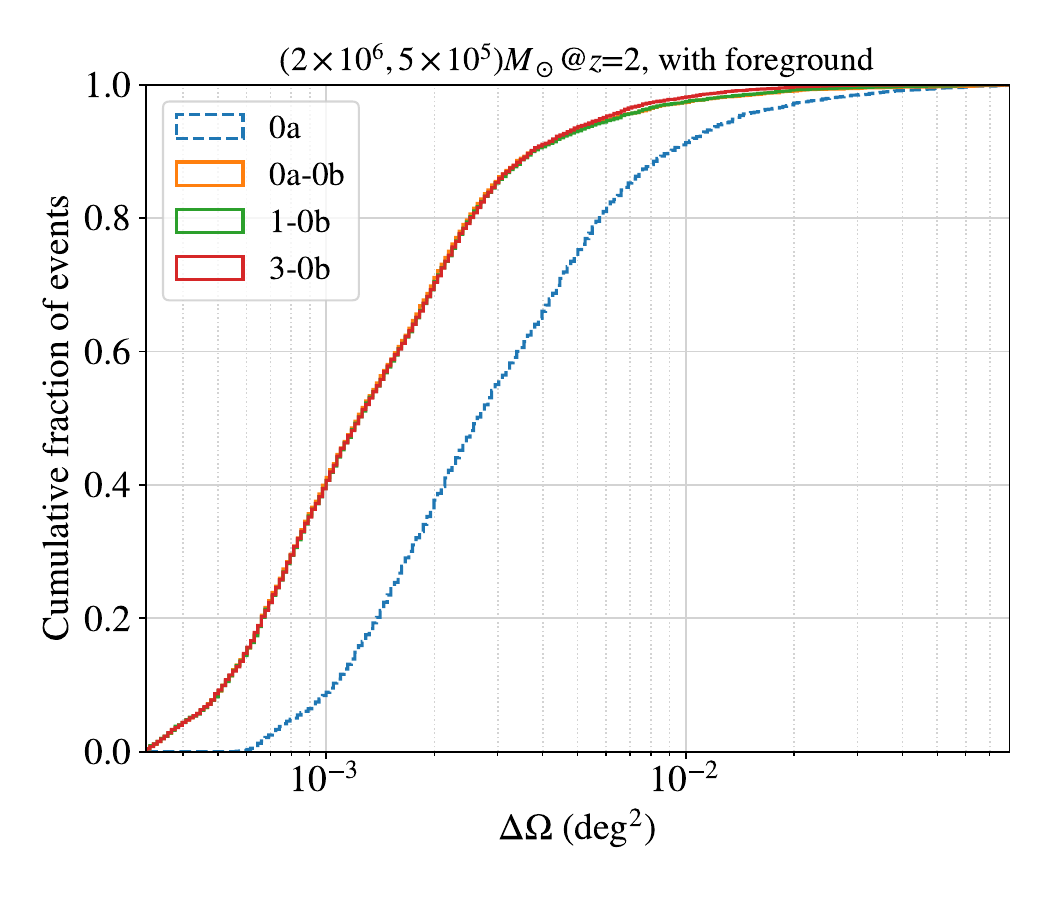}
     \includegraphics[width=0.46\textwidth]{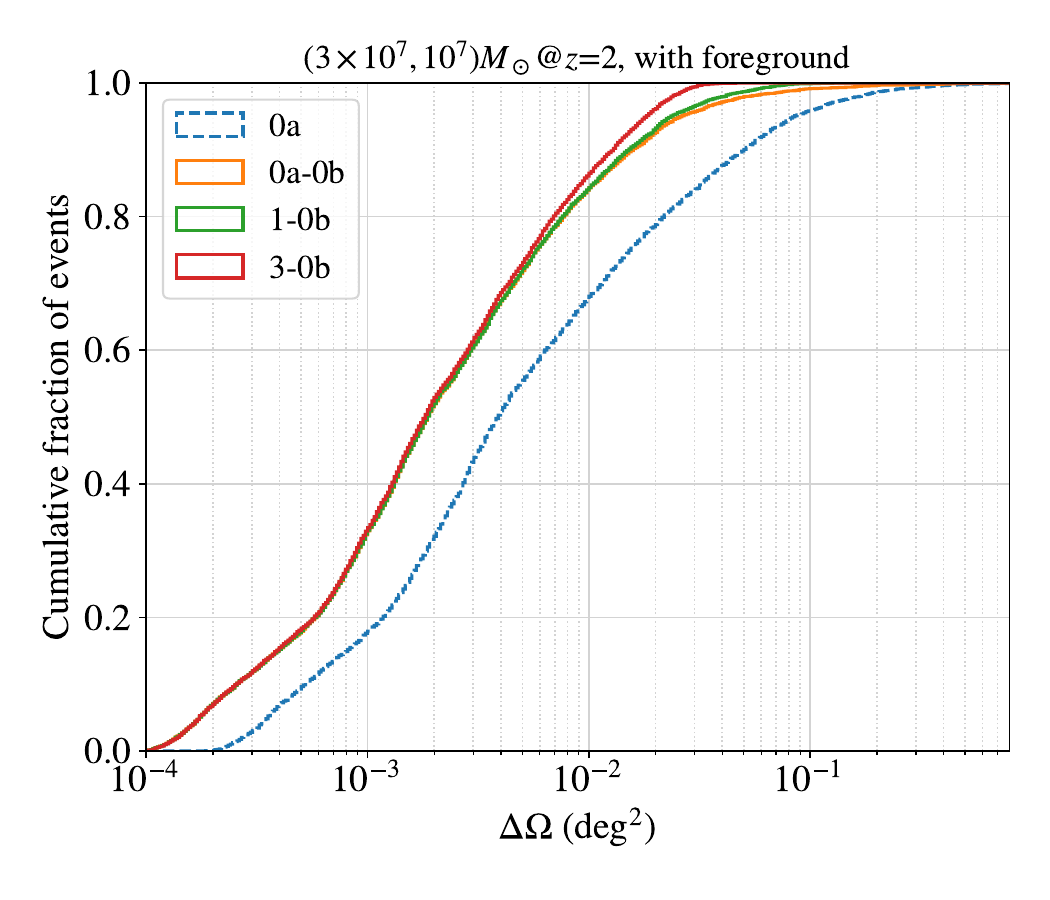}
     \includegraphics[width=0.46\textwidth]{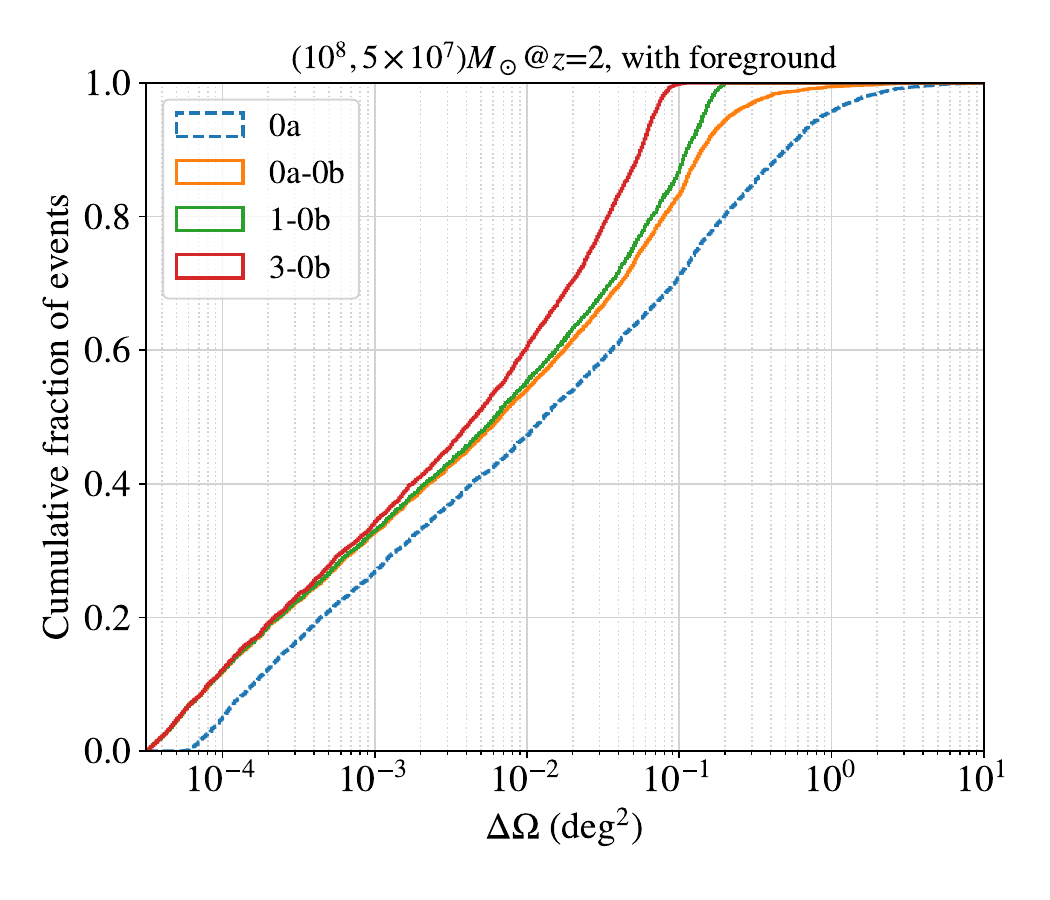}
     \caption{Angular resolutions for three source populations using joint observations from dual-detector networks. The plots from top to bottom correspond to light, medium, and heavy sources, respectively. For each network configuration, solid curves represent results obtained using only the GW (2,2) mode, while dashed curves of the same color indicate results that include additional higher-order modes.
     }
     \label{fig:histogram_22_vs_hm_joint}
 \end{figure}

Several sub-mHz GW missions have been proposed, including ASTROD-GW \cite{Ni:2012eh,Ni:2016wcv}, the Folkner mission \cite{Baker:2019pnp}, and LISAmax \cite{Martens:2023mgm}. Joint observations from dual-detector networks can further enhance angular resolution. Fig. \ref{fig:histogram_22_vs_hm_joint} presents the angular resolutions for three source populations using a dual-detector network under foreground noise.
The cumulative histograms, from top to bottom, correspond to the light, medium, and heavy sources, respectively. Three network configurations are considered: 1) 0a–0b: a joint observation using detectors in the 0a and 0b orbital configurations, as described in Section \ref{subsec:orbit}, 2) 1–0b: a joint observation with orbits 1 and 0b, 3) 3–0b: a joint observation with orbits 3 and 0b.
For comparison, the results from a single detector using the 0a configuration are shown by the dashed blue curves. The evaluation includes the dominant (2,2) mode and higher-order modes.

As shown in the top panel of Fig. \ref{fig:histogram_22_vs_hm_joint}, the angular resolutions from the three dual-detector network configurations are largely overlapping, indicating comparable capabilities in localizing MBBHs. Compared to the resolution from a single mission using the orbit 0a configuration, the joint observations achieve an improvement of approximately a factor of 2, primarily due to the increased SNR from the combined detectors.
For medium-mass sources, the three networks exhibit slight performance differences for the least well-localized events, as seen in the middle panel.
Owing to orbital precession, the 3–0b network (red solid curve) achieves a noticeable improvement in localization precision. This improvement becomes even more pronounced for heavy MBBHs, as shown in the bottom panel. By contrast, the non-precessing dual-detector network, 0a–0b, shows more limited effectiveness, achieving only an approximate twofold reduction in sky localization uncertainty.

\section{Conclusions and Discussion} \label{sec:conclusion}

In this work, we have investigated the capability of future sub-mHz GW missions to detect and localize MBBH coalescences. We began by evaluating the sensitivity of such missions while fully accounting for the galactic foreground noise. Despite the presence of this confusion-limited noise, we find that SNRs for MBBHs can still reach values as high as several thousand across a wide range of redshifts, confirming the strong detection potential of sub-mHz detectors.
To assess localization performance, we considered three mission orbit configurations: a non-precessing configuration and two precessing configurations. Our results indicate that, when only the (2,2) mode is considered, the non-precessing orbit suffers from a two-fold degeneracy in sky localization between opposite hemispheres. This degeneracy is resolved either by using precessing orbits or by including higher harmonics. Moreover, the inclusion of higher-order modes significantly improves localization precision, especially for sources near the ecliptic plane, and reduces the dependence on orbital precession. As a result, the three orbital configurations perform comparably when higher-order modes are present.

The sensitivity band of the sub-mHz mission investigated here spans [10 $\mu$Hz, 10 mHz], overlapping partially with the LISA frequency range. However, due to its longer interferometric arms, the sub-mHz configuration offers enhanced response to lower-frequency GWs, allowing for more precise localization of high-mass MBBHs. This advantage makes sub-mHz missions especially promising for enabling multi-messenger observations and improving measurements in GW cosmology. Additionally, the extremely high SNRs achievable, potentially exceeding 10,000, provide a unique opportunity to probe general relativity and alternative theories of gravity with unprecedented precision.

We also examined the benefits of joint observations using two identical sub-mHz detectors. While the improvement in sky localization for individual MBBHs is limited due to the nearly aligned antenna patterns, the dual-detector configuration will be crucial for detecting a SGWB. In particular, cross-correlation of two interferometers placed in the ecliptic plane provides an optimal setup for isolating the SGWB signal from instrumental noise. A comprehensive investigation of the SGWB detection potential of sub-mHz missions will be pursued in future work.

\section*{Acknowledgements}
We thank Prof. Wei-Tou Ni and Prof. Zhoujian Cao for their helpful comments and suggestions. 
G.W. was supported by the National Key Research and Development Program of China under Grant No. 2021YFC2201903.
R.J.W. was supported by the National Natural Science Foundation of China No.12235019.
B.H. was supported by the National Key Research and Development Program of China Grant No. 2021YFC2203001.
R.G.C. was supported by the National Natural Science Foundation of China Grants No. 12235019.
Y.M.X. was supported by the Universitat de les Illes Balears (UIB); the Spanish Agencia Estatal de Investigación grants PID2022-138626NB-I00, RED2022-134204-E, RED2022-134411-T, funded by MICIU/AEI/10.13039/501100011033 and the ERDF/EU; and the Comunitat Autònoma de les Illes Balears through the Conselleria d'Educació i Universitats with funds from the European Union - NextGenerationEU/PRTR-C17.I1 (SINCO2022/6719) and from the European Union - European Regional Development Fund (ERDF) (SINCO2022/18146).

This work are performed by using the python packages \textsf{numpy} \cite{harris2020array}, \textsf{scipy} \cite{2020SciPy-NMeth}, \textsf{pandas} \cite{pandas}, \textsf{MultiNest} \cite{Feroz:2008xx} and \textsf{PyMultiNest} \cite{Buchner:2014nha}, and the plots are make by utilizing \textsf{matplotlib} \cite{Hunter:2007ouj}, \textsf{GetDist} \cite{Lewis:2019xzd}. The GW signals are generated using \textsf{LALSuite} \cite{lalsuite,swiglal} and \textsf{PyCBC} \cite{alex_nitz_2024_10473621}.


\bibliography{ref}

\begin{thebibliography}{79}%
\makeatletter
\providecommand \@ifxundefined [1]{%
 \@ifx{#1\undefined}
}%
\providecommand \@ifnum [1]{%
 \ifnum #1\expandafter \@firstoftwo
 \else \expandafter \@secondoftwo
 \fi
}%
\providecommand \@ifx [1]{%
 \ifx #1\expandafter \@firstoftwo
 \else \expandafter \@secondoftwo
 \fi
}%
\providecommand \natexlab [1]{#1}%
\providecommand \enquote  [1]{``#1''}%
\providecommand \bibnamefont  [1]{#1}%
\providecommand \bibfnamefont [1]{#1}%
\providecommand \citenamefont [1]{#1}%
\providecommand \href@noop [0]{\@secondoftwo}%
\providecommand \href [0]{\begingroup \@sanitize@url \@href}%
\providecommand \@href[1]{\@@startlink{#1}\@@href}%
\providecommand \@@href[1]{\endgroup#1\@@endlink}%
\providecommand \@sanitize@url [0]{\catcode `\\12\catcode `\$12\catcode
  `\&12\catcode `\#12\catcode `\^12\catcode `\_12\catcode `\%12\relax}%
\providecommand \@@startlink[1]{}%
\providecommand \@@endlink[0]{}%
\providecommand \url  [0]{\begingroup\@sanitize@url \@url }%
\providecommand \@url [1]{\endgroup\@href {#1}{\urlprefix }}%
\providecommand \urlprefix  [0]{URL }%
\providecommand \Eprint [0]{\href }%
\providecommand \doibase [0]{https://doi.org/}%
\providecommand \selectlanguage [0]{\@gobble}%
\providecommand \bibinfo  [0]{\@secondoftwo}%
\providecommand \bibfield  [0]{\@secondoftwo}%
\providecommand \translation [1]{[#1]}%
\providecommand \BibitemOpen [0]{}%
\providecommand \bibitemStop [0]{}%
\providecommand \bibitemNoStop [0]{.\EOS\space}%
\providecommand \EOS [0]{\spacefactor3000\relax}%
\providecommand \BibitemShut  [1]{\csname bibitem#1\endcsname}%
\let\auto@bib@innerbib\@empty
\bibitem [{\citenamefont {Abbott}\ \emph {et~al.}(2016)\citenamefont {Abbott}
  \emph {et~al.}}]{LIGOScientific:2016aoc}%
  \BibitemOpen
  \bibfield  {author} {\bibinfo {author} {\bibfnamefont {B.~P.}\ \bibnamefont
  {Abbott}} \emph {et~al.} (\bibinfo {collaboration} {LIGO Scientific,
  Virgo}),\ }\bibfield  {title} {\bibinfo {title} {{Observation of
  Gravitational Waves from a Binary Black Hole Merger}},\ }\href
  {https://doi.org/10.1103/PhysRevLett.116.061102} {\bibfield  {journal}
  {\bibinfo  {journal} {Phys. Rev. Lett.}\ }\textbf {\bibinfo {volume} {116}},\
  \bibinfo {pages} {061102} (\bibinfo {year} {2016})},\ \Eprint
  {https://arxiv.org/abs/1602.03837} {arXiv:1602.03837 [gr-qc]} \BibitemShut
  {NoStop}%
\bibitem [{\citenamefont {Abbott}\ \emph {et~al.}(2019)\citenamefont {Abbott}
  \emph {et~al.}}]{LIGOScientific:2018mvr}%
  \BibitemOpen
  \bibfield  {author} {\bibinfo {author} {\bibfnamefont {B.~P.}\ \bibnamefont
  {Abbott}} \emph {et~al.} (\bibinfo {collaboration} {LIGO Scientific,
  Virgo}),\ }\bibfield  {title} {\bibinfo {title} {{GWTC-1: A
  Gravitational-Wave Transient Catalog of Compact Binary Mergers Observed by
  LIGO and Virgo during the First and Second Observing Runs}},\ }\href
  {https://doi.org/10.1103/PhysRevX.9.031040} {\bibfield  {journal} {\bibinfo
  {journal} {Phys. Rev. X}\ }\textbf {\bibinfo {volume} {9}},\ \bibinfo {pages}
  {031040} (\bibinfo {year} {2019})},\ \Eprint
  {https://arxiv.org/abs/1811.12907} {arXiv:1811.12907 [astro-ph.HE]}
  \BibitemShut {NoStop}%
\bibitem [{\citenamefont {Abbott}\ \emph
  {et~al.}(2021{\natexlab{a}})\citenamefont {Abbott} \emph
  {et~al.}}]{LIGOScientific:2020ibl}%
  \BibitemOpen
  \bibfield  {author} {\bibinfo {author} {\bibfnamefont {R.}~\bibnamefont
  {Abbott}} \emph {et~al.} (\bibinfo {collaboration} {LIGO Scientific,
  Virgo}),\ }\bibfield  {title} {\bibinfo {title} {{GWTC-2: Compact Binary
  Coalescences Observed by LIGO and Virgo During the First Half of the Third
  Observing Run}},\ }\href {https://doi.org/10.1103/PhysRevX.11.021053}
  {\bibfield  {journal} {\bibinfo  {journal} {Phys. Rev. X}\ }\textbf {\bibinfo
  {volume} {11}},\ \bibinfo {pages} {021053} (\bibinfo {year}
  {2021}{\natexlab{a}})},\ \Eprint {https://arxiv.org/abs/2010.14527}
  {arXiv:2010.14527 [gr-qc]} \BibitemShut {NoStop}%
\bibitem [{\citenamefont {Abbott}\ \emph
  {et~al.}(2021{\natexlab{b}})\citenamefont {Abbott} \emph
  {et~al.}}]{LIGOScientific:2021djp}%
  \BibitemOpen
  \bibfield  {author} {\bibinfo {author} {\bibfnamefont {R.}~\bibnamefont
  {Abbott}} \emph {et~al.} (\bibinfo {collaboration} {LIGO Scientific, VIRGO,
  KAGRA}),\ }\bibfield  {title} {\bibinfo {title} {{GWTC-3: Compact Binary
  Coalescences Observed by LIGO and Virgo During the Second Part of the Third
  Observing Run}},\ }\href@noop {} {\  (\bibinfo {year}
  {2021}{\natexlab{b}})},\ \Eprint {https://arxiv.org/abs/2111.03606}
  {arXiv:2111.03606 [gr-qc]} \BibitemShut {NoStop}%
\bibitem [{\citenamefont {Akutsu}\ \emph {et~al.}(2019)\citenamefont {Akutsu}
  \emph {et~al.}}]{KAGRA:2018plz}%
  \BibitemOpen
  \bibfield  {author} {\bibinfo {author} {\bibfnamefont {T.}~\bibnamefont
  {Akutsu}} \emph {et~al.} (\bibinfo {collaboration} {KAGRA}),\ }\bibfield
  {title} {\bibinfo {title} {{KAGRA: 2.5 Generation Interferometric
  Gravitational Wave Detector}},\ }\href
  {https://doi.org/10.1038/s41550-018-0658-y} {\bibfield  {journal} {\bibinfo
  {journal} {Nature Astron.}\ }\textbf {\bibinfo {volume} {3}},\ \bibinfo
  {pages} {35} (\bibinfo {year} {2019})},\ \Eprint
  {https://arxiv.org/abs/1811.08079} {arXiv:1811.08079 [gr-qc]} \BibitemShut
  {NoStop}%
\bibitem [{\citenamefont {{LIGO Scientific Collaboration, Virgo Collaboration,
  KAGRA Collaboration}}()}]{gracedb}%
  \BibitemOpen
  \bibfield  {author} {\bibinfo {author} {\bibnamefont {{LIGO Scientific
  Collaboration, Virgo Collaboration, KAGRA Collaboration}}},\ }\href@noop {}
  {\bibinfo {title} {{Gravitational-Wave Candidate Event Database
  (GraceDB)}}},\ \bibinfo {howpublished}
  {\url{https://gracedb.ligo.org}}\BibitemShut {NoStop}%
\bibitem [{\citenamefont {Punturo}\ \emph {et~al.}(2010)\citenamefont {Punturo}
  \emph {et~al.}}]{Punturo:2010zz}%
  \BibitemOpen
  \bibfield  {author} {\bibinfo {author} {\bibfnamefont {M.}~\bibnamefont
  {Punturo}} \emph {et~al.},\ }\bibfield  {title} {\bibinfo {title} {{The
  Einstein Telescope: A third-generation gravitational wave observatory}},\
  }\href {https://doi.org/10.1088/0264-9381/27/19/194002} {\bibfield  {journal}
  {\bibinfo  {journal} {Class. Quant. Grav.}\ }\textbf {\bibinfo {volume}
  {27}},\ \bibinfo {pages} {194002} (\bibinfo {year} {2010})}\BibitemShut
  {NoStop}%
\bibitem [{\citenamefont {Reitze}\ \emph {et~al.}(2019)\citenamefont {Reitze}
  \emph {et~al.}}]{Reitze:2019iox}%
  \BibitemOpen
  \bibfield  {author} {\bibinfo {author} {\bibfnamefont {D.}~\bibnamefont
  {Reitze}} \emph {et~al.},\ }\bibfield  {title} {\bibinfo {title} {{Cosmic
  Explorer: The U.S. Contribution to Gravitational-Wave Astronomy beyond
  LIGO}},\ }\href@noop {} {\bibfield  {journal} {\bibinfo  {journal} {Bull. Am.
  Astron. Soc.}\ }\textbf {\bibinfo {volume} {51}},\ \bibinfo {pages} {035}
  (\bibinfo {year} {2019})},\ \Eprint {https://arxiv.org/abs/1907.04833}
  {arXiv:1907.04833 [astro-ph.IM]} \BibitemShut {NoStop}%
\bibitem [{\citenamefont {Amaro-Seoane}\ \emph {et~al.}(2017)\citenamefont
  {Amaro-Seoane} \emph {et~al.}}]{LISA:2017pwj}%
  \BibitemOpen
  \bibfield  {author} {\bibinfo {author} {\bibfnamefont {P.}~\bibnamefont
  {Amaro-Seoane}} \emph {et~al.} (\bibinfo {collaboration} {LISA}),\ }\bibfield
   {title} {\bibinfo {title} {{Laser Interferometer Space Antenna}},\
  }\href@noop {} {\  (\bibinfo {year} {2017})},\ \Eprint
  {https://arxiv.org/abs/1702.00786} {arXiv:1702.00786 [astro-ph.IM]}
  \BibitemShut {NoStop}%
\bibitem [{\citenamefont {Hu}\ and\ \citenamefont {Wu}(2017)}]{Hu:2017mde}%
  \BibitemOpen
  \bibfield  {author} {\bibinfo {author} {\bibfnamefont {W.-R.}\ \bibnamefont
  {Hu}}\ and\ \bibinfo {author} {\bibfnamefont {Y.-L.}\ \bibnamefont {Wu}},\
  }\bibfield  {title} {\bibinfo {title} {{The Taiji Program in Space for
  gravitational wave physics and the nature of gravity}},\ }\href
  {https://doi.org/10.1093/nsr/nwx116} {\bibfield  {journal} {\bibinfo
  {journal} {Natl. Sci. Rev.}\ }\textbf {\bibinfo {volume} {4}},\ \bibinfo
  {pages} {685} (\bibinfo {year} {2017})}\BibitemShut {NoStop}%
\bibitem [{\citenamefont {Luo}\ \emph {et~al.}(2020)\citenamefont {Luo} \emph
  {et~al.}}]{Luo:2020bls}%
  \BibitemOpen
  \bibfield  {author} {\bibinfo {author} {\bibfnamefont {J.}~\bibnamefont
  {Luo}} \emph {et~al.},\ }\bibfield  {title} {\bibinfo {title} {{The first
  round result from the TianQin-1 satellite}},\ }\href
  {https://doi.org/10.1088/1361-6382/aba66a} {\bibfield  {journal} {\bibinfo
  {journal} {Class. Quant. Grav.}\ }\textbf {\bibinfo {volume} {37}},\ \bibinfo
  {pages} {185013} (\bibinfo {year} {2020})},\ \Eprint
  {https://arxiv.org/abs/2008.09534} {arXiv:2008.09534 [physics.ins-det]}
  \BibitemShut {NoStop}%
\bibitem [{\citenamefont {Barausse}\ and\ \citenamefont
  {Lapi}(2021)}]{Barausse:2020kjy}%
  \BibitemOpen
  \bibfield  {author} {\bibinfo {author} {\bibfnamefont {E.}~\bibnamefont
  {Barausse}}\ and\ \bibinfo {author} {\bibfnamefont {A.}~\bibnamefont
  {Lapi}},\ }\bibinfo {title} {{Massive Black-Hole Mergers}}\ (\bibinfo {year}
  {2021})\ \Eprint {https://arxiv.org/abs/2011.01994} {arXiv:2011.01994
  [astro-ph.GA]} \BibitemShut {NoStop}%
\bibitem [{\citenamefont {Babak}\ \emph {et~al.}(2017)\citenamefont {Babak},
  \citenamefont {Gair}, \citenamefont {Sesana}, \citenamefont {Barausse},
  \citenamefont {Sopuerta}, \citenamefont {Berry}, \citenamefont {Berti},
  \citenamefont {Amaro-Seoane}, \citenamefont {Petiteau},\ and\ \citenamefont
  {Klein}}]{Babak:2017tow}%
  \BibitemOpen
  \bibfield  {author} {\bibinfo {author} {\bibfnamefont {S.}~\bibnamefont
  {Babak}}, \bibinfo {author} {\bibfnamefont {J.}~\bibnamefont {Gair}},
  \bibinfo {author} {\bibfnamefont {A.}~\bibnamefont {Sesana}}, \bibinfo
  {author} {\bibfnamefont {E.}~\bibnamefont {Barausse}}, \bibinfo {author}
  {\bibfnamefont {C.~F.}\ \bibnamefont {Sopuerta}}, \bibinfo {author}
  {\bibfnamefont {C.~P.~L.}\ \bibnamefont {Berry}}, \bibinfo {author}
  {\bibfnamefont {E.}~\bibnamefont {Berti}}, \bibinfo {author} {\bibfnamefont
  {P.}~\bibnamefont {Amaro-Seoane}}, \bibinfo {author} {\bibfnamefont
  {A.}~\bibnamefont {Petiteau}},\ and\ \bibinfo {author} {\bibfnamefont
  {A.}~\bibnamefont {Klein}},\ }\bibfield  {title} {\bibinfo {title} {{Science
  with the space-based interferometer LISA. V: Extreme mass-ratio inspirals}},\
  }\href {https://doi.org/10.1103/PhysRevD.95.103012} {\bibfield  {journal}
  {\bibinfo  {journal} {Phys. Rev. D}\ }\textbf {\bibinfo {volume} {95}},\
  \bibinfo {pages} {103012} (\bibinfo {year} {2017})},\ \Eprint
  {https://arxiv.org/abs/1703.09722} {arXiv:1703.09722 [gr-qc]} \BibitemShut
  {NoStop}%
\bibitem [{\citenamefont {Littenberg}\ \emph {et~al.}(2020)\citenamefont
  {Littenberg}, \citenamefont {Cornish}, \citenamefont {Lackeos},\ and\
  \citenamefont {Robson}}]{Littenberg:2020bxy}%
  \BibitemOpen
  \bibfield  {author} {\bibinfo {author} {\bibfnamefont {T.}~\bibnamefont
  {Littenberg}}, \bibinfo {author} {\bibfnamefont {N.}~\bibnamefont {Cornish}},
  \bibinfo {author} {\bibfnamefont {K.}~\bibnamefont {Lackeos}},\ and\ \bibinfo
  {author} {\bibfnamefont {T.}~\bibnamefont {Robson}},\ }\bibfield  {title}
  {\bibinfo {title} {{Global Analysis of the Gravitational Wave Signal from
  Galactic Binaries}},\ }\href {https://doi.org/10.1103/PhysRevD.101.123021}
  {\bibfield  {journal} {\bibinfo  {journal} {Phys. Rev. D}\ }\textbf {\bibinfo
  {volume} {101}},\ \bibinfo {pages} {123021} (\bibinfo {year} {2020})},\
  \Eprint {https://arxiv.org/abs/2004.08464} {arXiv:2004.08464 [gr-qc]}
  \BibitemShut {NoStop}%
\bibitem [{\citenamefont {Schutz}(1986)}]{Schutz:1986gp}%
  \BibitemOpen
  \bibfield  {author} {\bibinfo {author} {\bibfnamefont {B.~F.}\ \bibnamefont
  {Schutz}},\ }\bibfield  {title} {\bibinfo {title} {{Determining the Hubble
  Constant from Gravitational Wave Observations}},\ }\href
  {https://doi.org/10.1038/323310a0} {\bibfield  {journal} {\bibinfo  {journal}
  {Nature}\ }\textbf {\bibinfo {volume} {323}},\ \bibinfo {pages} {310}
  (\bibinfo {year} {1986})}\BibitemShut {NoStop}%
\bibitem [{\citenamefont {Zhu}\ \emph {et~al.}(2022)\citenamefont {Zhu},
  \citenamefont {Xie}, \citenamefont {Hu}, \citenamefont {Liu}, \citenamefont
  {Li}, \citenamefont {Napolitano}, \citenamefont {Tang}, \citenamefont
  {Zhang},\ and\ \citenamefont {Mei}}]{Zhu:2021bpp}%
  \BibitemOpen
  \bibfield  {author} {\bibinfo {author} {\bibfnamefont {L.-G.}\ \bibnamefont
  {Zhu}}, \bibinfo {author} {\bibfnamefont {L.-H.}\ \bibnamefont {Xie}},
  \bibinfo {author} {\bibfnamefont {Y.-M.}\ \bibnamefont {Hu}}, \bibinfo
  {author} {\bibfnamefont {S.}~\bibnamefont {Liu}}, \bibinfo {author}
  {\bibfnamefont {E.-K.}\ \bibnamefont {Li}}, \bibinfo {author} {\bibfnamefont
  {N.~R.}\ \bibnamefont {Napolitano}}, \bibinfo {author} {\bibfnamefont
  {B.-T.}\ \bibnamefont {Tang}}, \bibinfo {author} {\bibfnamefont {J.-d.}\
  \bibnamefont {Zhang}},\ and\ \bibinfo {author} {\bibfnamefont
  {J.}~\bibnamefont {Mei}},\ }\bibfield  {title} {\bibinfo {title}
  {{Constraining the Hubble constant to a precision of about 1\% using
  multi-band dark standard siren detections}},\ }\href
  {https://doi.org/10.1007/s11433-021-1859-9} {\bibfield  {journal} {\bibinfo
  {journal} {Sci. China Phys. Mech. Astron.}\ }\textbf {\bibinfo {volume}
  {65}},\ \bibinfo {pages} {259811} (\bibinfo {year} {2022})},\ \Eprint
  {https://arxiv.org/abs/2110.05224} {arXiv:2110.05224 [astro-ph.CO]}
  \BibitemShut {NoStop}%
\bibitem [{\citenamefont {Yang}\ \emph {et~al.}(2020)\citenamefont {Yang},
  \citenamefont {Gayathri}, \citenamefont {arka}, \citenamefont {arka},\ and\
  \citenamefont {Bartos}}]{Yang:2020yoc}%
  \BibitemOpen
  \bibfield  {author} {\bibinfo {author} {\bibfnamefont {Y.}~\bibnamefont
  {Yang}}, \bibinfo {author} {\bibfnamefont {V.}~\bibnamefont {Gayathri}},
  \bibinfo {author} {\bibfnamefont {S.~M.}\ \bibnamefont {arka}}, \bibinfo
  {author} {\bibfnamefont {Z.~M.}\ \bibnamefont {arka}},\ and\ \bibinfo
  {author} {\bibfnamefont {I.}~\bibnamefont {Bartos}},\ }\bibfield  {title}
  {\bibinfo {title} {{Determining the Hubble Constant with Black Hole Mergers
  in Active Galactic Nuclei}},\ }\href@noop {} {\  (\bibinfo {year} {2020})},\
  \Eprint {https://arxiv.org/abs/2009.13739} {arXiv:2009.13739 [astro-ph.HE]}
  \BibitemShut {NoStop}%
\bibitem [{\citenamefont {Holz}\ and\ \citenamefont
  {Hughes}(2005)}]{Holz:2005df}%
  \BibitemOpen
  \bibfield  {author} {\bibinfo {author} {\bibfnamefont {D.~E.}\ \bibnamefont
  {Holz}}\ and\ \bibinfo {author} {\bibfnamefont {S.~A.}\ \bibnamefont
  {Hughes}},\ }\bibfield  {title} {\bibinfo {title} {{Using gravitational-wave
  standard sirens}},\ }\href {https://doi.org/10.1086/431341} {\bibfield
  {journal} {\bibinfo  {journal} {Astrophys. J.}\ }\textbf {\bibinfo {volume}
  {629}},\ \bibinfo {pages} {15} (\bibinfo {year} {2005})},\ \Eprint
  {https://arxiv.org/abs/astro-ph/0504616} {arXiv:astro-ph/0504616}
  \BibitemShut {NoStop}%
\bibitem [{\citenamefont {Agazie}\ \emph {et~al.}(2023)\citenamefont {Agazie}
  \emph {et~al.}}]{NANOGrav:2023gor}%
  \BibitemOpen
  \bibfield  {author} {\bibinfo {author} {\bibfnamefont {G.}~\bibnamefont
  {Agazie}} \emph {et~al.} (\bibinfo {collaboration} {NANOGrav}),\ }\bibfield
  {title} {\bibinfo {title} {{The NANOGrav 15 yr Data Set: Evidence for a
  Gravitational-wave Background}},\ }\href
  {https://doi.org/10.3847/2041-8213/acdac6} {\bibfield  {journal} {\bibinfo
  {journal} {Astrophys. J. Lett.}\ }\textbf {\bibinfo {volume} {951}},\
  \bibinfo {pages} {L8} (\bibinfo {year} {2023})},\ \Eprint
  {https://arxiv.org/abs/2306.16213} {arXiv:2306.16213 [astro-ph.HE]}
  \BibitemShut {NoStop}%
\bibitem [{\citenamefont {Antoniadis}\ \emph {et~al.}(2023)\citenamefont
  {Antoniadis} \emph {et~al.}}]{EPTA:2023fyk}%
  \BibitemOpen
  \bibfield  {author} {\bibinfo {author} {\bibfnamefont {J.}~\bibnamefont
  {Antoniadis}} \emph {et~al.} (\bibinfo {collaboration} {EPTA, InPTA:}),\
  }\bibfield  {title} {\bibinfo {title} {{The second data release from the
  European Pulsar Timing Array - III. Search for gravitational wave signals}},\
  }\href {https://doi.org/10.1051/0004-6361/202346844} {\bibfield  {journal}
  {\bibinfo  {journal} {Astron. Astrophys.}\ }\textbf {\bibinfo {volume}
  {678}},\ \bibinfo {pages} {A50} (\bibinfo {year} {2023})},\ \Eprint
  {https://arxiv.org/abs/2306.16214} {arXiv:2306.16214 [astro-ph.HE]}
  \BibitemShut {NoStop}%
\bibitem [{\citenamefont {Reardon}\ \emph {et~al.}(2023)\citenamefont {Reardon}
  \emph {et~al.}}]{Reardon:2023gzh}%
  \BibitemOpen
  \bibfield  {author} {\bibinfo {author} {\bibfnamefont {D.~J.}\ \bibnamefont
  {Reardon}} \emph {et~al.},\ }\bibfield  {title} {\bibinfo {title} {{Search
  for an Isotropic Gravitational-wave Background with the Parkes Pulsar Timing
  Array}},\ }\href {https://doi.org/10.3847/2041-8213/acdd02} {\bibfield
  {journal} {\bibinfo  {journal} {Astrophys. J. Lett.}\ }\textbf {\bibinfo
  {volume} {951}},\ \bibinfo {pages} {L6} (\bibinfo {year} {2023})},\ \Eprint
  {https://arxiv.org/abs/2306.16215} {arXiv:2306.16215 [astro-ph.HE]}
  \BibitemShut {NoStop}%
\bibitem [{\citenamefont {Xu}\ \emph {et~al.}(2023)\citenamefont {Xu} \emph
  {et~al.}}]{Xu:2023wog}%
  \BibitemOpen
  \bibfield  {author} {\bibinfo {author} {\bibfnamefont {H.}~\bibnamefont {Xu}}
  \emph {et~al.},\ }\bibfield  {title} {\bibinfo {title} {{Searching for the
  Nano-Hertz Stochastic Gravitational Wave Background with the Chinese Pulsar
  Timing Array Data Release I}},\ }\href
  {https://doi.org/10.1088/1674-4527/acdfa5} {\bibfield  {journal} {\bibinfo
  {journal} {Res. Astron. Astrophys.}\ }\textbf {\bibinfo {volume} {23}},\
  \bibinfo {pages} {075024} (\bibinfo {year} {2023})},\ \Eprint
  {https://arxiv.org/abs/2306.16216} {arXiv:2306.16216 [astro-ph.HE]}
  \BibitemShut {NoStop}%
\bibitem [{\citenamefont {Ni}(2013)}]{Ni:2012eh}%
  \BibitemOpen
  \bibfield  {author} {\bibinfo {author} {\bibfnamefont {W.-T.}\ \bibnamefont
  {Ni}},\ }\bibfield  {title} {\bibinfo {title} {{ASTROD-GW: Overview and
  Progress}},\ }\href {https://doi.org/10.1142/S0218271813410046} {\bibfield
  {journal} {\bibinfo  {journal} {Int. J. Mod. Phys. D}\ }\textbf {\bibinfo
  {volume} {22}},\ \bibinfo {pages} {1341004} (\bibinfo {year} {2013})},\
  \Eprint {https://arxiv.org/abs/1212.2816} {arXiv:1212.2816 [astro-ph.IM]}
  \BibitemShut {NoStop}%
\bibitem [{\citenamefont {Ni}(2016)}]{Ni:2016wcv}%
  \BibitemOpen
  \bibfield  {author} {\bibinfo {author} {\bibfnamefont {W.-T.}\ \bibnamefont
  {Ni}},\ }\bibfield  {title} {\bibinfo {title} {{Gravitational wave detection
  in space}},\ }\href {https://doi.org/10.1142/S0218271816300019} {\bibfield
  {journal} {\bibinfo  {journal} {Int. J. Mod. Phys. D}\ }\textbf {\bibinfo
  {volume} {25}},\ \bibinfo {pages} {1630001} (\bibinfo {year} {2016})},\
  \Eprint {https://arxiv.org/abs/1610.01148} {arXiv:1610.01148 [astro-ph.IM]}
  \BibitemShut {NoStop}%
\bibitem [{\citenamefont {Baker}\ \emph {et~al.}(2019)\citenamefont {Baker}
  \emph {et~al.}}]{Baker:2019pnp}%
  \BibitemOpen
  \bibfield  {author} {\bibinfo {author} {\bibfnamefont {J.}~\bibnamefont
  {Baker}} \emph {et~al.},\ }\bibfield  {title} {\bibinfo {title} {{Space Based
  Gravitational Wave Astronomy Beyond LISA}},\ }\href@noop {} {\bibfield
  {journal} {\bibinfo  {journal} {Bull. Am. Astron. Soc.}\ }\textbf {\bibinfo
  {volume} {51}},\ \bibinfo {pages} {243} (\bibinfo {year} {2019})},\ \Eprint
  {https://arxiv.org/abs/1907.11305} {arXiv:1907.11305 [astro-ph.IM]}
  \BibitemShut {NoStop}%
\bibitem [{\citenamefont {Martens}\ \emph {et~al.}(2023)\citenamefont
  {Martens}, \citenamefont {Khan},\ and\ \citenamefont
  {Bayle}}]{Martens:2023mgm}%
  \BibitemOpen
  \bibfield  {author} {\bibinfo {author} {\bibfnamefont {W.}~\bibnamefont
  {Martens}}, \bibinfo {author} {\bibfnamefont {M.}~\bibnamefont {Khan}},\ and\
  \bibinfo {author} {\bibfnamefont {J.-B.}\ \bibnamefont {Bayle}},\ }\bibfield
  {title} {\bibinfo {title} {{LISAmax: improving the low-frequency
  gravitational-wave sensitivity by two orders of magnitude}},\ }\href
  {https://doi.org/10.1088/1361-6382/acf3c7} {\bibfield  {journal} {\bibinfo
  {journal} {Class. Quant. Grav.}\ }\textbf {\bibinfo {volume} {40}},\ \bibinfo
  {pages} {195022} (\bibinfo {year} {2023})},\ \Eprint
  {https://arxiv.org/abs/2304.08287} {arXiv:2304.08287 [gr-qc]} \BibitemShut
  {NoStop}%
\bibitem [{\citenamefont {Sesana}\ \emph {et~al.}(2021)\citenamefont {Sesana}
  \emph {et~al.}}]{Sesana:2019vho}%
  \BibitemOpen
  \bibfield  {author} {\bibinfo {author} {\bibfnamefont {A.}~\bibnamefont
  {Sesana}} \emph {et~al.},\ }\bibfield  {title} {\bibinfo {title} {{Unveiling
  the gravitational universe at $\mu$-Hz frequencies}},\ }\href
  {https://doi.org/10.1007/s10686-021-09709-9} {\bibfield  {journal} {\bibinfo
  {journal} {Exper. Astron.}\ }\textbf {\bibinfo {volume} {51}},\ \bibinfo
  {pages} {1333} (\bibinfo {year} {2021})},\ \Eprint
  {https://arxiv.org/abs/1908.11391} {arXiv:1908.11391 [astro-ph.IM]}
  \BibitemShut {NoStop}%
\bibitem [{\citenamefont {Ni}(2009)}]{Ni:2008bj}%
  \BibitemOpen
  \bibfield  {author} {\bibinfo {author} {\bibfnamefont {W.-T.}\ \bibnamefont
  {Ni}},\ }\bibfield  {title} {\bibinfo {title} {{Super-ASTROD: Probing
  primordial gravitational waves and mapping the outer solar system}},\ }\href
  {https://doi.org/10.1088/0264-9381/26/7/075021} {\bibfield  {journal}
  {\bibinfo  {journal} {Class. Quant. Grav.}\ }\textbf {\bibinfo {volume}
  {26}},\ \bibinfo {pages} {075021} (\bibinfo {year} {2009})},\ \Eprint
  {https://arxiv.org/abs/0812.0887} {arXiv:0812.0887 [astro-ph]} \BibitemShut
  {NoStop}%
\bibitem [{\citenamefont {Vecchio}\ and\ \citenamefont
  {Wickham}(2004)}]{Vecchio:2004ec}%
  \BibitemOpen
  \bibfield  {author} {\bibinfo {author} {\bibfnamefont {A.}~\bibnamefont
  {Vecchio}}\ and\ \bibinfo {author} {\bibfnamefont {E.~D.~L.}\ \bibnamefont
  {Wickham}},\ }\bibfield  {title} {\bibinfo {title} {{The Effect of the LISA
  response function on observations of monochromatic sources}},\ }\href
  {https://doi.org/10.1103/PhysRevD.70.082002} {\bibfield  {journal} {\bibinfo
  {journal} {Phys. Rev. D}\ }\textbf {\bibinfo {volume} {70}},\ \bibinfo
  {pages} {082002} (\bibinfo {year} {2004})},\ \Eprint
  {https://arxiv.org/abs/gr-qc/0406039} {arXiv:gr-qc/0406039} \BibitemShut
  {NoStop}%
\bibitem [{\citenamefont {Nissanke}\ \emph {et~al.}(2012)\citenamefont
  {Nissanke}, \citenamefont {Vallisneri}, \citenamefont {Nelemans},\ and\
  \citenamefont {Prince}}]{Nissanke:2012eh}%
  \BibitemOpen
  \bibfield  {author} {\bibinfo {author} {\bibfnamefont {S.}~\bibnamefont
  {Nissanke}}, \bibinfo {author} {\bibfnamefont {M.}~\bibnamefont
  {Vallisneri}}, \bibinfo {author} {\bibfnamefont {G.}~\bibnamefont
  {Nelemans}},\ and\ \bibinfo {author} {\bibfnamefont {T.~A.}\ \bibnamefont
  {Prince}},\ }\bibfield  {title} {\bibinfo {title} {{Gravitational-wave
  emission from compact Galactic binaries}},\ }\href
  {https://doi.org/10.1088/0004-637X/758/2/131} {\bibfield  {journal} {\bibinfo
   {journal} {Astrophys. J.}\ }\textbf {\bibinfo {volume} {758}},\ \bibinfo
  {pages} {131} (\bibinfo {year} {2012})},\ \Eprint
  {https://arxiv.org/abs/1201.4613} {arXiv:1201.4613 [astro-ph.GA]}
  \BibitemShut {NoStop}%
\bibitem [{\citenamefont {Cornish}\ and\ \citenamefont
  {Robson}(2017)}]{Cornish:2017vip}%
  \BibitemOpen
  \bibfield  {author} {\bibinfo {author} {\bibfnamefont {N.}~\bibnamefont
  {Cornish}}\ and\ \bibinfo {author} {\bibfnamefont {T.}~\bibnamefont
  {Robson}},\ }\bibfield  {title} {\bibinfo {title} {{Galactic binary science
  with the new LISA design}},\ }\href
  {https://doi.org/10.1088/1742-6596/840/1/012024} {\bibfield  {journal}
  {\bibinfo  {journal} {J. Phys. Conf. Ser.}\ }\textbf {\bibinfo {volume}
  {840}},\ \bibinfo {pages} {012024} (\bibinfo {year} {2017})},\ \Eprint
  {https://arxiv.org/abs/1703.09858} {arXiv:1703.09858 [astro-ph.IM]}
  \BibitemShut {NoStop}%
\bibitem [{\citenamefont {Korol}\ \emph {et~al.}(2017)\citenamefont {Korol},
  \citenamefont {Rossi}, \citenamefont {Groot}, \citenamefont {Nelemans},
  \citenamefont {Toonen},\ and\ \citenamefont {Brown}}]{Korol:2017qcx}%
  \BibitemOpen
  \bibfield  {author} {\bibinfo {author} {\bibfnamefont {V.}~\bibnamefont
  {Korol}}, \bibinfo {author} {\bibfnamefont {E.~M.}\ \bibnamefont {Rossi}},
  \bibinfo {author} {\bibfnamefont {P.~J.}\ \bibnamefont {Groot}}, \bibinfo
  {author} {\bibfnamefont {G.}~\bibnamefont {Nelemans}}, \bibinfo {author}
  {\bibfnamefont {S.}~\bibnamefont {Toonen}},\ and\ \bibinfo {author}
  {\bibfnamefont {A.~G.~A.}\ \bibnamefont {Brown}},\ }\bibfield  {title}
  {\bibinfo {title} {{Prospects for detection of detached double white dwarf
  binaries with Gaia, LSST and LISA}},\ }\href
  {https://doi.org/10.1093/mnras/stx1285} {\bibfield  {journal} {\bibinfo
  {journal} {Mon. Not. Roy. Astron. Soc.}\ }\textbf {\bibinfo {volume} {470}},\
  \bibinfo {pages} {1894} (\bibinfo {year} {2017})},\ \Eprint
  {https://arxiv.org/abs/1703.02555} {arXiv:1703.02555 [astro-ph.HE]}
  \BibitemShut {NoStop}%
\bibitem [{\citenamefont {Seoane}\ \emph {et~al.}(2023)\citenamefont {Seoane}
  \emph {et~al.}}]{LISA:2022yao}%
  \BibitemOpen
  \bibfield  {author} {\bibinfo {author} {\bibfnamefont {P.~A.}\ \bibnamefont
  {Seoane}} \emph {et~al.} (\bibinfo {collaboration} {LISA}),\ }\bibfield
  {title} {\bibinfo {title} {{Astrophysics with the Laser Interferometer Space
  Antenna}},\ }\href {https://doi.org/10.1007/s41114-022-00041-y} {\bibfield
  {journal} {\bibinfo  {journal} {Living Rev. Rel.}\ }\textbf {\bibinfo
  {volume} {26}},\ \bibinfo {pages} {2} (\bibinfo {year} {2023})},\ \Eprint
  {https://arxiv.org/abs/2203.06016} {arXiv:2203.06016 [gr-qc]} \BibitemShut
  {NoStop}%
\bibitem [{\citenamefont {Korol}\ \emph {et~al.}(2018)\citenamefont {Korol},
  \citenamefont {Koop},\ and\ \citenamefont {Rossi}}]{Korol:2018ulo}%
  \BibitemOpen
  \bibfield  {author} {\bibinfo {author} {\bibfnamefont {V.}~\bibnamefont
  {Korol}}, \bibinfo {author} {\bibfnamefont {O.}~\bibnamefont {Koop}},\ and\
  \bibinfo {author} {\bibfnamefont {E.~M.}\ \bibnamefont {Rossi}},\ }\bibfield
  {title} {\bibinfo {title} {{Detectability of double white dwarfs in the Local
  Group with LISA}},\ }\href {https://doi.org/10.3847/2041-8213/aae587}
  {\bibfield  {journal} {\bibinfo  {journal} {Astrophys. J. Lett.}\ }\textbf
  {\bibinfo {volume} {866}},\ \bibinfo {pages} {L20} (\bibinfo {year}
  {2018})},\ \Eprint {https://arxiv.org/abs/1808.05959} {arXiv:1808.05959
  [astro-ph.HE]} \BibitemShut {NoStop}%
\bibitem [{\citenamefont {Nelemans}\ \emph {et~al.}(2001)\citenamefont
  {Nelemans}, \citenamefont {Yungelson},\ and\ \citenamefont
  {Portegies~Zwart}}]{Nelemans:2001hp}%
  \BibitemOpen
  \bibfield  {author} {\bibinfo {author} {\bibfnamefont {G.}~\bibnamefont
  {Nelemans}}, \bibinfo {author} {\bibfnamefont {L.~R.}\ \bibnamefont
  {Yungelson}},\ and\ \bibinfo {author} {\bibfnamefont {S.~F.}\ \bibnamefont
  {Portegies~Zwart}},\ }\bibfield  {title} {\bibinfo {title} {{The
  gravitational wave signal from the galactic disk population of binaries
  containing two compact objects}},\ }\href
  {https://doi.org/10.1051/0004-6361:20010683} {\bibfield  {journal} {\bibinfo
  {journal} {Astron. Astrophys.}\ }\textbf {\bibinfo {volume} {375}},\ \bibinfo
  {pages} {890} (\bibinfo {year} {2001})},\ \Eprint
  {https://arxiv.org/abs/astro-ph/0105221} {arXiv:astro-ph/0105221}
  \BibitemShut {NoStop}%
\bibitem [{\citenamefont {Wang}\ \emph {et~al.}(2023)\citenamefont {Wang},
  \citenamefont {Yan}, \citenamefont {Hu},\ and\ \citenamefont
  {Ni}}]{Wang:2023jct}%
  \BibitemOpen
  \bibfield  {author} {\bibinfo {author} {\bibfnamefont {G.}~\bibnamefont
  {Wang}}, \bibinfo {author} {\bibfnamefont {Z.}~\bibnamefont {Yan}}, \bibinfo
  {author} {\bibfnamefont {B.}~\bibnamefont {Hu}},\ and\ \bibinfo {author}
  {\bibfnamefont {W.-T.}\ \bibnamefont {Ni}},\ }\bibfield  {title} {\bibinfo
  {title} {{Investigating galactic double white dwarfs for the sub-mHz
  gravitational wave mission ASTROD-GW}},\ }\href
  {https://doi.org/10.1103/PhysRevD.107.124022} {\bibfield  {journal} {\bibinfo
   {journal} {Phys. Rev. D}\ }\textbf {\bibinfo {volume} {107}},\ \bibinfo
  {pages} {124022} (\bibinfo {year} {2023})},\ \Eprint
  {https://arxiv.org/abs/2302.07625} {arXiv:2302.07625 [gr-qc]} \BibitemShut
  {NoStop}%
\bibitem [{\citenamefont {rui Men}\ \emph {et~al.}(2010)\citenamefont {rui
  Men}, \citenamefont {tou Ni},\ and\ \citenamefont {Wang}}]{MEN:2010434}%
  \BibitemOpen
  \bibfield  {author} {\bibinfo {author} {\bibfnamefont {J.}~\bibnamefont {rui
  Men}}, \bibinfo {author} {\bibfnamefont {W.}~\bibnamefont {tou Ni}},\ and\
  \bibinfo {author} {\bibfnamefont {G.}~\bibnamefont {Wang}},\ }\bibfield
  {title} {\bibinfo {title} {Design of astrod-gw orbit},\ }\href
  {https://doi.org/https://doi.org/10.1016/j.chinastron.2010.10.008} {\bibfield
   {journal} {\bibinfo  {journal} {Chinese Astronomy and Astrophysics}\
  }\textbf {\bibinfo {volume} {34}},\ \bibinfo {pages} {434} (\bibinfo {year}
  {2010})}\BibitemShut {NoStop}%
\bibitem [{\citenamefont {Wang}\ and\ \citenamefont {tou
  Ni}(2012)}]{WANG:2012211}%
  \BibitemOpen
  \bibfield  {author} {\bibinfo {author} {\bibfnamefont {G.}~\bibnamefont
  {Wang}}\ and\ \bibinfo {author} {\bibfnamefont {W.}~\bibnamefont {tou Ni}},\
  }\bibfield  {title} {\bibinfo {title} {Time-delay interferometry for
  astrod-gw},\ }\href
  {https://doi.org/https://doi.org/10.1016/j.chinastron.2012.04.009} {\bibfield
   {journal} {\bibinfo  {journal} {Chinese Astronomy and Astrophysics}\
  }\textbf {\bibinfo {volume} {36}},\ \bibinfo {pages} {211} (\bibinfo {year}
  {2012})}\BibitemShut {NoStop}%
\bibitem [{\citenamefont {Wang}(2011)}]{Wang:2011tlj}%
  \BibitemOpen
  \bibfield  {author} {\bibinfo {author} {\bibfnamefont {G.}~\bibnamefont
  {Wang}},\ }\href@noop {} {\bibinfo {title} {{Time-Delay Interferometry for
  ASTROD-GW}}} (\bibinfo {year} {2011}),\ \Eprint
  {https://arxiv.org/abs/2406.14173} {arXiv:2406.14173 [gr-qc]} \BibitemShut
  {NoStop}%
\bibitem [{\citenamefont {Wang}\ and\ \citenamefont {Ni}(2012)}]{Wang:2012fqs}%
  \BibitemOpen
  \bibfield  {author} {\bibinfo {author} {\bibfnamefont {G.}~\bibnamefont
  {Wang}}\ and\ \bibinfo {author} {\bibfnamefont {W.-t.}\ \bibnamefont {Ni}},\
  }\bibfield  {title} {\bibinfo {title} {{Time-delay Interferometry for
  ASTROD-GW}},\ }\href {https://doi.org/10.1016/j.chinastron.2012.04.009}
  {\bibfield  {journal} {\bibinfo  {journal} {Chin. Astron. Astrophys.}\
  }\textbf {\bibinfo {volume} {36}},\ \bibinfo {pages} {211} (\bibinfo {year}
  {2012})}\BibitemShut {NoStop}%
\bibitem [{\citenamefont {Gang}\ and\ \citenamefont
  {Wei-Tou}(2015)}]{Wang:2014uaw}%
  \BibitemOpen
  \bibfield  {author} {\bibinfo {author} {\bibfnamefont {W.}~\bibnamefont
  {Gang}}\ and\ \bibinfo {author} {\bibfnamefont {N.}~\bibnamefont {Wei-Tou}},\
  }\bibfield  {title} {\bibinfo {title} {{Orbit optimization and time delay
  interferometry for inclined ASTROD-GW formation with half-year
  precession-period}},\ }\href {https://doi.org/10.1088/1674-1056/24/5/059501}
  {\bibfield  {journal} {\bibinfo  {journal} {Chin. Phys. B}\ }\textbf
  {\bibinfo {volume} {24}},\ \bibinfo {pages} {059501} (\bibinfo {year}
  {2015})},\ \Eprint {https://arxiv.org/abs/1409.4162} {arXiv:1409.4162
  [gr-qc]} \BibitemShut {NoStop}%
\bibitem [{\citenamefont {Wang}\ and\ \citenamefont {Ni}(2013)}]{Wang:2012te}%
  \BibitemOpen
  \bibfield  {author} {\bibinfo {author} {\bibfnamefont {G.}~\bibnamefont
  {Wang}}\ and\ \bibinfo {author} {\bibfnamefont {W.~T.}\ \bibnamefont {Ni}},\
  }\bibfield  {title} {\bibinfo {title} {{Orbit optimization for ASTROD-GW and
  its time delay interferometry with two arms using CGC ephemeris}},\ }\href
  {https://doi.org/10.1088/1674-1056/22/4/049501} {\bibfield  {journal}
  {\bibinfo  {journal} {Chin. Phys. B}\ }\textbf {\bibinfo {volume} {22}},\
  \bibinfo {pages} {049501} (\bibinfo {year} {2013})},\ \Eprint
  {https://arxiv.org/abs/1205.5175} {arXiv:1205.5175 [gr-qc]} \BibitemShut
  {NoStop}%
\bibitem [{\citenamefont {Vallisneri}\ and\ \citenamefont
  {Galley}(2012)}]{Vallisneri:2012np}%
  \BibitemOpen
  \bibfield  {author} {\bibinfo {author} {\bibfnamefont {M.}~\bibnamefont
  {Vallisneri}}\ and\ \bibinfo {author} {\bibfnamefont {C.~R.}\ \bibnamefont
  {Galley}},\ }\bibfield  {title} {\bibinfo {title} {{Non-sky-averaged
  sensitivity curves for space-based gravitational-wave observatories}},\
  }\href {https://doi.org/10.1088/0264-9381/29/12/124015} {\bibfield  {journal}
  {\bibinfo  {journal} {Class. Quant. Grav.}\ }\textbf {\bibinfo {volume}
  {29}},\ \bibinfo {pages} {124015} (\bibinfo {year} {2012})},\ \Eprint
  {https://arxiv.org/abs/1201.3684} {arXiv:1201.3684 [gr-qc]} \BibitemShut
  {NoStop}%
\bibitem [{\citenamefont {Vallisneri}(2005)}]{Vallisneri:2004bn}%
  \BibitemOpen
  \bibfield  {author} {\bibinfo {author} {\bibfnamefont {M.}~\bibnamefont
  {Vallisneri}},\ }\bibfield  {title} {\bibinfo {title} {{Synthetic LISA:
  Simulating time delay interferometry in a model LISA}},\ }\href
  {https://doi.org/10.1103/PhysRevD.71.022001} {\bibfield  {journal} {\bibinfo
  {journal} {Phys. Rev. D}\ }\textbf {\bibinfo {volume} {71}},\ \bibinfo
  {pages} {022001} (\bibinfo {year} {2005})},\ \Eprint
  {https://arxiv.org/abs/gr-qc/0407102} {arXiv:gr-qc/0407102} \BibitemShut
  {NoStop}%
\bibitem [{\citenamefont {Vallisneri}\ \emph {et~al.}(2008)\citenamefont
  {Vallisneri}, \citenamefont {Crowder},\ and\ \citenamefont
  {Tinto}}]{Vallisneri:2007xa}%
  \BibitemOpen
  \bibfield  {author} {\bibinfo {author} {\bibfnamefont {M.}~\bibnamefont
  {Vallisneri}}, \bibinfo {author} {\bibfnamefont {J.}~\bibnamefont
  {Crowder}},\ and\ \bibinfo {author} {\bibfnamefont {M.}~\bibnamefont
  {Tinto}},\ }\bibfield  {title} {\bibinfo {title} {{Sensitivity and
  parameter-estimation precision for alternate LISA configurations}},\ }\href
  {https://doi.org/10.1088/0264-9381/25/6/065005} {\bibfield  {journal}
  {\bibinfo  {journal} {Class. Quant. Grav.}\ }\textbf {\bibinfo {volume}
  {25}},\ \bibinfo {pages} {065005} (\bibinfo {year} {2008})},\ \Eprint
  {https://arxiv.org/abs/0710.4369} {arXiv:0710.4369 [gr-qc]} \BibitemShut
  {NoStop}%
\bibitem [{\citenamefont {Ni}(1996)}]{Ni:1996ns}%
  \BibitemOpen
  \bibfield  {author} {\bibinfo {author} {\bibfnamefont {W.-T.}\ \bibnamefont
  {Ni}},\ }\bibfield  {title} {\bibinfo {title} {{ASTROD and gravitational
  waves}},\ }in\ \href@noop {} {\emph {\bibinfo {booktitle} {{TAMA Workshop on
  Gravitational Wave Detection}}}}\ (\bibinfo {year} {1996})\ pp.\ \bibinfo
  {pages} {117--129}\BibitemShut {NoStop}%
\bibitem [{\citenamefont {{Armstrong}}\ \emph {et~al.}(1999)\citenamefont
  {{Armstrong}}, \citenamefont {{Estabrook}},\ and\ \citenamefont
  {{Tinto}}}]{1999ApJ...527..814A}%
  \BibitemOpen
  \bibfield  {author} {\bibinfo {author} {\bibfnamefont {J.~W.}\ \bibnamefont
  {{Armstrong}}}, \bibinfo {author} {\bibfnamefont {F.~B.}\ \bibnamefont
  {{Estabrook}}},\ and\ \bibinfo {author} {\bibfnamefont {M.}~\bibnamefont
  {{Tinto}}},\ }\bibfield  {title} {\bibinfo {title} {{Time-Delay
  Interferometry for Space-based Gravitational Wave Searches}},\ }\href
  {https://doi.org/10.1086/308110} {\bibfield  {journal} {\bibinfo  {journal}
  {\apj}\ }\textbf {\bibinfo {volume} {527}},\ \bibinfo {pages} {814} (\bibinfo
  {year} {1999})}\BibitemShut {NoStop}%
\bibitem [{\citenamefont {Wang}(2024{\natexlab{a}})}]{Wang:2024alm}%
  \BibitemOpen
  \bibfield  {author} {\bibinfo {author} {\bibfnamefont {G.}~\bibnamefont
  {Wang}},\ }\bibfield  {title} {\bibinfo {title} {{Time delay interferometry
  with minimal null frequencies}},\ }\href
  {https://doi.org/10.1103/PhysRevD.110.042005} {\bibfield  {journal} {\bibinfo
   {journal} {Phys. Rev. D}\ }\textbf {\bibinfo {volume} {110}},\ \bibinfo
  {pages} {042005} (\bibinfo {year} {2024}{\natexlab{a}})},\ \Eprint
  {https://arxiv.org/abs/2403.01490} {arXiv:2403.01490 [gr-qc]} \BibitemShut
  {NoStop}%
\bibitem [{\citenamefont {Wang}(2024{\natexlab{b}})}]{Wang:2024hgv}%
  \BibitemOpen
  \bibfield  {author} {\bibinfo {author} {\bibfnamefont {G.}~\bibnamefont
  {Wang}},\ }\bibfield  {title} {\bibinfo {title} {{Enhancing noise
  characterization with robust time delay interferometry combination}},\ }\href
  {https://doi.org/10.1103/PhysRevD.110.064085} {\bibfield  {journal} {\bibinfo
   {journal} {Phys. Rev. D}\ }\textbf {\bibinfo {volume} {110}},\ \bibinfo
  {pages} {064085} (\bibinfo {year} {2024}{\natexlab{b}})},\ \Eprint
  {https://arxiv.org/abs/2406.11305} {arXiv:2406.11305 [gr-qc]} \BibitemShut
  {NoStop}%
\bibitem [{\citenamefont {Wang}(2025)}]{Wang:2025mee}%
  \BibitemOpen
  \bibfield  {author} {\bibinfo {author} {\bibfnamefont {G.}~\bibnamefont
  {Wang}},\ }\bibfield  {title} {\bibinfo {title} {{Time delay interferometry
  with minimal null frequencies and shortened time span}},\ }\href@noop {} {\
  (\bibinfo {year} {2025})},\ \Eprint {https://arxiv.org/abs/2502.03983}
  {arXiv:2502.03983 [gr-qc]} \BibitemShut {NoStop}%
\bibitem [{\citenamefont {Hartwig}\ and\ \citenamefont
  {Muratore}(2022)}]{Hartwig:2021mzw}%
  \BibitemOpen
  \bibfield  {author} {\bibinfo {author} {\bibfnamefont {O.}~\bibnamefont
  {Hartwig}}\ and\ \bibinfo {author} {\bibfnamefont {M.}~\bibnamefont
  {Muratore}},\ }\bibfield  {title} {\bibinfo {title} {{Characterization of
  time delay interferometry combinations for the LISA instrument noise}},\
  }\href {https://doi.org/10.1103/PhysRevD.105.062006} {\bibfield  {journal}
  {\bibinfo  {journal} {Phys. Rev. D}\ }\textbf {\bibinfo {volume} {105}},\
  \bibinfo {pages} {062006} (\bibinfo {year} {2022})},\ \Eprint
  {https://arxiv.org/abs/2111.00975} {arXiv:2111.00975 [gr-qc]} \BibitemShut
  {NoStop}%
\bibitem [{\citenamefont {Lamberts}\ \emph {et~al.}(2019)\citenamefont
  {Lamberts}, \citenamefont {Blunt}, \citenamefont {Littenberg}, \citenamefont
  {Garrison-Kimmel}, \citenamefont {Kupfer},\ and\ \citenamefont
  {Sanderson}}]{Lamberts:2019nyk}%
  \BibitemOpen
  \bibfield  {author} {\bibinfo {author} {\bibfnamefont {A.}~\bibnamefont
  {Lamberts}}, \bibinfo {author} {\bibfnamefont {S.}~\bibnamefont {Blunt}},
  \bibinfo {author} {\bibfnamefont {T.~B.}\ \bibnamefont {Littenberg}},
  \bibinfo {author} {\bibfnamefont {S.}~\bibnamefont {Garrison-Kimmel}},
  \bibinfo {author} {\bibfnamefont {T.}~\bibnamefont {Kupfer}},\ and\ \bibinfo
  {author} {\bibfnamefont {R.~E.}\ \bibnamefont {Sanderson}},\ }\bibfield
  {title} {\bibinfo {title} {{Predicting the LISA white dwarf binary population
  in the Milky Way with cosmological simulations}},\ }\href
  {https://doi.org/10.1093/mnras/stz2834} {\bibfield  {journal} {\bibinfo
  {journal} {Mon. Not. Roy. Astron. Soc.}\ }\textbf {\bibinfo {volume} {490}},\
  \bibinfo {pages} {5888} (\bibinfo {year} {2019})},\ \Eprint
  {https://arxiv.org/abs/1907.00014} {arXiv:1907.00014 [astro-ph.HE]}
  \BibitemShut {NoStop}%
\bibitem [{\citenamefont {Li}\ \emph {et~al.}(2020)\citenamefont {Li},
  \citenamefont {Chen}, \citenamefont {Chen}, \citenamefont {Li}, \citenamefont
  {Yu},\ and\ \citenamefont {Han}}]{Li:2020voo}%
  \BibitemOpen
  \bibfield  {author} {\bibinfo {author} {\bibfnamefont {Z.}~\bibnamefont
  {Li}}, \bibinfo {author} {\bibfnamefont {X.}~\bibnamefont {Chen}}, \bibinfo
  {author} {\bibfnamefont {H.-L.}\ \bibnamefont {Chen}}, \bibinfo {author}
  {\bibfnamefont {J.}~\bibnamefont {Li}}, \bibinfo {author} {\bibfnamefont
  {S.}~\bibnamefont {Yu}},\ and\ \bibinfo {author} {\bibfnamefont
  {Z.}~\bibnamefont {Han}},\ }\bibfield  {title} {\bibinfo {title}
  {{Gravitational Wave Radiation of Double Degenerates with Extremely low-mass
  WD companions}}\ }\href {https://doi.org/10.3847/1538-4357/ab7dc2}
  {10.3847/1538-4357/ab7dc2} (\bibinfo {year} {2020}),\ \Eprint
  {https://arxiv.org/abs/2003.02480} {arXiv:2003.02480 [astro-ph.SR]}
  \BibitemShut {NoStop}%
\bibitem [{\citenamefont {Zhang}\ \emph {et~al.}(2021)\citenamefont {Zhang},
  \citenamefont {Mohanty}, \citenamefont {Zou},\ and\ \citenamefont
  {Liu}}]{Zhang:2021htc}%
  \BibitemOpen
  \bibfield  {author} {\bibinfo {author} {\bibfnamefont {X.}~\bibnamefont
  {Zhang}}, \bibinfo {author} {\bibfnamefont {S.~D.}\ \bibnamefont {Mohanty}},
  \bibinfo {author} {\bibfnamefont {X.}~\bibnamefont {Zou}},\ and\ \bibinfo
  {author} {\bibfnamefont {Y.}~\bibnamefont {Liu}},\ }\bibfield  {title}
  {\bibinfo {title} {{Resolving Galactic binaries in LISA data using particle
  swarm optimization and cross-validation}},\ }\href
  {https://doi.org/10.1103/PhysRevD.104.024023} {\bibfield  {journal} {\bibinfo
   {journal} {Phys. Rev. D}\ }\textbf {\bibinfo {volume} {104}},\ \bibinfo
  {pages} {024023} (\bibinfo {year} {2021})},\ \Eprint
  {https://arxiv.org/abs/2103.09391} {arXiv:2103.09391 [gr-qc]} \BibitemShut
  {NoStop}%
\bibitem [{\citenamefont {Thiele}\ \emph {et~al.}(2023)\citenamefont {Thiele},
  \citenamefont {Breivik}, \citenamefont {Sanderson},\ and\ \citenamefont
  {Luger}}]{Thiele:2021yyb}%
  \BibitemOpen
  \bibfield  {author} {\bibinfo {author} {\bibfnamefont {S.}~\bibnamefont
  {Thiele}}, \bibinfo {author} {\bibfnamefont {K.}~\bibnamefont {Breivik}},
  \bibinfo {author} {\bibfnamefont {R.~E.}\ \bibnamefont {Sanderson}},\ and\
  \bibinfo {author} {\bibfnamefont {R.}~\bibnamefont {Luger}},\ }\bibfield
  {title} {\bibinfo {title} {{Applying the Metallicity-dependent Binary
  Fraction to Double White Dwarf Formation: Implications for LISA}},\ }\href
  {https://doi.org/10.3847/1538-4357/aca7be} {\bibfield  {journal} {\bibinfo
  {journal} {Astrophys. J.}\ }\textbf {\bibinfo {volume} {945}},\ \bibinfo
  {pages} {162} (\bibinfo {year} {2023})},\ \Eprint
  {https://arxiv.org/abs/2111.13700} {arXiv:2111.13700 [astro-ph.HE]}
  \BibitemShut {NoStop}%
\bibitem [{\citenamefont {Wu}\ and\ \citenamefont {Shen}(2022)}]{Wu:2022knm}%
  \BibitemOpen
  \bibfield  {author} {\bibinfo {author} {\bibfnamefont {Q.}~\bibnamefont
  {Wu}}\ and\ \bibinfo {author} {\bibfnamefont {Y.}~\bibnamefont {Shen}},\
  }\bibfield  {title} {\bibinfo {title} {{A Catalog of Quasar Properties from
  Sloan Digital Sky Survey Data Release 16}},\ }\href
  {https://doi.org/10.3847/1538-4365/ac9ead} {\bibfield  {journal} {\bibinfo
  {journal} {Astrophys. J. Suppl.}\ }\textbf {\bibinfo {volume} {263}},\
  \bibinfo {pages} {42} (\bibinfo {year} {2022})},\ \Eprint
  {https://arxiv.org/abs/2209.03987} {arXiv:2209.03987 [astro-ph.GA]}
  \BibitemShut {NoStop}%
\bibitem [{\citenamefont {Colpi}\ \emph {et~al.}(2024)\citenamefont {Colpi}
  \emph {et~al.}}]{LISA:2024hlh}%
  \BibitemOpen
  \bibfield  {author} {\bibinfo {author} {\bibfnamefont {M.}~\bibnamefont
  {Colpi}} \emph {et~al.} (\bibinfo {collaboration} {LISA}),\ }\bibfield
  {title} {\bibinfo {title} {{LISA Definition Study Report}},\ }\href@noop {}
  {\  (\bibinfo {year} {2024})},\ \Eprint {https://arxiv.org/abs/2402.07571}
  {arXiv:2402.07571 [astro-ph.CO]} \BibitemShut {NoStop}%
\bibitem [{\citenamefont {{Flesch}}(2023)}]{2023OJAp....6E..49F}%
  \BibitemOpen
  \bibfield  {author} {\bibinfo {author} {\bibfnamefont {E.~W.}\ \bibnamefont
  {{Flesch}}},\ }\bibfield  {title} {\bibinfo {title} {{The Million Quasars
  (Milliquas) Catalogue, v8}},\ }\href
  {https://doi.org/10.21105/astro.2308.01505} {\bibfield  {journal} {\bibinfo
  {journal} {The Open Journal of Astrophysics}\ }\textbf {\bibinfo {volume}
  {6}},\ \bibinfo {eid} {49} (\bibinfo {year} {2023})},\ \Eprint
  {https://arxiv.org/abs/2308.01505} {arXiv:2308.01505 [astro-ph.GA]}
  \BibitemShut {NoStop}%
\bibitem [{\citenamefont {Wang}(2024{\natexlab{c}})}]{Wang:2024ssp}%
  \BibitemOpen
  \bibfield  {author} {\bibinfo {author} {\bibfnamefont {G.}~\bibnamefont
  {Wang}},\ }\bibfield  {title} {\bibinfo {title} {{SATDI: Simulation and
  Analysis for Time-Delay Interferometry}},\ }\href@noop {} {\  (\bibinfo
  {year} {2024}{\natexlab{c}})},\ \Eprint {https://arxiv.org/abs/2403.01726}
  {arXiv:2403.01726 [gr-qc]} \BibitemShut {NoStop}%
\bibitem [{\citenamefont {Pompili}\ \emph {et~al.}(2023)\citenamefont {Pompili}
  \emph {et~al.}}]{Pompili:2023tna}%
  \BibitemOpen
  \bibfield  {author} {\bibinfo {author} {\bibfnamefont {L.}~\bibnamefont
  {Pompili}} \emph {et~al.},\ }\bibfield  {title} {\bibinfo {title} {{Laying
  the foundation of the effective-one-body waveform models SEOBNRv5: Improved
  accuracy and efficiency for spinning nonprecessing binary black holes}},\
  }\href {https://doi.org/10.1103/PhysRevD.108.124035} {\bibfield  {journal}
  {\bibinfo  {journal} {Phys. Rev. D}\ }\textbf {\bibinfo {volume} {108}},\
  \bibinfo {pages} {124035} (\bibinfo {year} {2023})},\ \Eprint
  {https://arxiv.org/abs/2303.18039} {arXiv:2303.18039 [gr-qc]} \BibitemShut
  {NoStop}%
\bibitem [{\citenamefont {Feroz}\ \emph {et~al.}(2009)\citenamefont {Feroz},
  \citenamefont {Hobson},\ and\ \citenamefont {Bridges}}]{Feroz:2008xx}%
  \BibitemOpen
  \bibfield  {author} {\bibinfo {author} {\bibfnamefont {F.}~\bibnamefont
  {Feroz}}, \bibinfo {author} {\bibfnamefont {M.~P.}\ \bibnamefont {Hobson}},\
  and\ \bibinfo {author} {\bibfnamefont {M.}~\bibnamefont {Bridges}},\
  }\bibfield  {title} {\bibinfo {title} {{MultiNest: an efficient and robust
  Bayesian inference tool for cosmology and particle physics}},\ }\href
  {https://doi.org/10.1111/j.1365-2966.2009.14548.x} {\bibfield  {journal}
  {\bibinfo  {journal} {Mon. Not. Roy. Astron. Soc.}\ }\textbf {\bibinfo
  {volume} {398}},\ \bibinfo {pages} {1601} (\bibinfo {year} {2009})},\ \Eprint
  {https://arxiv.org/abs/0809.3437} {arXiv:0809.3437 [astro-ph]} \BibitemShut
  {NoStop}%
\bibitem [{\citenamefont {Buchner}\ \emph {et~al.}(2014)\citenamefont
  {Buchner}, \citenamefont {Georgakakis}, \citenamefont {Nandra}, \citenamefont
  {Hsu}, \citenamefont {Rangel}, \citenamefont {Brightman}, \citenamefont
  {Merloni}, \citenamefont {Salvato}, \citenamefont {Donley},\ and\
  \citenamefont {Kocevski}}]{Buchner:2014nha}%
  \BibitemOpen
  \bibfield  {author} {\bibinfo {author} {\bibfnamefont {J.}~\bibnamefont
  {Buchner}}, \bibinfo {author} {\bibfnamefont {A.}~\bibnamefont
  {Georgakakis}}, \bibinfo {author} {\bibfnamefont {K.}~\bibnamefont {Nandra}},
  \bibinfo {author} {\bibfnamefont {L.}~\bibnamefont {Hsu}}, \bibinfo {author}
  {\bibfnamefont {C.}~\bibnamefont {Rangel}}, \bibinfo {author} {\bibfnamefont
  {M.}~\bibnamefont {Brightman}}, \bibinfo {author} {\bibfnamefont
  {A.}~\bibnamefont {Merloni}}, \bibinfo {author} {\bibfnamefont
  {M.}~\bibnamefont {Salvato}}, \bibinfo {author} {\bibfnamefont
  {J.}~\bibnamefont {Donley}},\ and\ \bibinfo {author} {\bibfnamefont
  {D.}~\bibnamefont {Kocevski}},\ }\bibfield  {title} {\bibinfo {title} {{X-ray
  spectral modelling of the AGN obscuring region in the CDFS: Bayesian model
  selection and catalogue}},\ }\href
  {https://doi.org/10.1051/0004-6361/201322971} {\bibfield  {journal} {\bibinfo
   {journal} {Astron. Astrophys.}\ }\textbf {\bibinfo {volume} {564}},\
  \bibinfo {pages} {A125} (\bibinfo {year} {2014})},\ \Eprint
  {https://arxiv.org/abs/1402.0004} {arXiv:1402.0004 [astro-ph.HE]}
  \BibitemShut {NoStop}%
\bibitem [{\citenamefont {Thorne}(1980)}]{Thorne:1980ru}%
  \BibitemOpen
  \bibfield  {author} {\bibinfo {author} {\bibfnamefont {K.~S.}\ \bibnamefont
  {Thorne}},\ }\bibfield  {title} {\bibinfo {title} {{Multipole Expansions of
  Gravitational Radiation}},\ }\href
  {https://doi.org/10.1103/RevModPhys.52.299} {\bibfield  {journal} {\bibinfo
  {journal} {Rev. Mod. Phys.}\ }\textbf {\bibinfo {volume} {52}},\ \bibinfo
  {pages} {299} (\bibinfo {year} {1980})}\BibitemShut {NoStop}%
\bibitem [{\citenamefont {Ohme}\ \emph {et~al.}(2013)\citenamefont {Ohme},
  \citenamefont {Nielsen}, \citenamefont {Keppel},\ and\ \citenamefont
  {Lundgren}}]{Ohme:2013nsa}%
  \BibitemOpen
  \bibfield  {author} {\bibinfo {author} {\bibfnamefont {F.}~\bibnamefont
  {Ohme}}, \bibinfo {author} {\bibfnamefont {A.~B.}\ \bibnamefont {Nielsen}},
  \bibinfo {author} {\bibfnamefont {D.}~\bibnamefont {Keppel}},\ and\ \bibinfo
  {author} {\bibfnamefont {A.}~\bibnamefont {Lundgren}},\ }\bibfield  {title}
  {\bibinfo {title} {{Statistical and systematic errors for gravitational-wave
  inspiral signals: A principal component analysis}},\ }\href
  {https://doi.org/10.1103/PhysRevD.88.042002} {\bibfield  {journal} {\bibinfo
  {journal} {Phys. Rev. D}\ }\textbf {\bibinfo {volume} {88}},\ \bibinfo
  {pages} {042002} (\bibinfo {year} {2013})},\ \Eprint
  {https://arxiv.org/abs/1304.7017} {arXiv:1304.7017 [gr-qc]} \BibitemShut
  {NoStop}%
\bibitem [{\citenamefont {Usman}\ \emph {et~al.}(2019)\citenamefont {Usman},
  \citenamefont {Mills},\ and\ \citenamefont {Fairhurst}}]{Usman:2018imj}%
  \BibitemOpen
  \bibfield  {author} {\bibinfo {author} {\bibfnamefont {S.~A.}\ \bibnamefont
  {Usman}}, \bibinfo {author} {\bibfnamefont {J.~C.}\ \bibnamefont {Mills}},\
  and\ \bibinfo {author} {\bibfnamefont {S.}~\bibnamefont {Fairhurst}},\
  }\bibfield  {title} {\bibinfo {title} {{Constraining the Inclinations of
  Binary Mergers from Gravitational-wave Observations}},\ }\href
  {https://doi.org/10.3847/1538-4357/ab0b3e} {\bibfield  {journal} {\bibinfo
  {journal} {Astrophys. J.}\ }\textbf {\bibinfo {volume} {877}},\ \bibinfo
  {pages} {82} (\bibinfo {year} {2019})},\ \Eprint
  {https://arxiv.org/abs/1809.10727} {arXiv:1809.10727 [gr-qc]} \BibitemShut
  {NoStop}%
\bibitem [{\citenamefont {Mills}\ and\ \citenamefont
  {Fairhurst}(2021)}]{Mills:2020thr}%
  \BibitemOpen
  \bibfield  {author} {\bibinfo {author} {\bibfnamefont {C.}~\bibnamefont
  {Mills}}\ and\ \bibinfo {author} {\bibfnamefont {S.}~\bibnamefont
  {Fairhurst}},\ }\bibfield  {title} {\bibinfo {title} {{Measuring
  gravitational-wave higher-order multipoles}},\ }\href
  {https://doi.org/10.1103/PhysRevD.103.024042} {\bibfield  {journal} {\bibinfo
   {journal} {Phys. Rev. D}\ }\textbf {\bibinfo {volume} {103}},\ \bibinfo
  {pages} {024042} (\bibinfo {year} {2021})},\ \Eprint
  {https://arxiv.org/abs/2007.04313} {arXiv:2007.04313 [gr-qc]} \BibitemShut
  {NoStop}%
\bibitem [{\citenamefont {{Cutler}}\ and\ \citenamefont
  {{Flanagan}}(1994)}]{1994PhRvD..49.2658C}%
  \BibitemOpen
  \bibfield  {author} {\bibinfo {author} {\bibfnamefont {C.}~\bibnamefont
  {{Cutler}}}\ and\ \bibinfo {author} {\bibfnamefont {{\'E}.~E.}\ \bibnamefont
  {{Flanagan}}},\ }\bibfield  {title} {\bibinfo {title} {{Gravitational waves
  from merging compact binaries: How accurately can one extract the binary's
  parameters from the inspiral waveform?}},\ }\href
  {https://doi.org/10.1103/PhysRevD.49.2658} {\bibfield  {journal} {\bibinfo
  {journal} {\prd}\ }\textbf {\bibinfo {volume} {49}},\ \bibinfo {pages} {2658}
  (\bibinfo {year} {1994})},\ \Eprint {https://arxiv.org/abs/gr-qc/9402014}
  {arXiv:gr-qc/9402014 [gr-qc]} \BibitemShut {NoStop}%
\bibitem [{\citenamefont {Cutler}(1998)}]{Cutler:1997ta}%
  \BibitemOpen
  \bibfield  {author} {\bibinfo {author} {\bibfnamefont {C.}~\bibnamefont
  {Cutler}},\ }\bibfield  {title} {\bibinfo {title} {{Angular resolution of the
  LISA gravitational wave detector}},\ }\href
  {https://doi.org/10.1103/PhysRevD.57.7089} {\bibfield  {journal} {\bibinfo
  {journal} {Phys. Rev.}\ }\textbf {\bibinfo {volume} {D57}},\ \bibinfo {pages}
  {7089} (\bibinfo {year} {1998})},\ \Eprint
  {https://arxiv.org/abs/gr-qc/9703068} {arXiv:gr-qc/9703068 [gr-qc]}
  \BibitemShut {NoStop}%
\bibitem [{\citenamefont {Vallisneri}(2008)}]{Vallisneri:2007ev}%
  \BibitemOpen
  \bibfield  {author} {\bibinfo {author} {\bibfnamefont {M.}~\bibnamefont
  {Vallisneri}},\ }\bibfield  {title} {\bibinfo {title} {{Use and abuse of the
  Fisher information matrix in the assessment of gravitational-wave
  parameter-estimation prospects}},\ }\href
  {https://doi.org/10.1103/PhysRevD.77.042001} {\bibfield  {journal} {\bibinfo
  {journal} {Phys. Rev.}\ }\textbf {\bibinfo {volume} {D77}},\ \bibinfo {pages}
  {042001} (\bibinfo {year} {2008})},\ \Eprint
  {https://arxiv.org/abs/gr-qc/0703086} {arXiv:gr-qc/0703086 [GR-QC]}
  \BibitemShut {NoStop}%
\bibitem [{\citenamefont {Kuns}\ \emph {et~al.}(2019)\citenamefont {Kuns},
  \citenamefont {Yu}, \citenamefont {Chen},\ and\ \citenamefont
  {Adhikari}}]{Kuns:2019upi}%
  \BibitemOpen
  \bibfield  {author} {\bibinfo {author} {\bibfnamefont {K.~A.}\ \bibnamefont
  {Kuns}}, \bibinfo {author} {\bibfnamefont {H.}~\bibnamefont {Yu}}, \bibinfo
  {author} {\bibfnamefont {Y.}~\bibnamefont {Chen}},\ and\ \bibinfo {author}
  {\bibfnamefont {R.~X.}\ \bibnamefont {Adhikari}},\ }\bibfield  {title}
  {\bibinfo {title} {{Astrophysics and cosmology with a deci-hertz
  gravitational-wave detector: TianGO}},\ }\href@noop {} {\  (\bibinfo {year}
  {2019})},\ \Eprint {https://arxiv.org/abs/1908.06004} {arXiv:1908.06004
  [gr-qc]} \BibitemShut {NoStop}%
\bibitem [{\citenamefont {Garc\'\i{}a-Quir\'os}\ \emph
  {et~al.}(2020)\citenamefont {Garc\'\i{}a-Quir\'os}, \citenamefont {Colleoni},
  \citenamefont {Husa}, \citenamefont {Estell\'es}, \citenamefont {Pratten},
  \citenamefont {Ramos-Buades}, \citenamefont {Mateu-Lucena},\ and\
  \citenamefont {Jaume}}]{Garcia-Quiros:2020qpx}%
  \BibitemOpen
  \bibfield  {author} {\bibinfo {author} {\bibfnamefont {C.}~\bibnamefont
  {Garc\'\i{}a-Quir\'os}}, \bibinfo {author} {\bibfnamefont {M.}~\bibnamefont
  {Colleoni}}, \bibinfo {author} {\bibfnamefont {S.}~\bibnamefont {Husa}},
  \bibinfo {author} {\bibfnamefont {H.}~\bibnamefont {Estell\'es}}, \bibinfo
  {author} {\bibfnamefont {G.}~\bibnamefont {Pratten}}, \bibinfo {author}
  {\bibfnamefont {A.}~\bibnamefont {Ramos-Buades}}, \bibinfo {author}
  {\bibfnamefont {M.}~\bibnamefont {Mateu-Lucena}},\ and\ \bibinfo {author}
  {\bibfnamefont {R.}~\bibnamefont {Jaume}},\ }\bibfield  {title} {\bibinfo
  {title} {{Multimode frequency-domain model for the gravitational wave signal
  from nonprecessing black-hole binaries}},\ }\href
  {https://doi.org/10.1103/PhysRevD.102.064002} {\bibfield  {journal} {\bibinfo
   {journal} {Phys. Rev. D}\ }\textbf {\bibinfo {volume} {102}},\ \bibinfo
  {pages} {064002} (\bibinfo {year} {2020})},\ \Eprint
  {https://arxiv.org/abs/2001.10914} {arXiv:2001.10914 [gr-qc]} \BibitemShut
  {NoStop}%
\bibitem [{\citenamefont {{LIGO Scientific Collaboration}}(2018)}]{lalsuite}%
  \BibitemOpen
  \bibfield  {author} {\bibinfo {author} {\bibnamefont {{LIGO Scientific
  Collaboration}}},\ }\href@noop {} {\bibinfo {title} {{LIGO} {A}lgorithm
  {L}ibrary - {LALS}uite}},\ \bibinfo {howpublished} {free software (GPL)}
  (\bibinfo {year} {2018}),\ \bibinfo {note} {doi:
  \href{https://doi.org/10.7935/GT1W-FZ16}{10.7935/GT1W-FZ16}}\BibitemShut
  {NoStop}%
\bibitem [{\citenamefont {Harris}\ \emph {et~al.}(2020)\citenamefont {Harris},
  \citenamefont {Millman}, \citenamefont {van~der Walt}, \citenamefont
  {Gommers}, \citenamefont {Virtanen}, \citenamefont {Cournapeau},
  \citenamefont {Wieser}, \citenamefont {Taylor}, \citenamefont {Berg},
  \citenamefont {Smith}, \citenamefont {Kern}, \citenamefont {Picus},
  \citenamefont {Hoyer}, \citenamefont {van Kerkwijk}, \citenamefont {Brett},
  \citenamefont {Haldane}, \citenamefont {del R{\'{i}}o}, \citenamefont
  {Wiebe}, \citenamefont {Peterson}, \citenamefont {G{\'{e}}rard-Marchant},
  \citenamefont {Sheppard}, \citenamefont {Reddy}, \citenamefont {Weckesser},
  \citenamefont {Abbasi}, \citenamefont {Gohlke},\ and\ \citenamefont
  {Oliphant}}]{harris2020array}%
  \BibitemOpen
  \bibfield  {author} {\bibinfo {author} {\bibfnamefont {C.~R.}\ \bibnamefont
  {Harris}}, \bibinfo {author} {\bibfnamefont {K.~J.}\ \bibnamefont {Millman}},
  \bibinfo {author} {\bibfnamefont {S.~J.}\ \bibnamefont {van~der Walt}},
  \bibinfo {author} {\bibfnamefont {R.}~\bibnamefont {Gommers}}, \bibinfo
  {author} {\bibfnamefont {P.}~\bibnamefont {Virtanen}}, \bibinfo {author}
  {\bibfnamefont {D.}~\bibnamefont {Cournapeau}}, \bibinfo {author}
  {\bibfnamefont {E.}~\bibnamefont {Wieser}}, \bibinfo {author} {\bibfnamefont
  {J.}~\bibnamefont {Taylor}}, \bibinfo {author} {\bibfnamefont
  {S.}~\bibnamefont {Berg}}, \bibinfo {author} {\bibfnamefont {N.~J.}\
  \bibnamefont {Smith}}, \bibinfo {author} {\bibfnamefont {R.}~\bibnamefont
  {Kern}}, \bibinfo {author} {\bibfnamefont {M.}~\bibnamefont {Picus}},
  \bibinfo {author} {\bibfnamefont {S.}~\bibnamefont {Hoyer}}, \bibinfo
  {author} {\bibfnamefont {M.~H.}\ \bibnamefont {van Kerkwijk}}, \bibinfo
  {author} {\bibfnamefont {M.}~\bibnamefont {Brett}}, \bibinfo {author}
  {\bibfnamefont {A.}~\bibnamefont {Haldane}}, \bibinfo {author} {\bibfnamefont
  {J.~F.}\ \bibnamefont {del R{\'{i}}o}}, \bibinfo {author} {\bibfnamefont
  {M.}~\bibnamefont {Wiebe}}, \bibinfo {author} {\bibfnamefont
  {P.}~\bibnamefont {Peterson}}, \bibinfo {author} {\bibfnamefont
  {P.}~\bibnamefont {G{\'{e}}rard-Marchant}}, \bibinfo {author} {\bibfnamefont
  {K.}~\bibnamefont {Sheppard}}, \bibinfo {author} {\bibfnamefont
  {T.}~\bibnamefont {Reddy}}, \bibinfo {author} {\bibfnamefont
  {W.}~\bibnamefont {Weckesser}}, \bibinfo {author} {\bibfnamefont
  {H.}~\bibnamefont {Abbasi}}, \bibinfo {author} {\bibfnamefont
  {C.}~\bibnamefont {Gohlke}},\ and\ \bibinfo {author} {\bibfnamefont {T.~E.}\
  \bibnamefont {Oliphant}},\ }\bibfield  {title} {\bibinfo {title} {Array
  programming with {NumPy}},\ }\href
  {https://doi.org/10.1038/s41586-020-2649-2} {\bibfield  {journal} {\bibinfo
  {journal} {Nature}\ }\textbf {\bibinfo {volume} {585}},\ \bibinfo {pages}
  {357} (\bibinfo {year} {2020})}\BibitemShut {NoStop}%
\bibitem [{\citenamefont {Virtanen}\ \emph {et~al.}(2020)\citenamefont
  {Virtanen}, \citenamefont {Gommers}, \citenamefont {Oliphant}, \citenamefont
  {Haberland}, \citenamefont {Reddy}, \citenamefont {Cournapeau}, \citenamefont
  {Burovski}, \citenamefont {Peterson}, \citenamefont {Weckesser},
  \citenamefont {Bright}, \citenamefont {{van der Walt}}, \citenamefont
  {Brett}, \citenamefont {Wilson}, \citenamefont {Millman}, \citenamefont
  {Mayorov}, \citenamefont {Nelson}, \citenamefont {Jones}, \citenamefont
  {Kern}, \citenamefont {Larson}, \citenamefont {Carey}, \citenamefont {Polat},
  \citenamefont {Feng}, \citenamefont {Moore}, \citenamefont {{VanderPlas}},
  \citenamefont {Laxalde}, \citenamefont {Perktold}, \citenamefont {Cimrman},
  \citenamefont {Henriksen}, \citenamefont {Quintero}, \citenamefont {Harris},
  \citenamefont {Archibald}, \citenamefont {Ribeiro}, \citenamefont
  {Pedregosa}, \citenamefont {{van Mulbregt}},\ and\ \citenamefont {{SciPy 1.0
  Contributors}}}]{2020SciPy-NMeth}%
  \BibitemOpen
  \bibfield  {author} {\bibinfo {author} {\bibfnamefont {P.}~\bibnamefont
  {Virtanen}}, \bibinfo {author} {\bibfnamefont {R.}~\bibnamefont {Gommers}},
  \bibinfo {author} {\bibfnamefont {T.~E.}\ \bibnamefont {Oliphant}}, \bibinfo
  {author} {\bibfnamefont {M.}~\bibnamefont {Haberland}}, \bibinfo {author}
  {\bibfnamefont {T.}~\bibnamefont {Reddy}}, \bibinfo {author} {\bibfnamefont
  {D.}~\bibnamefont {Cournapeau}}, \bibinfo {author} {\bibfnamefont
  {E.}~\bibnamefont {Burovski}}, \bibinfo {author} {\bibfnamefont
  {P.}~\bibnamefont {Peterson}}, \bibinfo {author} {\bibfnamefont
  {W.}~\bibnamefont {Weckesser}}, \bibinfo {author} {\bibfnamefont
  {J.}~\bibnamefont {Bright}}, \bibinfo {author} {\bibfnamefont {S.~J.}\
  \bibnamefont {{van der Walt}}}, \bibinfo {author} {\bibfnamefont
  {M.}~\bibnamefont {Brett}}, \bibinfo {author} {\bibfnamefont
  {J.}~\bibnamefont {Wilson}}, \bibinfo {author} {\bibfnamefont {K.~J.}\
  \bibnamefont {Millman}}, \bibinfo {author} {\bibfnamefont {N.}~\bibnamefont
  {Mayorov}}, \bibinfo {author} {\bibfnamefont {A.~R.~J.}\ \bibnamefont
  {Nelson}}, \bibinfo {author} {\bibfnamefont {E.}~\bibnamefont {Jones}},
  \bibinfo {author} {\bibfnamefont {R.}~\bibnamefont {Kern}}, \bibinfo {author}
  {\bibfnamefont {E.}~\bibnamefont {Larson}}, \bibinfo {author} {\bibfnamefont
  {C.~J.}\ \bibnamefont {Carey}}, \bibinfo {author} {\bibfnamefont
  {{\.I}.}~\bibnamefont {Polat}}, \bibinfo {author} {\bibfnamefont
  {Y.}~\bibnamefont {Feng}}, \bibinfo {author} {\bibfnamefont {E.~W.}\
  \bibnamefont {Moore}}, \bibinfo {author} {\bibfnamefont {J.}~\bibnamefont
  {{VanderPlas}}}, \bibinfo {author} {\bibfnamefont {D.}~\bibnamefont
  {Laxalde}}, \bibinfo {author} {\bibfnamefont {J.}~\bibnamefont {Perktold}},
  \bibinfo {author} {\bibfnamefont {R.}~\bibnamefont {Cimrman}}, \bibinfo
  {author} {\bibfnamefont {I.}~\bibnamefont {Henriksen}}, \bibinfo {author}
  {\bibfnamefont {E.~A.}\ \bibnamefont {Quintero}}, \bibinfo {author}
  {\bibfnamefont {C.~R.}\ \bibnamefont {Harris}}, \bibinfo {author}
  {\bibfnamefont {A.~M.}\ \bibnamefont {Archibald}}, \bibinfo {author}
  {\bibfnamefont {A.~H.}\ \bibnamefont {Ribeiro}}, \bibinfo {author}
  {\bibfnamefont {F.}~\bibnamefont {Pedregosa}}, \bibinfo {author}
  {\bibfnamefont {P.}~\bibnamefont {{van Mulbregt}}},\ and\ \bibinfo {author}
  {\bibnamefont {{SciPy 1.0 Contributors}}},\ }\bibfield  {title} {\bibinfo
  {title} {{{SciPy} 1.0: Fundamental Algorithms for Scientific Computing in
  Python}},\ }\href {https://doi.org/10.1038/s41592-019-0686-2} {\bibfield
  {journal} {\bibinfo  {journal} {Nature Methods}\ }\textbf {\bibinfo {volume}
  {17}},\ \bibinfo {pages} {261} (\bibinfo {year} {2020})}\BibitemShut
  {NoStop}%
\bibitem [{\citenamefont {pandas~development team}(2020)}]{pandas}%
  \BibitemOpen
  \bibfield  {author} {\bibinfo {author} {\bibfnamefont {T.}~\bibnamefont
  {pandas~development team}},\ }\href {https://doi.org/10.5281/zenodo.3509134}
  {\bibinfo {title} {pandas-dev/pandas: Pandas}} (\bibinfo {year}
  {2020})\BibitemShut {NoStop}%
\bibitem [{\citenamefont {Hunter}(2007)}]{Hunter:2007ouj}%
  \BibitemOpen
  \bibfield  {author} {\bibinfo {author} {\bibfnamefont {J.~D.}\ \bibnamefont
  {Hunter}},\ }\bibfield  {title} {\bibinfo {title} {{Matplotlib: A 2D Graphics
  Environment}},\ }\href {https://doi.org/10.1109/MCSE.2007.55} {\bibfield
  {journal} {\bibinfo  {journal} {Comput. Sci. Eng.}\ }\textbf {\bibinfo
  {volume} {9}},\ \bibinfo {pages} {90} (\bibinfo {year} {2007})}\BibitemShut
  {NoStop}%
\bibitem [{\citenamefont {Lewis}(2019)}]{Lewis:2019xzd}%
  \BibitemOpen
  \bibfield  {author} {\bibinfo {author} {\bibfnamefont {A.}~\bibnamefont
  {Lewis}},\ }\bibfield  {title} {\bibinfo {title} {{GetDist: a Python package
  for analysing Monte Carlo samples}},\ }\href@noop {} {\  (\bibinfo {year}
  {2019})},\ \Eprint {https://arxiv.org/abs/1910.13970} {arXiv:1910.13970
  [astro-ph.IM]} \BibitemShut {NoStop}%
\bibitem [{\citenamefont {Wette}(2020)}]{swiglal}%
  \BibitemOpen
  \bibfield  {author} {\bibinfo {author} {\bibfnamefont {K.}~\bibnamefont
  {Wette}},\ }\bibfield  {title} {\bibinfo {title} {{SWIGLAL: Python and Octave
  interfaces to the LALSuite gravitational-wave data analysis libraries}},\
  }\href {https://doi.org/10.1016/j.softx.2020.100634} {\bibfield  {journal}
  {\bibinfo  {journal} {SoftwareX}\ }\textbf {\bibinfo {volume} {12}},\
  \bibinfo {pages} {100634} (\bibinfo {year} {2020})}\BibitemShut {NoStop}%
\bibitem [{\citenamefont {Nitz}\ \emph {et~al.}(2024)\citenamefont {Nitz},
  \citenamefont {Harry}, \citenamefont {Brown}, \citenamefont {Biwer},
  \citenamefont {Willis}, \citenamefont {Canton}, \citenamefont {Capano},
  \citenamefont {Dent}, \citenamefont {Pekowsky}, \citenamefont {Davies},
  \citenamefont {De}, \citenamefont {Cabero}, \citenamefont {Wu}, \citenamefont
  {Williamson}, \citenamefont {Machenschalk}, \citenamefont {Macleod},
  \citenamefont {Pannarale}, \citenamefont {Kumar}, \citenamefont {Reyes},
  \citenamefont {dfinstad}, \citenamefont {Kumar}, \citenamefont {Tápai},
  \citenamefont {Singer}, \citenamefont {Kumar}, \citenamefont {veronica
  villa}, \citenamefont {maxtrevor}, \citenamefont {Gadre}, \citenamefont
  {Khan}, \citenamefont {Fairhurst},\ and\ \citenamefont
  {Tolley}}]{alex_nitz_2024_10473621}%
  \BibitemOpen
  \bibfield  {author} {\bibinfo {author} {\bibfnamefont {A.}~\bibnamefont
  {Nitz}}, \bibinfo {author} {\bibfnamefont {I.}~\bibnamefont {Harry}},
  \bibinfo {author} {\bibfnamefont {D.}~\bibnamefont {Brown}}, \bibinfo
  {author} {\bibfnamefont {C.~M.}\ \bibnamefont {Biwer}}, \bibinfo {author}
  {\bibfnamefont {J.}~\bibnamefont {Willis}}, \bibinfo {author} {\bibfnamefont
  {T.~D.}\ \bibnamefont {Canton}}, \bibinfo {author} {\bibfnamefont
  {C.}~\bibnamefont {Capano}}, \bibinfo {author} {\bibfnamefont
  {T.}~\bibnamefont {Dent}}, \bibinfo {author} {\bibfnamefont {L.}~\bibnamefont
  {Pekowsky}}, \bibinfo {author} {\bibfnamefont {G.~S.~C.}\ \bibnamefont
  {Davies}}, \bibinfo {author} {\bibfnamefont {S.}~\bibnamefont {De}}, \bibinfo
  {author} {\bibfnamefont {M.}~\bibnamefont {Cabero}}, \bibinfo {author}
  {\bibfnamefont {S.}~\bibnamefont {Wu}}, \bibinfo {author} {\bibfnamefont
  {A.~R.}\ \bibnamefont {Williamson}}, \bibinfo {author} {\bibfnamefont
  {B.}~\bibnamefont {Machenschalk}}, \bibinfo {author} {\bibfnamefont
  {D.}~\bibnamefont {Macleod}}, \bibinfo {author} {\bibfnamefont
  {F.}~\bibnamefont {Pannarale}}, \bibinfo {author} {\bibfnamefont
  {P.}~\bibnamefont {Kumar}}, \bibinfo {author} {\bibfnamefont
  {S.}~\bibnamefont {Reyes}}, \bibinfo {author} {\bibnamefont {dfinstad}},
  \bibinfo {author} {\bibfnamefont {S.}~\bibnamefont {Kumar}}, \bibinfo
  {author} {\bibfnamefont {M.}~\bibnamefont {Tápai}}, \bibinfo {author}
  {\bibfnamefont {L.}~\bibnamefont {Singer}}, \bibinfo {author} {\bibfnamefont
  {P.}~\bibnamefont {Kumar}}, \bibinfo {author} {\bibnamefont {veronica
  villa}}, \bibinfo {author} {\bibnamefont {maxtrevor}}, \bibinfo {author}
  {\bibfnamefont {B.~U.~V.}\ \bibnamefont {Gadre}}, \bibinfo {author}
  {\bibfnamefont {S.}~\bibnamefont {Khan}}, \bibinfo {author} {\bibfnamefont
  {S.}~\bibnamefont {Fairhurst}},\ and\ \bibinfo {author} {\bibfnamefont
  {A.}~\bibnamefont {Tolley}},\ }\href
  {https://doi.org/10.5281/zenodo.10473621} {\bibinfo {title} {gwastro/pycbc:
  v2.3.3 release of pycbc}} (\bibinfo {year} {2024})\BibitemShut {NoStop}%
\end{thebibliography}%

\end{document}